\documentclass{aa}  

\usepackage{bm}
\usepackage{subfig}
\usepackage{rotating}
\usepackage{amssymb,amsmath}
\usepackage{graphicx}
\usepackage{hyperref}
\usepackage{placeins}

\usepackage{txfonts}
\usepackage{color}

\begin{document} 

        \titlerunning{SHARDDS Survey}
        \authorrunning{Dahlqvist et al.}
   \title{The SHARDDS survey: limits on planet occurrence rates based on point sources analysis via the Auto-RSM framework\thanks{Based on observations collected at the European Southern Observatory under ESO programmes 096.C-0388(A) and 097.C-0394(A)} }
   \author{C.-H. Dahlqvist\inst{1}, J. Milli\inst{2}, O. Absil\inst{1}\fnmsep\thanks{F.R.S.-FNRS Senior Research Associate}, F. Cantalloube\inst{3}, L. Matra\inst{8}, E. Choquet\inst{3}, C. del Burgo\inst{9}, J. P. Marshall\inst{4,5}, M.~Wyatt\inst{10}, S. Ertel\inst{6,7}}

   \institute{STAR Institute, Universit\'{e} de Li\`{e}ge, All\'{e}e du Six Ao\^{u}t 19c, 4000 Li\`{e}ge, Belgium\\
              \email{carl-henrik.dahlqvist@uliege.be} \and Universit\'{e} Grenoble-Alpes, CNRS, IPAG F-38000 Grenoble, France \and Aix Marseille Univ, CNRS, CNES, LAM, Marseille, France \and Academia Sinica, Institute of Astronomy and Astrophysics, 11F Astronomy-Mathematics Building, NTU/AS campus, No. 1, Section 4, Roosevelt Rd., Taipei 10617, Taiwan \and Centre for Astrophysics, University of Southern Queensland, Toowoomba, QLD 4350, Australia \and Large Binocular Telescope Observatory, University of Arizona, 933 North Cherry Avenue, Tucson, AZ 85721, USA \and Steward Observatory, Department of Astronomy, University of Arizona, 993 N. Cherry Ave, Tucson, AZ, 85721, US \and School of Physics, Trinity College Dublin, the University of Dublin, College Green, Dublin 2, Ireland  \and Instituto Nacional de Astrof\'{\i}sica, \'Optica y Electr\'onica, Luis Enrique Erro 1, Sta. Ma. Tonantzintla, Puebla, Mexico \and Institute of Astronomy, University of Cambridge, Madingley Road, Cambridge CB3 0HA, UK}
   \date{}

\abstract
   {In the past decade, high contrast imaging allowed the detection and characterisation of exoplanets, brown dwarfs, and circumstellar disks. Large surveys provided new insights about the frequency and properties of massive sub-stellar companions with separations from 5 to 300 au. }
   %These surveys relied on new-generation high-contrast imaging instruments to bring new clues about planetary formation and evolution mechanisms.
   {In this context, our study aims to detect and characterise potential exoplanets and brown dwarfs within debris disks, considering a diverse population of stars with respect to stellar age and spectral type. We present in this paper the analysis of a set of H-band images taken by the VLT/SPHERE instrument in the context of the SHARDDS survey. This survey gathers 55 main-sequence stars within 100\,pc, known to host a high-infrared-excess debris disk, allowing us to potentially better understand the complex interactions between substellar companions and disks.}
   {We rely on the Auto-RSM framework to perform an in-depth analysis of the considered targets, via the computation of detection maps and contrast curves. A clustering approach is used to divide the set of targets into multiple subsets, in order to reduce the computation time by estimating a single optimal parametrisation for each considered subset. Detection maps generated with different approaches are used along with contrast curves to identify potential planetary companions. Planet detection and planet occurrence frequencies are derived from the generated contrast curves, relying on two well-known evolutionary models, namely AMES-DUSTY and AMES-COND. Finally, we study the influence of the observing conditions and observing sequence characteristics on the performance measured in terms of contrast.}
   {The use of Auto-RSM allows us to reach high contrast at short separations, with a median contrast of $10^{-5}$ at 300 mas, for a completeness level of 95\%. A new planetary characterisation algorithm, based on the RSM framework, is developed and tested successfully, showing a higher astrometric and photometric precision for faint sources compared to standard approaches. Apart from the already known companion of HD206893 and two point-like sources around HD114082 which are most likely background stars, we did not detect any new companion around other stars. A correlation study between achievable contrasts and parameters characterising high contrast imaging sequences highlights the importance of the Strehl ratio, wind speed at a height of 30 meters, and presence of wind-driven halo to define the quality of high contrast images. Finally, planet detection and occurrence rate maps are generated and show, for the SHARDDS survey, a high sensitivity between 10 and 100 au for substellar companions with masses >10$M_J$. }
   {}


   \maketitle
%
%-------------------------------------------------------------------

\section{Introduction}
\label{sec:intro}

In our current understanding of planetary system formation, gas giant planets form in gas-rich protoplanetary disks that dissipate in a few million years \cite[e.g.][]{Williams2011}, leaving behind one or several planets as well as belts of smaller rocky bodies that never managed to grow to full-sized planets \cite[e.g. ][]{Krivov2010}. These belts, also known as debris disks because of their collisional activity, are composed of all sub-planetary rocky bodies, ranging from kilometer-sized planetesimals to micron-sized dust \cite[see for example][for a review]{Wyatt2008}. These dust particles are detectable by their reflected light or thermal emission, creating an infrared excess above the stellar photosphere. Current far-infrared surveys can detect debris disks with an infrared excess above $10^{-6}$ and identify debris disks in around 30\% of A stars and 20\% of FGK stars \cite[e.g. ][]{Eiroa13}, but the real occurrence rate could be much higher \citep{Pawellek2021}. 
%However, the asteroid belt or the Edgeworth-Kuiper belt have an infrared excess more than order of magnitude below the current sensitivity, therefore any planetary system might have a debris disk.  
Those disks are a natural place to look for exoplanets because planet formation succeeded at least to form large planetesimals in those systems. This is one of the reasons why direct imaging surveys generally include many debris disk host stars, such as in the SPHERE-SHINE survey \citep[SPHERE infrared survey for exoplanets, ][]{Desidera2021} or the GPI-GPIES survey \citep[Gemini Planet Imager Exoplanet Survey, ][]{Nielsen2019}. \citet{Meshkat17} found indeed a tentative evidence that giant planets have a higher occurrence rate in debris disks hosts, and the first emblematic directly imaged planets were found in the massive debris disks system $\beta$ Pic \citep{Lagrange2009} or HR\,8799 \citep{Marois2008}. Following this strategy, we present in this study a direct imaging survey of a sample of 55 main-sequence stars hosting high-infrared excess debris disks: the SPHERE High-Angular Resolution Debris Disks Survey (SHARDDS). This survey already revealed debris disks resolved for the first time in scattered light: HD\,114082 \citep{Wahhaj16,Engler2022}, 49 Ceti \citet{Choquet2017}, HD\,105 \citep{Marshall2018} as well as a substellar companion (HD\,206893 B) close to the deuterium burning limit \citep{Milli16b,Delorme2017_206,Romero2021}. Here, we use the homogeneous observations made in the context of this high-contrast survey to search for companions with the Regime Switching Model (hereafter RSM) post-processing algorithm \citep{Dahlqvist20} and provide detection maps and contrast curves. 

The RSM method focuses on the detection of point sources within high-contrast images, by making use of the angular diversity introduced via pupil tracking mode observations. The concept behind RSM is to model the spatio-temporal evolution of the pixel intensities contained in the cubes of residuals generated by several PSF-subtraction techniques. As each PSF-subtraction technique models the speckle field differently, combining multiple techniques helps to average out residual speckle noise while preserving potential planetary signals. The RSM approach relies on a two-state Markov chain to model annulus-wise the pixel intensities and estimate the probability to be either in a speckle noise regime or a planetary regime. The probability associated to the planetary regime is then used to compute a detection map. Compared to other state-of-the-art post-processing methods dedicated to high-contrast imaging, the \textit{Exoplanet Imaging Data Challenge}\footnote{\url{https://exoplanet-imaging-challenge.github.io/}} has shown that the RSM technique has a very low false positive rate and is among the best algorithms in terms of detection capabilities\citep{Cantalloube20}. 

More recently, the Auto-RSM framework \citep{Dahlqvist21b} was developed to reduce the burden of parameter selection and further optimise the performance of the RSM algorithm. This optimisation framework consists of three main steps: (i) the definition of the optimal set of parameters for the PSF-subtraction techniques, (ii) the optimisation of the RSM algorithm, and (iii) the selection of the optimal set of PSF-subtraction techniques and ADI sequences \citep[Angular Differential Imaging, ][]{Marois08} used to generate the final RSM probability map. The Auto-RSM framework being computationally expensive, a clustering approach is used to divide the set of targets into multiple subsets. For each subset, the cluster center is identified and the Auto-RSM framework is applied onto it to provide the optimal parametrisation for the entire cluster. The obtained optimal parametrisations are also compared to unveil potential commonalities and understand their relationship with the ADI sequence characteristics.

Detection maps are then computed via the RSM approach, relying on these optimal parametrisations. The detection maps are used to identify potential planetary companions, and a new companion characterisation framework based on the RSM approach is introduced. The detection maps are also used to compute contrast curves, which are used together to estimate detection probability maps and occurrence rate maps, based on well-known evolutionary models. The relationship existing between reachable contrasts and parameters characterising HCI observing sequences is also investigated.

The remainder of this paper is organised as follows. Section 2 describes the target selection for the SHARDDS survey. In Section~3, we present our data reduction pipeline involving the definition of clusters along with cluster centres on which the Auto-RSM optimisation procedure is applied. The computation of detection maps and contrast curves follows the estimation of the optimal parametrisations. Section 4 is devoted to the characterisation of potential planetary candidates. In Section~5, we consider the contrast curve as a performance metric and analyse the potential drivers of this performance. Section 6 focuses on the estimation of the planetary detection probability from which we derive an estimated planetary occurrence rate associated to the SHARDDS survey. Finally, Section~7 concludes this work.

\section{Survey description}
\label{sec:Survey}

The SHARDDS survey was designed to image circumstellar disks around bright nearby stars (within 100 pc from the Earth) in the near-infrared using the VLT/SPHERE instrument \citep[Very Large Telescope/Spectro-Polarimetric High-contrast Exoplanet REsearch,][]{Beuzit19}. The aim of the survey is to better understand the scarcity of debris disks detection in scattered light, by targeting disks without any scattered-light detection at the time of the survey design (2014), either because the target was not observed with high-contrast instruments, or because the disk might be too compact and faint to be accessible with first-generation high-contrast instruments such as HST/NICMOS \citep[Hubble Space Telescope/Near Infrared Camera and Multi-Object Spectrometer, ][]{Thompson98} or VLT/NaCo \citep[Very Large Telescope/Nasmyth Adaptive Optics System Near-Infrared Imager and Spectrograph, ][]{Lenzen03,Rousset2003}, having poor performance below $0.5"$. The underlying goals are to characterise the disks architecture and properties, and statistically link these properties to the stellar age, spectral type, and potential presence of companions. This paper contributes to the achievement of these objectives by applying the RSM detection algorithm \citep{Dahlqvist21b} on the datasets, to detect potential planetary candidates. The RSM detection algorithm was designed to unveil point-like sources and is therefore not fitted to detect extended features such as debris disks. The detection of companions can bring valuable information to better understand the secular interactions between debris disks and companions, and whether such interactions are always needed to explain particular signatures in disks such as azimuthal asymmetries, warps or sharp edges \citep[see ][ for emblematic examples of signatures within debris disks attributable to a companion]{Mouillet97,Lagrange12,Lestrade15}.

The SHARDDS survey includes 55 main-sequence stars visible from the Southern hemisphere, covering spectral types A-M and ages 10 Myr - 6 Gyr . This diverse sample of debris systems aims to provide a comprehensive view of planetary system properties and their time evolution. These stars were selected for the expected brightness of their disks (fractional luminosity above $10^{-4}$) and because they were not yet resolved in scattered light. All stars that were not observable from Paranal with an airmass below 2, were excluded from the sample. The SPHERE/IRDIS instrument \citep[InfraRed Dual-band Imager and Spectrograph]{Dohlen08} was used with the broad-band H filter ($\lambda=1.625\mu m, \Delta \lambda=0.290\mu m$), as well as an apodised Lyot coronagraph with a radius of 92 mas (N\_ALC\_YJH\_S) to reach a high contrast in the innermost regions. The broad-band H filter was selected for its wide spectral band-pass allowing to collect more disk photons, but also because the performance of the extreme adaptive optics system improves at longer wavelengths and the dust from debris disks typically displays a red colour, while the thermal background is not as high as in the K band and does not dominate the noise budget at large separations. The observations were made in pupil-stabilised mode, using the Angular Differential Imaging observing strategy. The targets were observed around meridian passage to ensure a large rotation of the field of view, with about 40 minutes long coronagraphic images. The observations were grouped in two programs, 46 sources were imaged during P96 (1 October 2015 - 31 March 2016) and 9 during P97 (1 April 2016 - 30 September 2016). Due to adverse observing conditions, multiple observation sessions were required for some targets, leading to an actual dataset of 73 ADI sequences. Table \ref{Datasets} provides details on the set of targets, including the number of observation sequences acquired for each target (epoch). The distances, magnitudes, and spectral types were taken from the Hippparcos and GAIA catalogues \citep{HIPPARCOS,GAIA}. The target Fomalhaut C, part of the SHARDDS sample, was excluded from our analysis because of poor observing conditions for all three epochs \cite[dataset published in][]{Cronin-Coltsmann2021}.

\begin{table*}[!htbp]
                        \caption{Name, coordinates, magnitude distribution, spectral-type, age and distance, along with the number of ADI sequences for each SHARDDS target. }
                        \label{Datasets}
\centering
\footnotesize
                        \begin{tabular}{lcccccccc}
                        
                        \hline
Name	&	RA	&	DEC	&	V mag	&	H mag	&	Sp. type	&	Age (My)	&	Distance (pc)	&	\# Epochs	\\
 \hline
HD\,105	&	00:05:53	&	-41:45:11	&	7.53	&	6.19	&	G0V	&	30$^1$	&	38.85	&	1	\\
HD\,203	&	00:06:50	&	-23:06:27	&	6.17	&	5.33	&	F3V	&	23$^2$	&	39.97	&	1	\\
HD\,377	&	00:08:26	&	+06:37:00	&	7.59	&	6.15	&	G2V	&	170$^3$	&	38.52	&	1	\\
HD\,3003	&	00:32:44	&	-63:01:53	&	5.09	&	5.16	&	A0V	&	30$^1$	&	45.89	&	1	\\
HD\,3670	&	00:38:57	&	-52:32:03	&	8.21	&	7.15	&	F5V	&	30$^4$	&	77.58	&	1	\\
HD\,9672	&	01:34:38	&	-15:40:34	&	5.61	&	5.53	&	A1V	&	40$^6$	&	57.08	&	1	\\
HD\,10472	&	01:40:24	&	-60:59:56	&	7.61	&	6.69	&	F2IV/V	&	30$^7$	&	71.17	&	2	\\
HD\,10638	&	01:44:23	&	+32:30:57	&	6.73	&	6.19	&	A3	&	100$^8$	&	68.68	&	1	\\
HD\,13246	&	02:07:26	&	-59:40:45	&	7.50	&	6.30	&	F7V	&	40$^9$	&	45.60	&	1	\\
HD\,14082B	&	02:17:25	&	+28:44:30	&	7.74	&	6.36	&	G2V	&	21$^9$	&	39.75	&	1	\\
AG-Tri	&	02:27:29	&	+30:58:24	&	10.12	&	7.24	&	K8	&	23$^1$	&	41.05	&	4	\\
HD\,15257	&	02:28:10	&	+29:40:09	&	5.29	&	4.82	&	F0III	&	1000$^8$	&	49.93	&	1	\\
HD\,16743	&	02:39:08	&	-52:56:05	&	6.77	&	5.97	&	F1III/IV	&	200$^8$	&	57.94	&	1	\\
HD\,17390	&	02:46:45	&	-21:38:22	&	6.47	&	5.63	&	F3IV/V	&	610$^{10}$	&	48.19	&	1	\\
HD\,21997	&	03:31:54	&	-25:36:50	&	6.37	&	6.12	&	A3IV/V	&	30$^{11}$	&	69.64	&	1	\\
HD\,22179	&	03:35:30	&	+31:13:37	&	8.93	&	7.49	&	G5IV	&	63$^{12}$	&	70.37	&	1	\\
HD\,24636	&	03:48:11	&	-74:41:38	&	7.13	&	6.22	&	F3IV/V	&	30$^{13}$	&	57.05	&	1	\\
HD\,25457	&	04:02:37	&	-00:16:08	&	5.38	&	4.34	&	F6V	&	70$^{13}$	&	18.77	&	1	\\
HD\,31392	&	04:54:04	&	-35:24:16	&	7.61	&	5.89	&	G9V	&	3690$^{10}$	&	25.77	&	1	\\
HD\,35650	&	05:24:30	&	-38:58:10	&	9.05	&	6.11	&	K6V	&	70$^1$	&	17.48	&	1	\\
HD\,274255	&	05:30:14	&	-42:41:50	&	9.71	&	6.47	&	M0V	&	1000$^{14}$	&	19.15	&	1	\\
HD\,37484	&	05:37:40	&	-28:37:34	&	7.25	&	6.29	&	F3V	&	30$^{15}$	&	59.10	&	2	\\
HD\,38207	&	05:43:21	&	-20:11:21	&	8.47	&	7.55	&	F2V	&	534$^{16}$	&	110.99	&	1	\\
HD\,38206	&	05:43:22	&	-18:33:26	&	5.73	&	5.84	&	A0V	&	30 $^{15}$	&	71.43	&	2	\\
HD\,40540	&	05:57:53	&	-34:28:34	&	7.54	&	6.93	&	A8IV	&	170$^3$	&	88.26	&	1	\\
HD\,53842	&	06:46:14	&	-83:59:29	&	8.62	&	6.40	&	F5V	&	30$^{17}$	&	57.87	&	1	\\
HD\,60491	&	07:34:26	&	-06:53:48	&	8.14	&	6.14	&	K2V	&	500$^{18}$	&	23.51	&	1	\\
HD\,69830	&	08:18:24	&	-12:37:55	&	5.95	&	4.36	&	G8V	&	5670$^{16}$	&	12.56	&	1	\\
HD\,71722	&	08:26:25	&	-52:48:26	&	6.04	&	5.91	&	A0V	&	324$^3$	&	69.35	&	1	\\
HD\,73350	&	08:37:50	&	-06:48:24	&	6.73	&	5.32	&	G5V	&	600$^{19}$	&	24.34	&	1	\\
HD\,76582	&	08:57:35	&	+15:34:52	&	5.68	&	5.21	&	F0IV	&	538$^{20}$	&	48.80	&	1	\\
HD\,80950	&	09:17:28	&	-74:44:04	&	5.86	&	5.92	&	A0V	&	138$^{20}$	&	77.34	&	1	\\
HD\,82943	&	09:34:51	&	-12:07:46	&	6.53	&	5.25	&	F9V	&	430$^3$	&	27.61	&	4	\\
HD\,84075	&	09:36:18	&	-78:20:41	&	8.59	&	7.24	&	G2V	&	40$^{13}$	&	64.10	&	1	\\
HD\,107649	&	12:22:25	&	-51:01:34	&	8.78	&	7.76	&	F5V	&	17$^{21}$	&	108.34	&	1	\\
HIP\,63942	&	13:06:15	&	+20:43:45	&	9.40	&	6.21	&	K5	&	4500$^{22}$	&	18.80	&	1	\\
HD\,114082	&	13:09:16	&	-60:18:30	&	8.21	&	7.23	&	F3V	&	17$^{21}$	&	95.69	&	1	\\
HD\,120534	&	13:50:40	&	-31:12:23	&	7.02	&	6.33	&	A5V	&	320$^{17}$	&	86.81	&	3	\\
HD\,122652	&	14:02:32	&	+31:39:39	&	7.15	&	5.94	&	F8	&	500$^8$	&	39.54	&	2	\\
HD\,133803	&	15:07:15	&	-29:30:16	&	8.12	&	7.36	&	A9V	&	16$^{21}$	&	110.74	&	2	\\
HD\,135599	&	15:15:59	&	+00:47:46	&	6.91	&	5.12	&	K0V	&	1300$^{23}$	&	15.82	&	2	\\
HD\,138965&15:40:11& -70:13:40 & 6.42 & 6.34 & A1V & 348$^{27}$ & 78.08 & 1\\
HD\,145229  &16:09:26&	+11:34:28&7.44 & 6.06 &G0& 650$^{26}$ &33.74&1\\
HD\,157728 & 17:24:06 & +22:57:37 & 5.72 & 5.22 & A7V & 100$^{17}$ & 42.74 & 1\\
HD\,164249A	& 18:03:03	& -51:38:56& 7.01& 6.02 &F6V&1800$^{25}$ &49.60&1\\
HD\,172555	&18:45:26	&-64:52:16& 4.77 &4.25 &A7V& 20$^{28}$ &28.79 &1\\
HD\,181296	& 19:22:51 	& -54:25:26 & 5.02& 5.15 &A0V&12$^{30}$& 47.37&1\\
HD\,182681 & 19:26:56 &-29:44:35 & 5.64 & 5.66 &  B8.5V & 107$^{31}$ &  71.42 & 1\\
HD\,192758	&	20:18:16	&	-42:51:36	&	7.03	&	6.30	&	A5V	&	45$^{17}$	&	66.53	&	2	\\
HD\,201219	&21:07:56&+07:25:58 & 0 & 46.5 &G5& 5370$^{10}$ & 37.89&1\\
HD\,205674	&	21:37:21	&	-18:26:28	&	7.17	&	6.25	&	F4IV	&	850$^{10}$	&	56.40	&	2	\\
HD\,206893	&	21:45:22	&	-12:47:00	&	6.67	&	5.69	&	F5V	&	250$^{29}$	&	40.80	&	1	\\
HD\,218340	&	23:08:12	&	-63:37:41	&	8.44	&	7.07	&	G3V	&	2050$^{24}$	&	56.18	&	1	\\
HD\,221853	&	23:35:36	&	+08:22:57	&	7.34	&	6.44	&	F0	&	20$^{17}$	&	65.40	&	1	\\

 \hline
                        \end{tabular}
                        \tablefoot{
For the definition of the star age multiple papers have been used: $^1$ \citep{Zuckerman04}, $^2$ \citep{Zuckerman01}, $^3$ \citep{Chen_2014}, $^4$ \citep{Moor11}, $^6$ \citep{Rodriguez12}, $^7$ \citep{Fernandez2008}, $^8$ \citep{Rhee_2007}, $^9$ \citep{Malo13}, $^{10}$ \citep{Casagrande11}, $^{11}$ \citep{Torres08}, $^{12}$ \citep{Metchev_2009}, $^{13}$ \citep{Zuckerman11}, $^{14}$ \citep{Meshkat17}, $^{15}$ \citep{dasilva09},  $^{16}$ \citep{Vican_2012}, $^{17}$ \citep{Moor06}, $^{18}$ \citep{King03}, $^{19}$ \citep{Taberno12}, $^{20}$ \citep{Zorec12},  $^{21}$ \citep{Mamajek02}, $^{22}$ \citep{West08}, $^{23}$ \citep{Mamajek08}, $^{24}$ \citep{Delgado14}, $^{25}$ \citep{HIPPARCOS},  $^{26}$ \citep{Kim05}, $^{27}$ \citep{Matthews18}, $^{28}$ \citep{Mamajek14},$^{29}$ \citep{Delorme2017}, $^{30}$ \citep{Smith08}, $^{31}$ \citep{Gullikson16}. }
                                \end{table*}

\section{Data reduction}

\subsection{Pre-processing and extraction of environmental data}

The first reduction steps consist in applying standard calibrations to the raw IRDIS images (sky subtraction, flat-field correction, and bad-pixel correction), and registering the frames. This was done using a dedicated pipeline in python \footnote{available at \href{https://github.com/jmilou/sphere\_pipeline.git}{https://github.com/jmilou/sphere\_pipeline.git}}. The frame registration was done using the four satellite spots imprinted on the IRDIS images by a specific waffle pattern applied on the deformable mirror of SPHERE \citep{Delorme2017,Galicher2018}. %In a few specific cases with promising astrophysical signals, we compared this pre-processing with that provided publicly by the SPHERE Data Center \citep{Delorme2017,Galicher2018}. 
The ouput of the pre-processing consists of a temporal cube of frames (individual detector integrations), cosmetically cleaned and recentered, called hereafter an ADI sequence. This cube is accompanied by the corresponding list of parallactic angles for the dedicated high-contrast image processing steps (see section \ref{subsec:Imageproc}).

For the clustering of data and to guide the interpretation, we also extracted environmental data from either the adaptive optics telemetry\footnote{The SPHERE real time controller called SPARTA stores a summary of the adaptive optics telemetry during each observation. Those files are available on the ESO archive as described in \citet{Milli2017_SPHERE}. We developed an automatic script to query and analyse the SPARTA and ASM data available at \href{https://github.com/jmilou/sparta.git}{https://github.com/jmilou/sparta.git}} or the Astronomical Site Monitor (ASM) of the Cerro Paranal Observatory\footnote{\href{http://archive.eso.org/cms/eso-data/ambient-conditions.html}{http://archive.eso.org/cms/eso-data/ambient-conditions.html}}. We collected, among other, data on the seeing, coherence time, relative humidity, temperature, wind speed, and direction at various heights above the platform, Strehl ratio, precipitable water vapour.

\subsection{Image processing}
\label{subsec:Imageproc}

The resulting corrected sets of ADI sequences have been cropped to a 199 $\times$ 199 pixels size, corresponding to the innermost region of the field of view (FOV). We consider angular separations below 1.25 arcsec to take advantage of the higher sensitivity of the RSM map algorithm in the region near the host star, while limiting the computation time. Indeed the increased performance of the RSM map algorithm compared to other PSF-subtraction techniques reduces above 1 arcsec \citep[see ][]{Dahlqvist20,Dahlqvist21,Cantalloube20}, which makes it less suitable for larger angular distances when considering its high computational cost. The computation time is also reduced by limiting the size of the ADI sequences to a maximum of 300 frames, relying on image binning when necessary. The binning procedure consists in the computation of a pixel-wise moving average of the derotated cube. The noise content of these ADI sequences should be reduced by the binning procedure via partial time-averaging.

\subsection{Clustering}
\label{subsec:Clustering}

In order to take full advantage of the RSM algorithm, we rely on the Auto-RSM optimisation framework \citep[see ][]{Dahlqvist21b} to define the optimal sets of parameters for the PSF-subtraction techniques and the RSM algorithm itself. This optimisation pipeline being computationally expensive, we propose to apply it on a subset of targets representative of the whole dataset. The obtained optimal parametrisations can then be used to compute the RSM detection maps for all targets. \cite{Dahlqvist21b} showed a relatively high degree of similarity in the optimal parametrisations of both the PSF-subtraction techniques and the RSM algorithm, when using ADI sequences generated with the VLT SPHERE instrument. Dividing the SHARDDS dataset into multiple subsets should nevertheless allow us to account for small variations in the optimal parametrisations depending on the ADI sequence characteristics.

The subdivision of the SHARDDS dataset in multiple subsets is based on a set of observables characterising the ADI sequences. The subdivision itself is done via the K-means algorithm \citep{Macqueen67}, a centroid-based clustering procedure aiming to find the centroids that minimise the within-cluster sum-of-squares, also called inertia. The K-means algorithm was selected as it provides a good estimate of the centroids position. This is a key element to define properly which ADI sequence within a cluster is the most representative. These centroids being often not associated to a sample, we define the most representative ADI sequences as the ones closest to the cluster centroids. Once defined, the Auto-RSM optimisation framework is applied on the selected set of ADI sequences. The optimal parametrisations are then used to compute the RSM detection maps for the remaining ADI sequences of each cluster, following the standard RSM map procedure.

\subsubsection{Clustering parameters}

The K-means algorithm needs to be applied on a set of parameters that characterise the properties of the ADI sequences. For our cluster analysis, we chose metrics providing information about the sequence, the observing conditions, and the noise distribution within the set of frames. This set of observables consists in the mean seeing, the Strehl ratio, the mean coherence time, the number of images, the total field rotation in term of parallactic angle, the raw contrast at 200, 500, and 700 mas, the autocorrelation timescale between images, the mean wind speed at 30 meters, and the wind driven halo strength and asymmetry\citep{Cantalloube20a}. 

The seeing, Strehl ratio, and coherence time are commonly used performance indicators to assess the observing conditions. Considering the 40 minutes integration time used for the SHARDDS survey, the number of images contained in the ADI sequence affects the sampling frequency, and therefore both the performance and the parametrisation of the PSF-subtraction algorithm (e.g. optimal number of principal components). The field rotation also impacts the performance because of the higher self-subtraction of the signal in the case of small field rotation. When mitigating self-subtraction, it translates into a reduced set of available images to compute the reference PSF.

The raw contrasts were estimated by placing apertures of 1 Full Width at Half Maximum (FWHM, 43\,mas) diameter in the selected annuli and computing the ratio between the mean encircled flux and the stellar flux. The autocorrelation timescale between the ADI sequence images was estimated by considering the region between 300-600 mas, where the adaptive optics is affecting the most the performance. The flux within a one FWHM aperture was computed for each pixel in the selected region and for each image. An exponential function was then fitted on the temporal autocorrelation of these fluxes and its exponential factor was kept as a measure of the autocorrelation decay rate. We expect that a slower autocorrelation decay will result in lower performance. 

The wind-driven halo (WHD) strength and asymmetry were computed using the method presented in \cite{Cantalloube20a}. The WDH is a bright elongated structure centred on the coronagraph in high contrast images, due to uncompensated atmospheric turbulence. The WDH cannot be easily treated with standard PSF-subtraction techniques and affects therefore the achievable contrast at small angular separations (below 1000 mas). Along with the WDH, the low wind effect \cite[LWE,][]{Milli2018_LWE} is also a wind-driven phenomenon degrading the performance of high contrast imaging. LWE arises from uncorrected wavefront aberrations due to air temperature inhomogeneities in large telescope pupil, caused by the radiative cooling of the spiders, which dominates in the absence of wind. We included the wind speed at 30 m to account for this potential effect. 

The number of images included in the ADI sequences was identified as a key metrics for the definition of the optimal parametrisation during the development of the Auto-RSM framework. We have therefore decided to divide the SHARDDS dataset into two subsets before applying the clustering algorithm. We defined a threshold of 151 frames to separate the two subsets, as this value ensures that the standard deviation of the number of images within each subset is equivalent. This ensures a similar distribution in terms of sequence size within the two subsets.

\subsubsection{Application and results of the K-means clustering}

The K-means algorithm being based on Euclidean distance, the selected set of metrics must be standardised before applying the clustering algorithm, to avoid that metrics with larger values dominate the calculation. Before applying the K-means algorithm, we looked for possible multicollinearity between the selected set of observables. Relying on the variance inflation factor \citep[VIF][]{Belsley05} and Pearson correlations \citep{benesty2009pearson}, we removed the contrast at 200 and 700 mas, which led to multicollinearity, affecting potentially the definition of the clusters. The initialisation of the K-means algorithm consisting in the random selection of initial centroids, the results may lack consistency and differ from one estimation to another. The algorithm can also be affected by the order of the observables. In order to tackle these two issues, we initialised our estimation by running the K-means algorithm 100 times, selecting at each iteration a different permutation of the parameters. We then took the mean of these centroids positions to initialise the final cluster definition.

        \begin{figure}[t]
\footnotesize
  \centering
    \subfloat[]{\includegraphics[width=220pt]{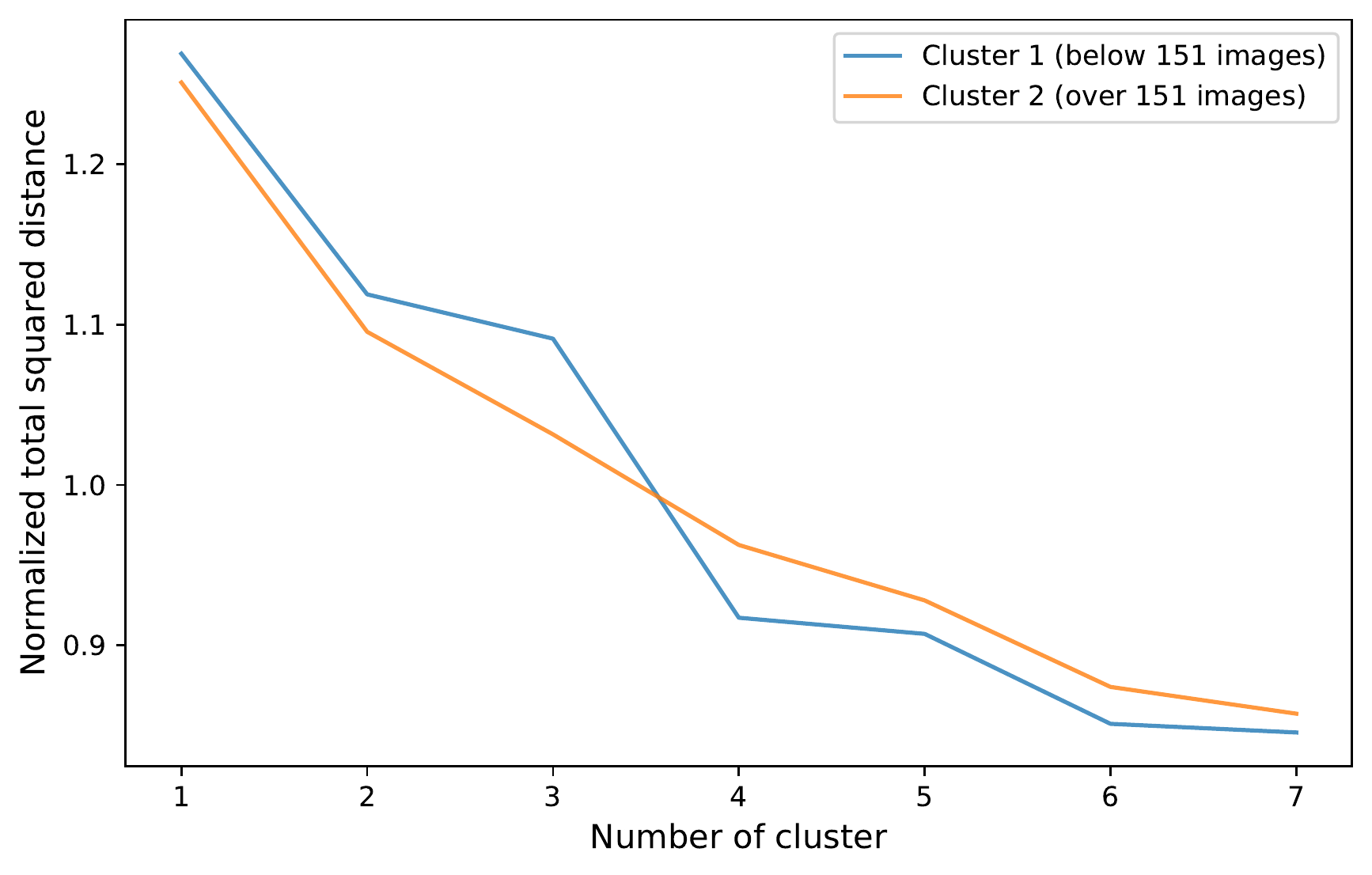}}
  \caption{\label{clustdist} Evolution of the standardised total square distance between every cluster member and their centroid, depending on the number of cluster for the two subsets (i.e. ADI sequences with a number of images higher or lower than 151 frames) }
\end{figure}

We finally defined the number of clusters. This definition was based on the analysis of the evolution of the total squared distance between cluster members and their centroid when changing the number of clusters. Looking at Fig. \ref{clustdist}, we see that the largest fraction of the total squared distance reduction occurs between one and four clusters. We therefore selected for both subsets a number of clusters equal to four, implying a total of eight ADI sequences on which Auto-RSM will be applied. The eight cluster centroids, as well as the composition of their respective clusters are presented in Table \ref{Clusters}.

\begin{table}[t]
                        \caption{Subdivision of the SHARDDS dataset into eight clusters.}
                        \label{Clusters}
\centering

                        \begin{tabular}{ll}
                        
                        \hline
Cluster center  & Cluster members \\                           
 \hline
\textbf{Cluster 1-1} &\\
HD\,192758 & HD\,38207, HD\,37484, HD\,10472, AG Tri,\\
&  HD\,84075,HD\,192758 $2^{nd}$ epoch, HD\,274255 \\
\textbf{Cluster 1-2} &\\
HD\,3670 & HD\,37484 $2^{nd}$ epoch , HD\,22179, \\
& AG Tri $2^{nd}$ epoch, AG Tri  $3^{rd}$ epoch, \\
&HD\,82943 $3^{rd}$ epoch ,HD\,114082 \\
\textbf{Cluster 1-3} &\\
HD\,201219 &  HD\,53842, AG Tri $4^{th}$ epoch, HD\,218340, \\
&HD\,221853\\
\textbf{Cluster 1-4} &\\
HD\,14082B & HD\,82943, HD\,107649\\
\textbf{Cluster 2-1} &\\
HD\,21997 & HD\,24636, HD\,15257, HD\,10472 $2^{nd}$ epoch,\\
& HD\,145229 $2^{nd}$ epoch\\ 
\textbf{Cluster 2-2} &\\
HD\,206893 & HD\,40540, HD\,35650, HD\,31392, HD\,25457,\\
& HD\,17390, HD\,16743, HD\,9672, HD\,105, \\
& HD\,69830, HD\,71722,HD\,120534, HD\,182681,\\
&  HD\,120534 $2^{nd}$ epoch, HD\,164249A\\
\textbf{Cluster 2-3} &\\
HD\,181296 & HD\,14082B $2^{nd}$ epoch, HD\,13246, HD\,203, \\
& HD\,60491, HD\,122652, HD\,135599 $3^{rd}$ epoch,\\
& HD\,145229, HD\,172555, HD\,181296\\
\textbf{Cluster 2-4} &\\
HD\,3003& HD\,377, HD\,73350, HD\,76582, HD\,80950,\\
 &HD\,82943 $2^{nd}$ epoch, HD\,82943 $4^{th}$ epoch,\\
 &HD\,138965, HD\,157728, HIP63942,\\
 & HD\,122652 $2^{nd}$ epoch\\
\hline
                        \end{tabular}
                                \end{table}

\begin{figure}[t]
\footnotesize
  \centering

  \subfloat[]{\includegraphics[width=250pt]{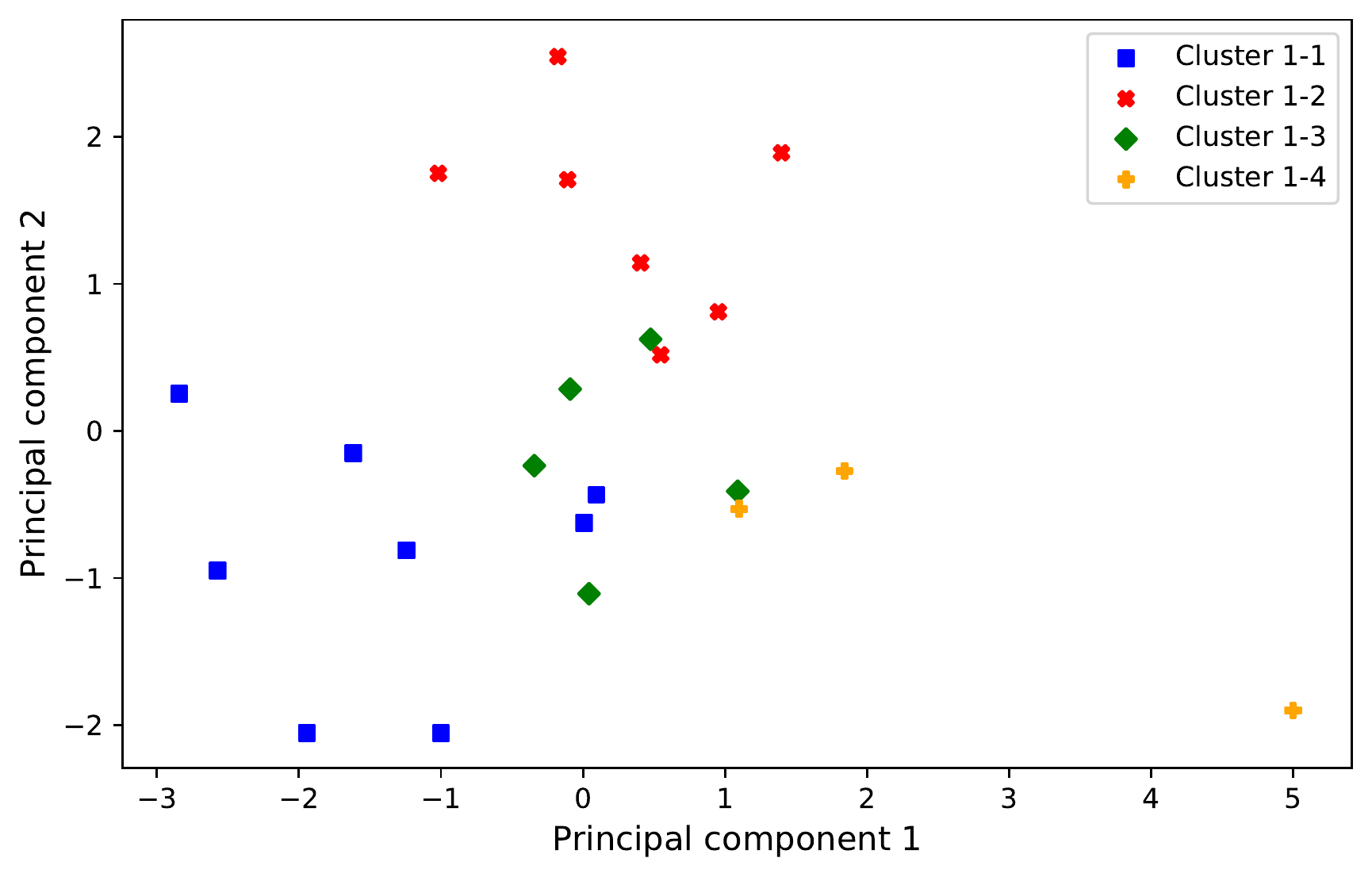}}\\
    \subfloat[]{\includegraphics[width=250pt]{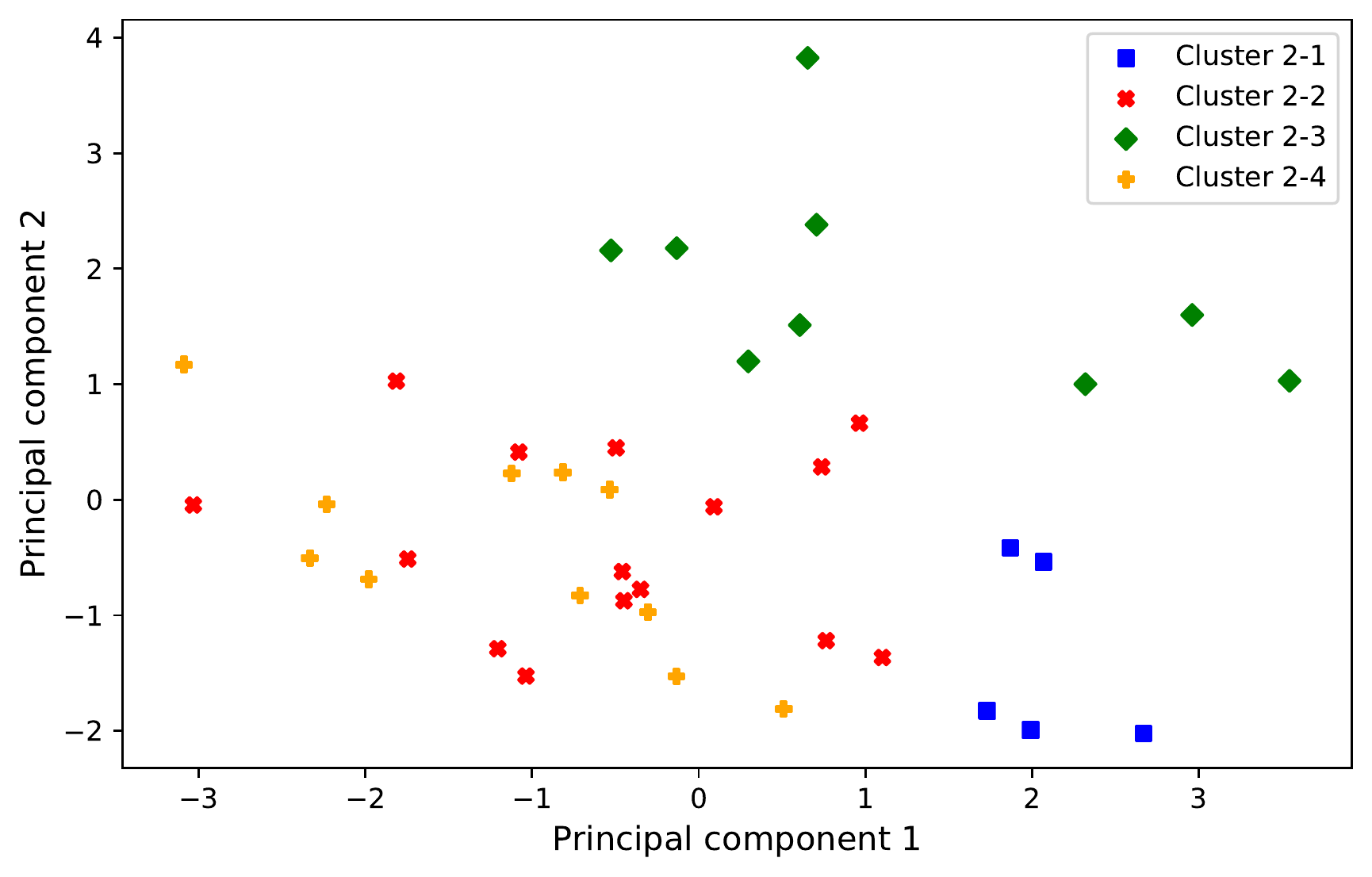}}

  \caption{\label{ClusterProj} Projection of the SHARDDS survey targets on the first two principal components computed based on their observational characteristics. The top (respectively bottom) graph provides the targets with a number of frames in their ADI sequence below 151 (respectively above 151). The colours indicates to which cluster the target has been assigned.}
\end{figure}

After the subdivision of the dataset into eight clusters, we made several consistency checks by relying on principal component analysis to reduce the dimensionality of our set of observables and eliminate residual correlations between the variables. We tested the K-means algorithm with different numbers of principal components and retrieved almost every time the same set of clusters. Figure \ref{ClusterProj} illustrates the repartition between the different clusters in the space formed by the first two principal components. As can be seen, the different clusters are relatively well defined except for cluster 2-2 and 2-4, for which a larger set of principal components are necessary to make a clear distinction. We finally applied a Gaussian mixture model instead of the K-means algorithm as a last consistency check. The Gaussian mixture model considers on top of the number of clusters and the centroid position, the standard deviation of the distance between cluster members to characterise clusters. The obtained cluster repartitions were very close although not exactly the same.

Two targets were excluded from these clusters, HD\,133803 and HD\,205674. They were treated separately as they were imaged at two epochs separated by only a couple of days. We therefore took advantage of the ability of the RSM algorithm to deal with multiple ADI sequences at once to generate a single detection map per target. This was not possible for the other multi-epoch targets due to the longer time span separating the image sequences, implying a potential movement of planetary candidates.

\subsection{High contrast image processing}
\label{subsec:Detectmap}

This section is devoted to the computation of RSM detection maps for all the targets included in the SHARDDS survey, as well as the computation of the contrast curves. This computation starts with the optimisation of the model parameters via the Auto-RSM framework for the eight selected targets (see cluster center in Table \ref{Clusters}). The Auto-RSM framework requires the selection of the PSF-subtraction techniques as well as the definition of the parameter ranges to be considered during the optimisation. We considered in this paper six different PSF-subtraction techniques: annular PCA \citep[APCA,][]{Gonzalez17}, non-negative matrix factorisation \citep[NMF,][]{Ren18}, the local low rank plus sparse plus Gaussian decomposition \citep[LLSG,][]{Gonzalez16}, locally optimised combination of images \citep[LOCI,][]{Lafreniere07}, and forward-model versions of KLIP \citep{Soummer12,Pueyo16}, and LOCI \citep[see ][for more details]{Dahlqvist21}. %In order to further improve the RSM map algorithm performance, we considered for APCA, NMF, LLSG and KLIP two different ranges for the number of principal components/ranks, which are considered as separate models by the algorithm. Planetary signals and residual speckle noise evolve differently with the number of principal components used to generate the reference PSF. We take advantage of this differentiated evolution to further boost the ability of the RSM map algorithm to disentangle faint planetary signals from residuals speckle noise, following to some extent the approach proposed in \cite{Gonzalez18}.

The considered ranges of principal components for APCA, NMF, and KLIP, the ranks for LLSG, and the tolerance for LOCI are selected by a new function of the PyRSM python package\footnote{\url{https://github.com/chdahlqvist/RSMmap}}, which regroups the different functions of the Auto-RSM framework. This function studies the evolution of the contrast at different angular separations when modifying the number of principal components, ranks, or tolerance. The upper boundary of the considered ranges is defined as the value for which the contrast, averaged over the different angular separations, reaches a peak. In the case of APCA, NMF, LLSG, and KLIP, the obtained ranges were divided in two equal size ranges, to form two separate models. This should provide more diversity to the RSM algorithm and increase the framework's performance as planetary signals and residual speckle noise evolve differently with the number of principal components used to generate the reference PSF. Regarding the other parameters of the PSF-subtraction techniques, a single range was defined for all cluster centroids. The range for the number of segments was fixed to $[1,4]$, the FOV rotation threshold to $[0.25,1]$ and the crop size to [3,5] for standard PSF-subtraction techniques and [7,9] for the forward model versions to account for the side lobes due to self-subtraction \citep[see ][ for more details about these parameters]{Dahlqvist21b}. The computation of the PSF forward model being computationally very intensive and side lobes due to self-subtraction becoming fainter for increasing angular separation, we considered the forward model versions for only the first 400 mas. %Indeed the main interest of PSF forward modelling is to account for the signature of over- and self-subtraction of the PSF due to the reference PSF subtraction. This effect becoming weaker when the angular separation increases, due to larger field rotation, it is not necessary to use such advanced model at large distance from the host star.

Having defined all the parameters, the Auto-RSM optimisation framework was applied on each centroid, using the full frame mode to optimise the PSF-subtraction techniques and RSM algorithm parameters, the forward model to compute the RSM detection maps, and the bottom-up approach to select the optimal set of likelihoods \citep[see ][]{Dahlqvist21b}. Following the original Auto-RSM framework, the parameters optimisation was performed using the reversed parallactic angles. Considering the low probability of detecting a planet, we also tried to use the original parallactic angle to optimise the parameters, but it did not lead to a performance increase in terms of contrast. We therefore relied on reversed parallactic angles to avoid any potential planetary signal suppression during the optimisation process.

We investigate in Appendix \ref{common} the similarities existing between the optimal parametrisations obtained for the eight cluster centroids, as well as the relationships between these optimal parameters and the set of metrics characterising the ADI sequences. The comparison of the optimal parametrisations is done via the computation of dissimilarity measures between cluster centroids, for both the PSF-subtraction techniques and the RSM algorithm. The results demonstrate a relatively high degree of similarity between the different parametrisations, confirming the conclusions drawn in \cite{Dahlqvist21b} about the high stability of the ADI sequence imaged by the VLT/SPHERE instrument. The Pearson correlations between the ten observables characterising our ADI sequences, and the PSF-subtraction techniques parameters show a sensible correlation for some observables, with the contrast at 500 mas showing the highest average correlation rate, and the exponent of the autocorrelation function the lowest one.

\subsubsection{Detection maps}

Following the definition of the optimal set of parameters for the cluster centroids, we computed the RSM detection maps for every target of the SHARDDS survey. Two sets of detection maps were computed using the original and the reverse parallactic angles. The detection maps with the reversed parallactic angles allowed the computation of a radially dependent residual noise estimate, which is subtracted from the detection map to account for the noise angular evolution (see Appendix~\ref{thresh} for more details about the radial threshold computation and its interpretation).

The resulting detection maps were then analysed to uncover potential planetary signals or other bright structures. From this analysis, we rejected HD\,107649 due to the presence of extended speckle-like bright structures. For other targets, some redundant epochs presenting a high degree of residual noise were also removed\footnote{These ADI sequences include AG Tri, AG Tri $2^{nd}$ epoch, AG Tri $3^{rd}$ epoch, HD\,82943 and HD\,82943 $3^{rd}$ epoch}. From the remaining ADI sequences, we identified 16 targets containing a point-like source or an extended bright structure above a probability threshold of 0.05. To insure that these detections were not the result of a sub-optimal parametrisation of the RSM algorithm, we applied the Auto-RSM algorithm to 15 of these targets. From the set of 16 targets including detections above a 0.05 probability threshold, one was a cluster centroid (HD\,206893) for which we kept the original RSM detection maps. 

We performed a correlation analysis similar to the one made in Appendix \ref{common} on these 15 targets, in order to assess the influence of a stronger speckle field on the optimal parametrisations. We found much lower correlation rates between these optimal parameters and the set of metrics characterising the ADI sequences.
%, with only 7\% of the raw correlation rates above the 50\% threshold. The parametrisation of these 15 targets also showed 
We also observed a higher degree of dissimilarity between the parametrisations of these 15 targets, especially for the PSF-subtraction techniques parameters. These results highlight the limits of a clustering approach based solely on the parameters characterising the ADI sequence, when facing noisier samples. They also demonstrate the necessity to adopt an empirical approach, such as the Auto-RSM optimisation framework, to optimise the parametrisation when the samples noise structure cannot be well captured by the set of ADI sequence characteristics. However, the low residual noise level in the detection maps shown in Figures \ref{Empty_map1}-\ref{Empty_map3}, as well as the large fraction of the survey dataset (70\%) that did not require the use of Auto-RSM, still favour the use of a limited number of optimal parameter sets computed for well chosen targets.

Following this individual optimisation, the analysis of the resulting 16 detection maps allowed the detection of three already known point-like sources that will be further analysed in the next section (see Figure \ref{Target_map}). The detection maps containing no plausible planetary candidates are shown in Appendix \ref{detmap}.  As can be seen from Figures \ref{Empty_map1}-\ref{Empty_map3} , the residual noise level is most of the time very low, except for bright structures observed in HD\,53842 and HD\,80950. These structures are diffraction patterns due to the presence of a bright companion just outside the 199 $\times$ 199 pixels window considered in this analysis. For HD\,80950, the companion is situated at a projected separation of 130 au with an apparent magnitude in H band of 9.97. HD\,53842 is a very young binary system, with a primary spectral type F5 star and a secondary M-dwarf situated at a projected separation of 82 au, with an estimated orbital period of 300 years (C. del Burgo, in prep).

\subsubsection{Contrast curves}
\label{subsec:Contrastcurve}

        \begin{figure*}[!h]
\footnotesize
  \centering

  \includegraphics[width=400pt]{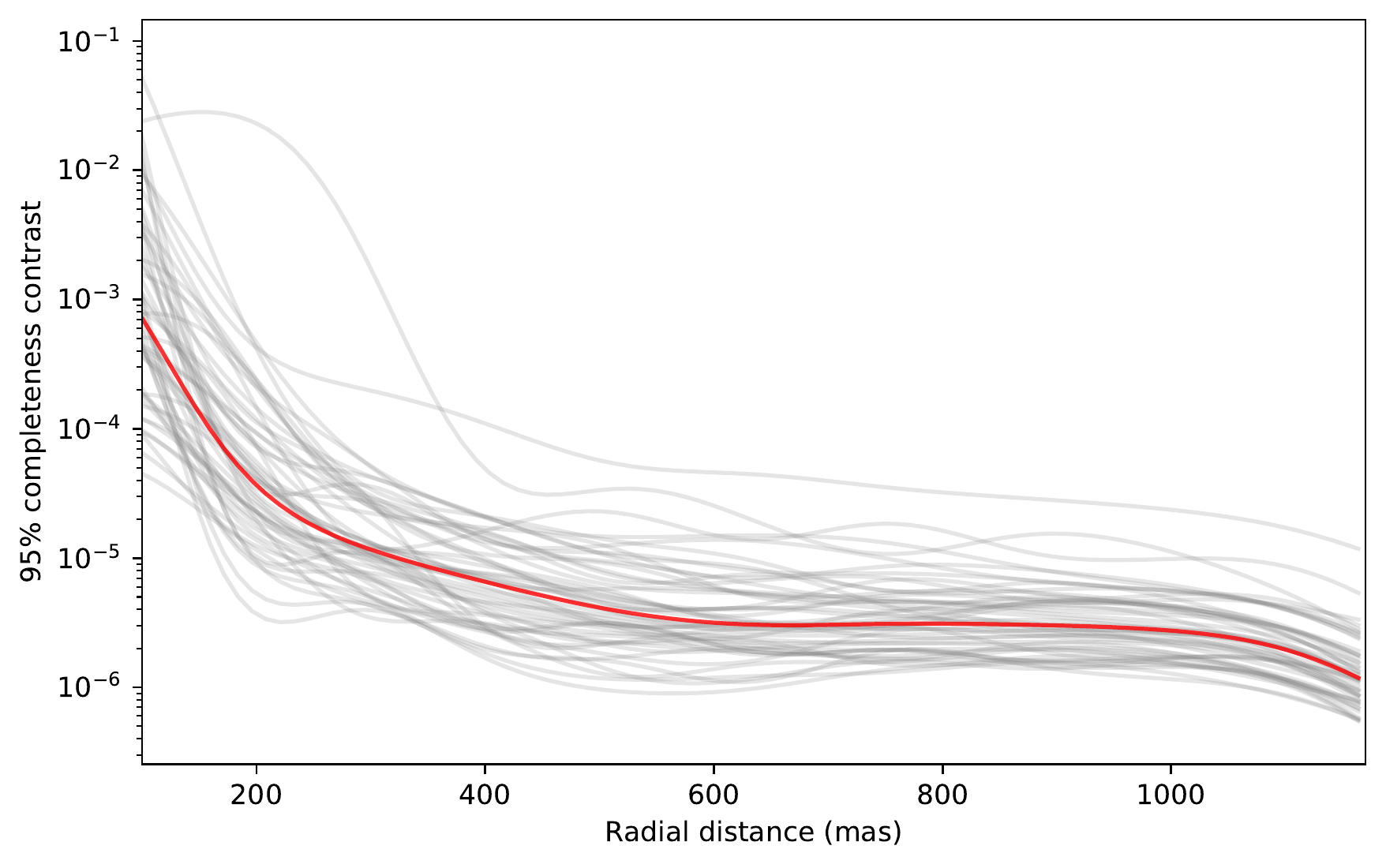}
 
  \caption{\label{ContrastCurve} Contrast curves computed for the set of considered targets (gray) and median contrast curve (thick red) computed at a 95\% completeness level.}
\end{figure*}

Following the computation of the detection maps, we relied on an optimised version of the approach proposed in \cite{Dahlqvist21} to compute contrast curves for every target. When relying on probability detection maps, standard signal-to-noise ratio (S/N) based approaches involving the estimation of the throughput and the noise standard deviation \citep{Mawet14} cannot be used. We replace this definition by an empirical estimation of the contrast corresponding to a predefined detection rate (also called true positive rate) computed at a specific threshold. As it is not possible to reach a 5$\sigma$ confidence level empirically, this threshold corresponds simply to the first detection of a false positive within the entire detection map. The detection rate is computed, for a given angular separation, via the injection of fake companions at different azimuths. The computation of the contrast follows an iterative procedure, where the contrast is increased or decreased depending on the obtained detection rate and the previously tested contrasts \citep[see ][for a detailed presentation of this iterative procedure]{Dahlqvist21}. 

We selected a detection rate of  95\%, which is the traditional completeness level for the computation of planet detection probability or occurrence rate (see Section \ref{sec:Sensitivity}). This detection rate requires the successive injection of 20 fake companions per considered annulus. We considered nine angular separations ranging from 60 to 1150 mas. From the original 73 ADI sequences forming the SHARDDS survey, we removed 13 ADI sequences because of poor observing conditions, and/or the existence of multiple epochs for several targets. For a few targets, several epochs were kept as they showed a similar level of residual noise. When multiple epochs where available, the lowest contrast was kept for each considered angular separation, to generate a single contrast curve per target. A radial basis multiquadric function (RBF) \citep{Hardy71} was then used to perform the interpolation between the nine angular separations for which a contrast was estimated. 

Figure \ref{ContrastCurve} provides a consolidated view of the contrast curves, with gray curves showing the individual contrast curves corresponding to each target and the red line providing the median. As can be seen from these curves, the contrast decreases quickly with the angular separation, with a median contrast below $10^{-5}$ at already 300 mas. %Considering the high completeness level we have selected, it demonstrates the good performance of the RSM map algorithm at angular separations below 1 arcsec.
However, we observe a relatively high dispersion of the contrasts at close separations, with the contrast ranging from $3\times 10^{-1}$ to $3\times 10^{-4}$ at 100 mas. This high dispersion can be directly linked to the observing conditions. This relationship between the performance in terms of contrasts and the observing conditions will be further investigated in Section~\ref{sec:CCanalysis}. 

%This will also be the opportunity to check the performance of cluster members versus cluster centroids. Considering the relatively high stability of the optimal parametrisations of the cluster centroids observed in the previous section, the RSM and PSF-subtraction techniques parametrisations do not seem to be very sensitive to the quality of the observations.

\section{Identification of planetary candidates }
\label{sec:Identification}

\begin{figure}[t]
\footnotesize
  \centering
        \subfloat[HD114082]{\includegraphics[width=240pt]{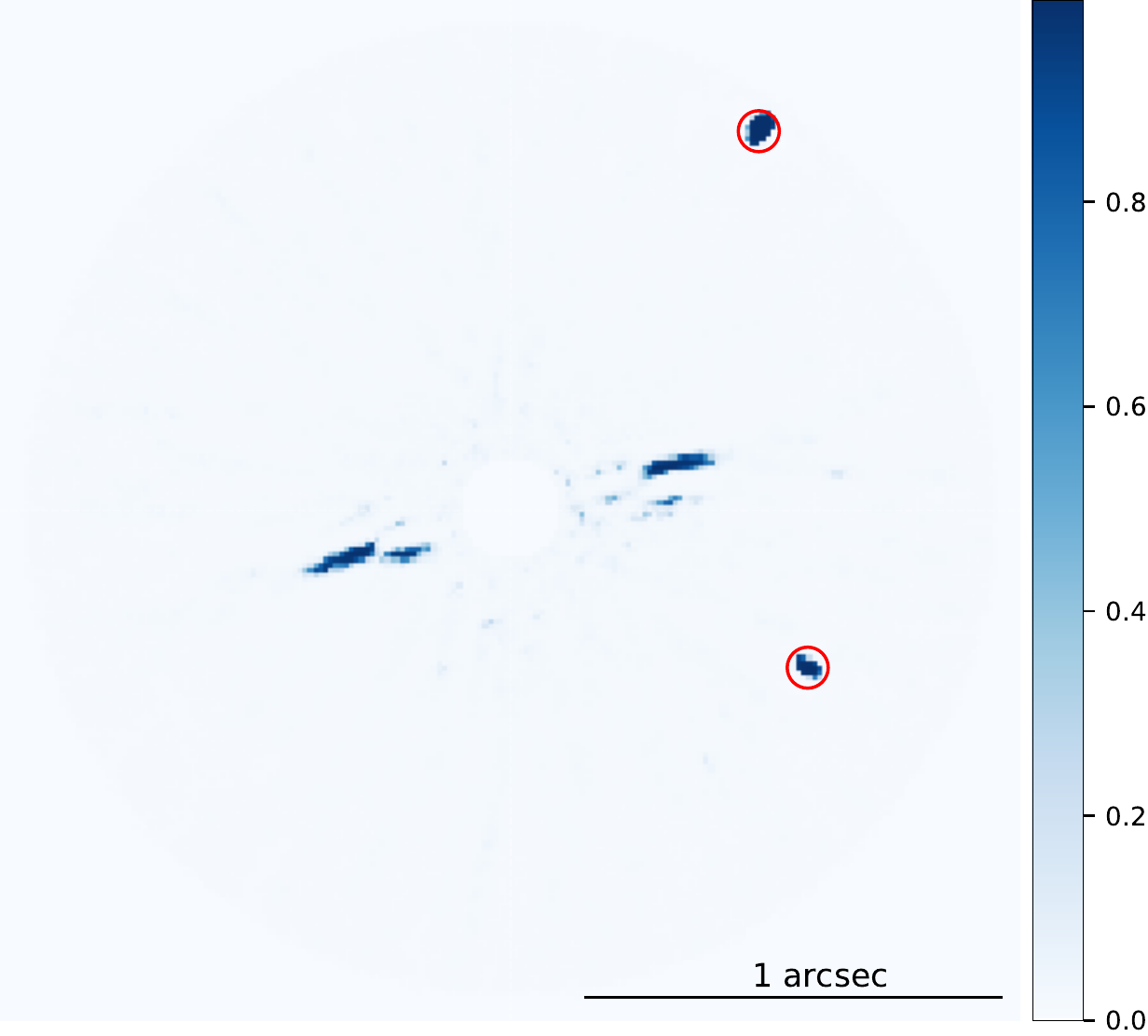}}\\
          \subfloat[HD206893]{\includegraphics[width=240pt]{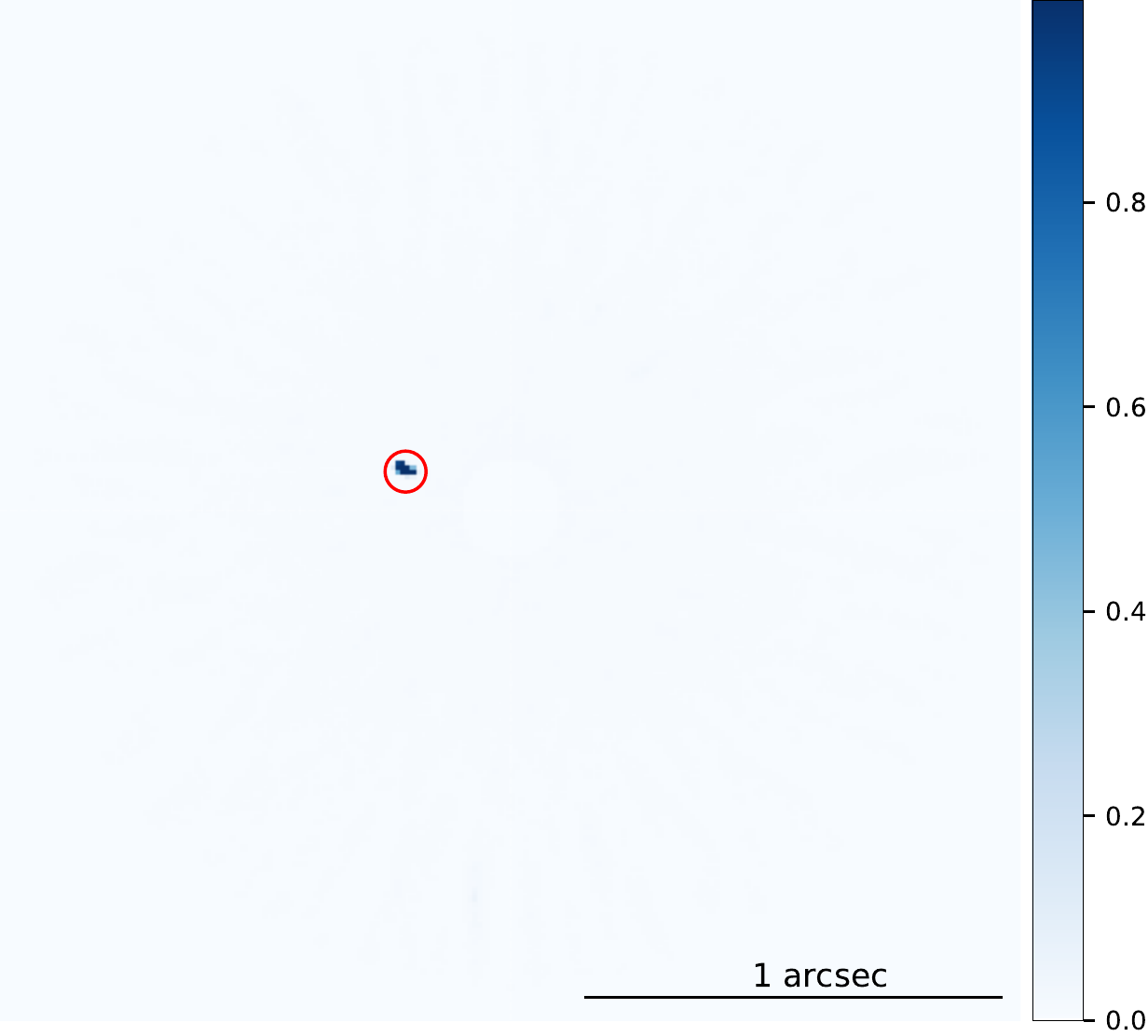}}

  \caption{\label{Target_map} RSM detection maps generated using Auto-RSM. These detection maps led to the detection of one or multiple planetary candidates. The color scale is expressed in terms of inferred detection probabilities. A square root scale has been selected to highlight potential residual speckle noise.}
\end{figure}

Figure \ref{Target_map} presents the two ADI sequences containing a signal above the previously defined threshold of 0.05, after having applied Auto-RSM on the 16 sequences for which a signal was previously detected. The two ADI sequences contain already known targets, with HD\,206893 B identified in \cite{Milli16b} , and the debris disk from HD\,114082 in \cite{Wahhaj16} which includes also two background stars. In the rest of the section, we propose a new way to extract the photometry and astrometry of point-like sources based on the RSM framework, and apply it to these two datasets.

\subsection{RSM NEGFC algorithm}
\label{subsec:RSMNEGFC}

Like in the negative fake companion (NEGFC) approach \citep{Lagrange10,Marois10,Wertz17}, the astrometry and photometry are determined by injecting a fake companion at the expected position of the planet, with a negative flux providing the photometry. Multiple positions and fluxes are tested and their optimum is defined as the values minimising a loss function defined as the average probability inside an aperture of two FWHM centred on the expected location of the planet. The minimisation relies on a particle swarm optimisation (PSO) framework \citep{Kennedy95}. A series of particles, each defining a set of parameters, travel within the parameter space following an iterative procedure. At each step the velocity of these particles in the parameter space is updated based on the knowledge of the particle's own optimum and the global optimum of the entire swarm. 

The PSO framework was chosen as it showed, during tests, a higher convergence rate than Bayesian optimisation and allowed multi-core estimation\footnote{Multi-core optimisation is not possible with usual minimisation algorithms such as Nelder-Mead, Newton or Broyden-Fletcher-Goldfarb-Shanno.}, reducing the computation time. More standard minimisation frameworks (Nelder-Mead, Newton, or Broyden-Fletcher-Goldfarb-Shanno) were tested without success because of the non linear behaviour of the selected loss function near the optimum and the presence of multiple local optima. The inertia, the cognitive, and social coefficient parametrising the PSO algorithm help defining the right balance between exploitation of known minima and exploration of the parameter space. Several sets of parameters were tested and the set $[\alpha=0.5,\beta_p=1,\beta_g=1]$ was selected, as it led to a high convergence rate while avoiding local minima. 

The algorithm is initialised by relying on a detection map generated with the RSM map algorithm using the forward-backward mode, which considers both past and future observations to infer the detection probability. This mode has demonstrated a higher precision in terms of astrometry \citep[see ][]{Dahlqvist21}. Once an initial astrometry has been defined, a range of fluxes is tested to get an initial estimation of the photometry. The PSO framework is then used to minimise the average probability in the two-FWHM aperture centred on the expected position. We relied on ten particles with a maximum number of iterations equal to 20. At the end of the PSO minimisation, the global minimum is kept and a confidence interval is computed based on the computation of the inverted Hessian matrix\footnote{The Hessian matrix is calculated with finite difference derivative approximation.}.

We tested additional versions of the planetary signal characterisation algorithm. We tried to subtract a local measure of the noise from the average probability within the two-FWHM aperture. This local noise was computed as the detection probabilities averaged over two sections of the annulus with a width of one FWHM containing the signal, situated at a distance of 1.5 FWHM on either sides of the expected target position. We did not consider the entire annulus, as local features may be observed in the detection map, leading to a potential bias. We also considered replacing the PSO minimisation by a Bayesian optimisation. We tested these different versions along with the NEGFC function provided by the VIP package \citep{Gonzalez17}, which relies on Nelder-Mead minimisation. 

We based our performance comparison on the ADI sequence obtained on HD\,3003, considering an intermediate angular separation of $4 \lambda/D$. We injected fake companions at eight different azimuths and considered eight different contrasts ranging from $1\times10^{-5}$ to $8\times10^{-5}$. This range goes from a non detection in a traditional S/N map (a detection just above the background with the RSM map) to a very bright planetary signal. This should allow us to investigate the behaviour of the planetary signal characterisation algorithms in two very different regimes. The astrometric error is computed as the root mean squared (rms) position error between the obtained position and the injected fake companion true position, averaged over the eight considered azimuths. The photometric error follows the same approach but comparing in terms of rms the estimated photometry and the true underlying one.

        \begin{figure}[t]
\footnotesize
  \centering

  \subfloat[]{\includegraphics[width=250pt]{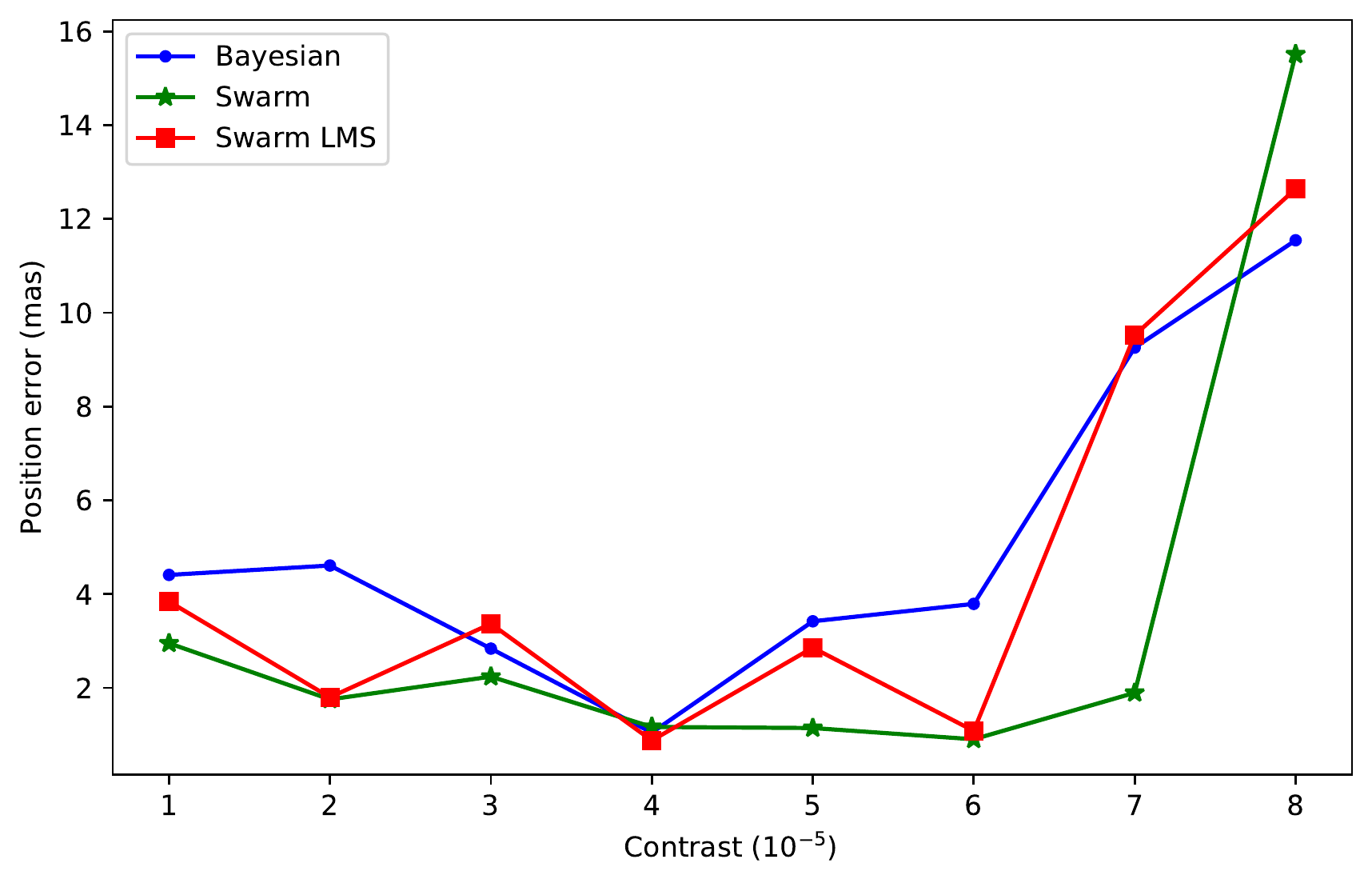}}\\
  \subfloat[]{\includegraphics[width=250pt]{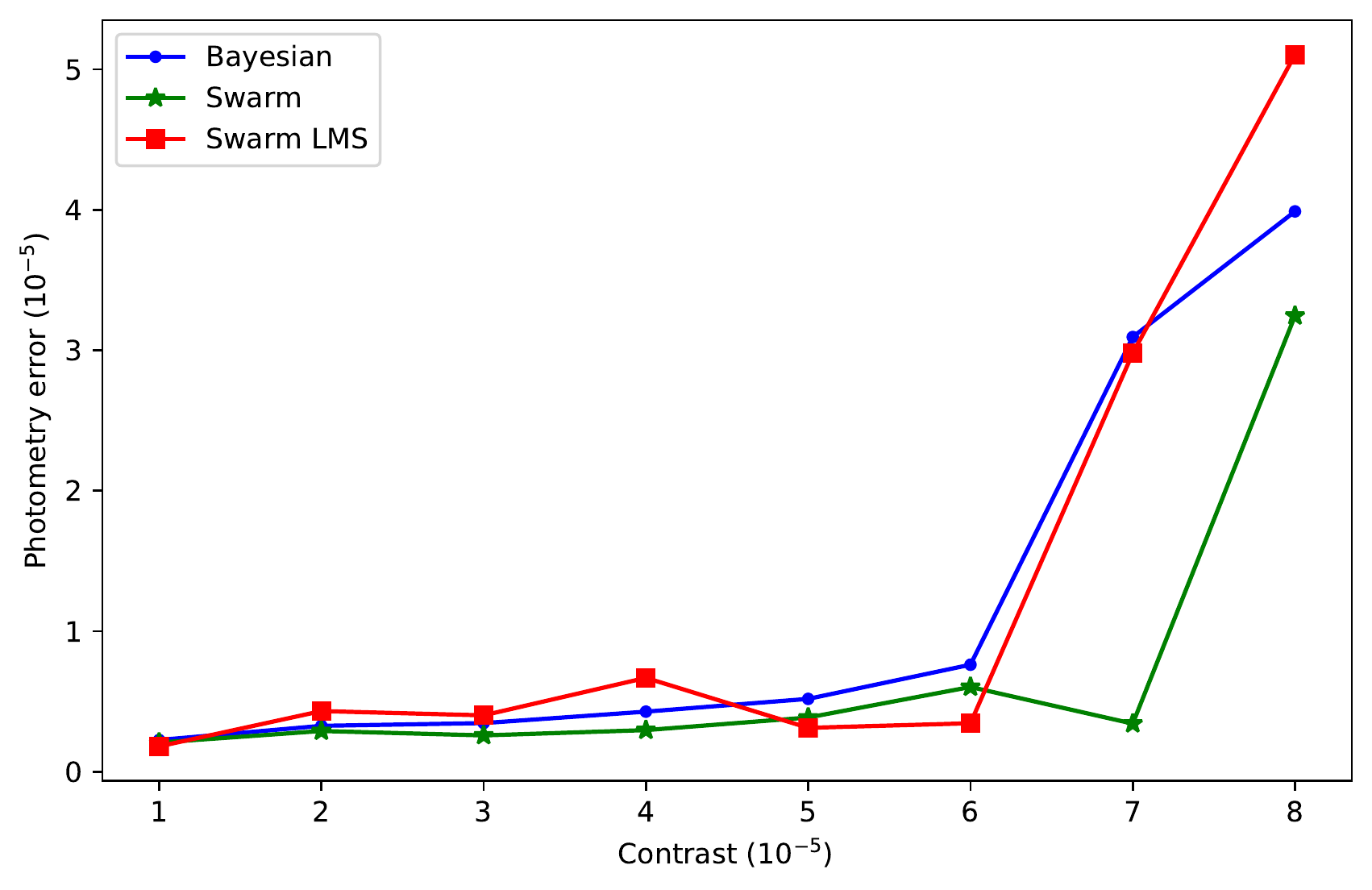}}\\

  \caption{\label{perf_algo} Astrometric and photometric errors for the RSM based planetary signal characterisation algorithm using the PSO approach with and without subtraction of the local mean noise (resp. red and green) and using Bayesian optimisation (blue). The upper graph shows the dependence of the averaged rms position error on the contrast, while the bottom one shows the dependence of the photometric rms error (computed at a angular separation of $4\lambda /D$).}
\end{figure}

\begin{figure}[t]
\footnotesize
  \centering

    \subfloat[]{\includegraphics[width=250pt]{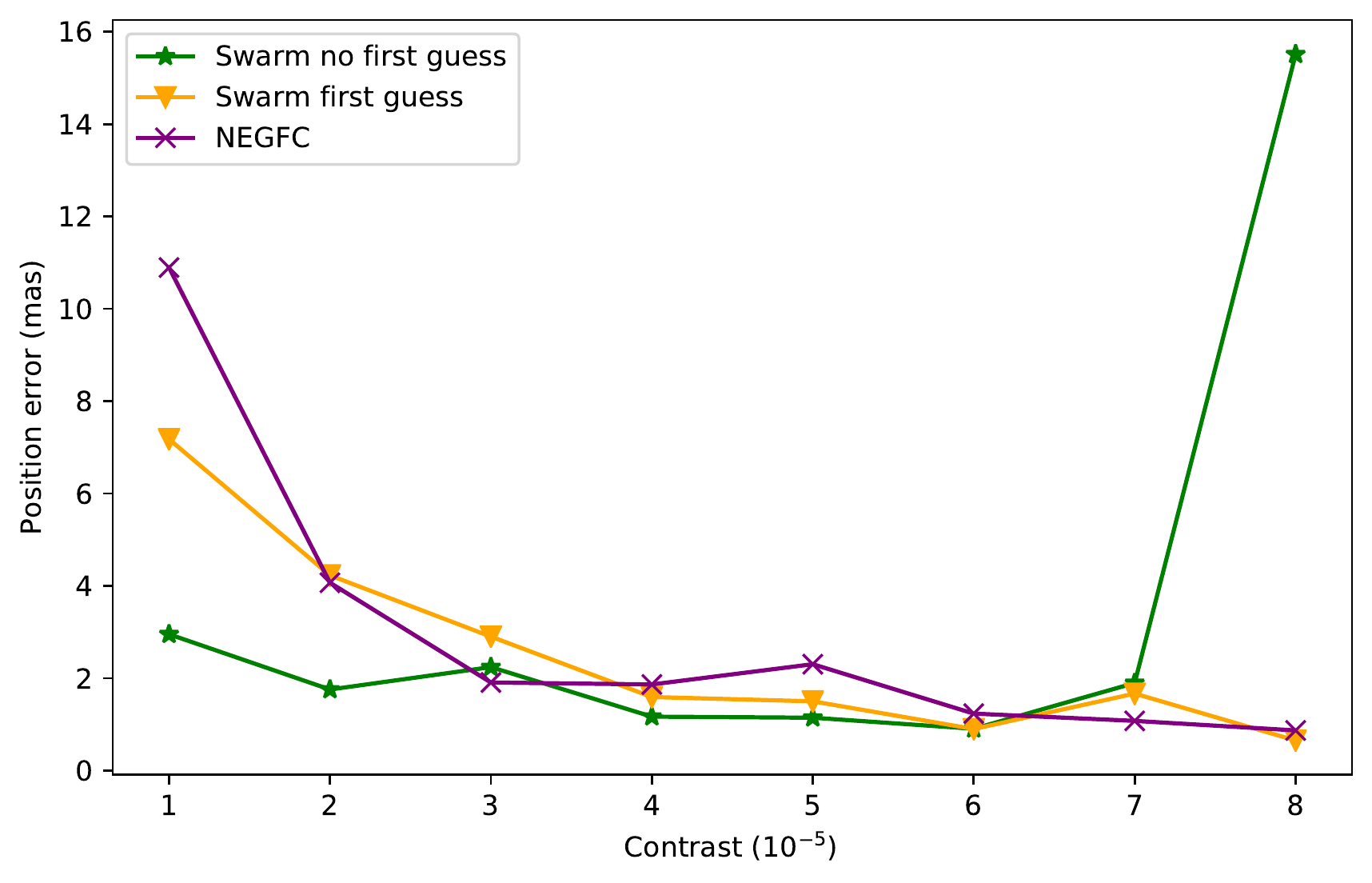}}\\
    \subfloat[]{\includegraphics[width=250pt]{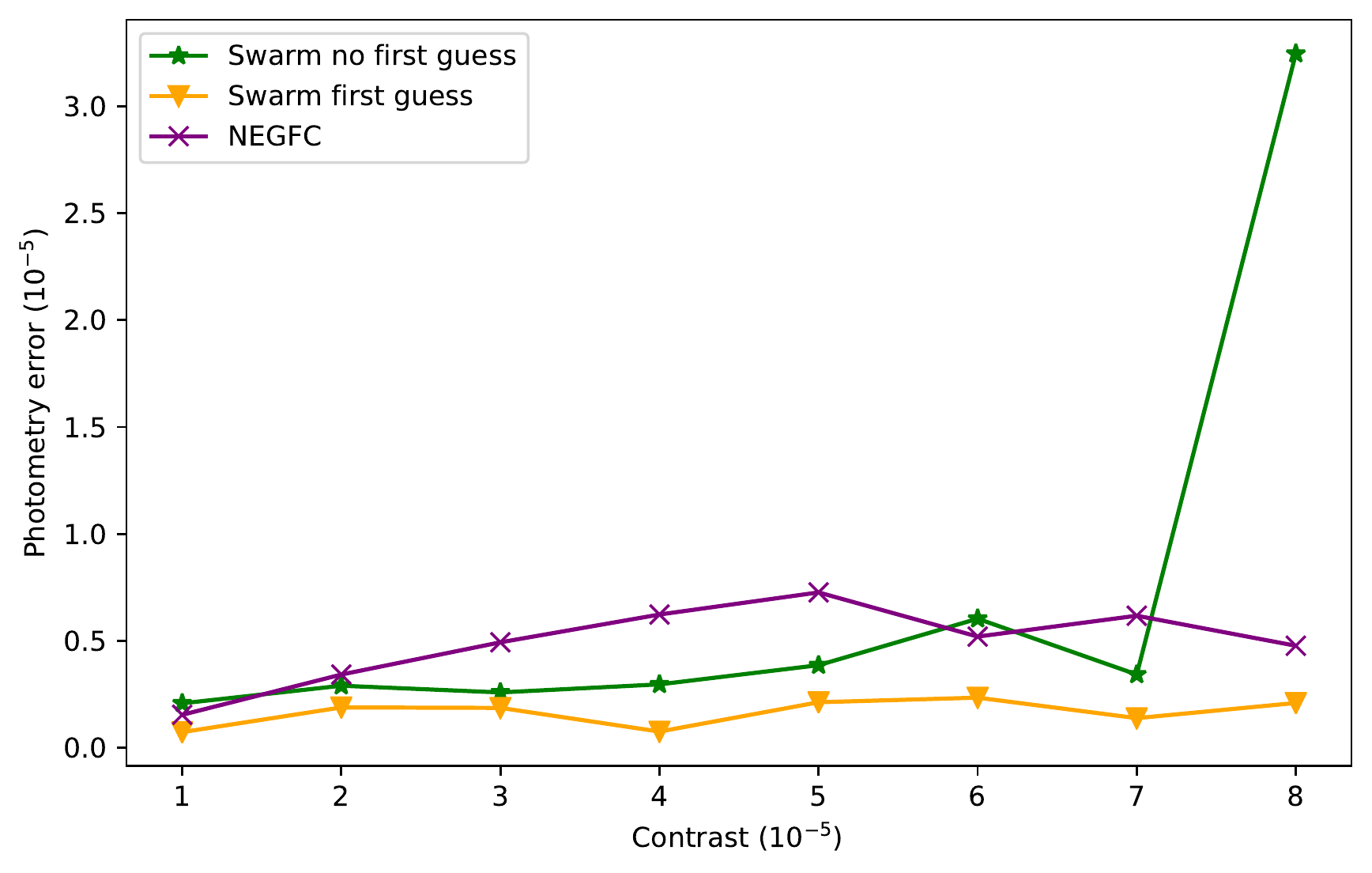}}\\

\caption{\label{perf_algo2} Astrometric and photometric errors for the NEGFC approach (purple) and for RSM based planetary signal characterisation algorithm using the PSO approach with and without the initialisation step relying on the NEGFC approach (resp. yellow and green). The upper graph shows the dependence of the averaged rms position error on the contrast, while the bottom one shows the dependence of the photometric rms error (computed at a radial distance of $4\lambda /D$).}
\end{figure}

Figure \ref{perf_algo} shows the evolution of the astrometric and photometric mean error with the contrast. The upper graph shows a higher performance of the PSO approach without local mean subtraction, except for the highest contrast value. When comparing with the NEGFC algorithm in Figures \ref{perf_algo2}, we see that our approach provides a more accurate estimation of the astromery and photometry for low contrast values, while breaking at high contrast values. This lower performance for very bright companions comes from the fact that a slight shift of the negative injected fake companion compared to the true underlying position, leads to the appearance of bright artefacts near the companion position, and therefore to a high loss function value which prevents its effective minimisation. This is explained by the very high sensitivity of the RSM map algorithm, which is a drawback in this particular case. A way to prevent this behaviour is to apply as an initialisation step the NEGFC algorithm and then use the RSM-based PSO approach. We see from Figures \ref{perf_algo2}, that this approach reduces drastically the error for very bright companions, while unfortunately decreasing the astrometric accuracy when facing faint signals (but increasing the overall photometric accuracy). The optimal solution would be one combining both approaches, relying on the NEGFC approach to initialise the PSO algorithm as from a given brightness threshold.

\subsection{Point-source characterisation}
\label{subsec:Characterisation}

\begin{figure}[t]
\footnotesize
  \centering
  \subfloat[HD206893]{\includegraphics[width=240pt]{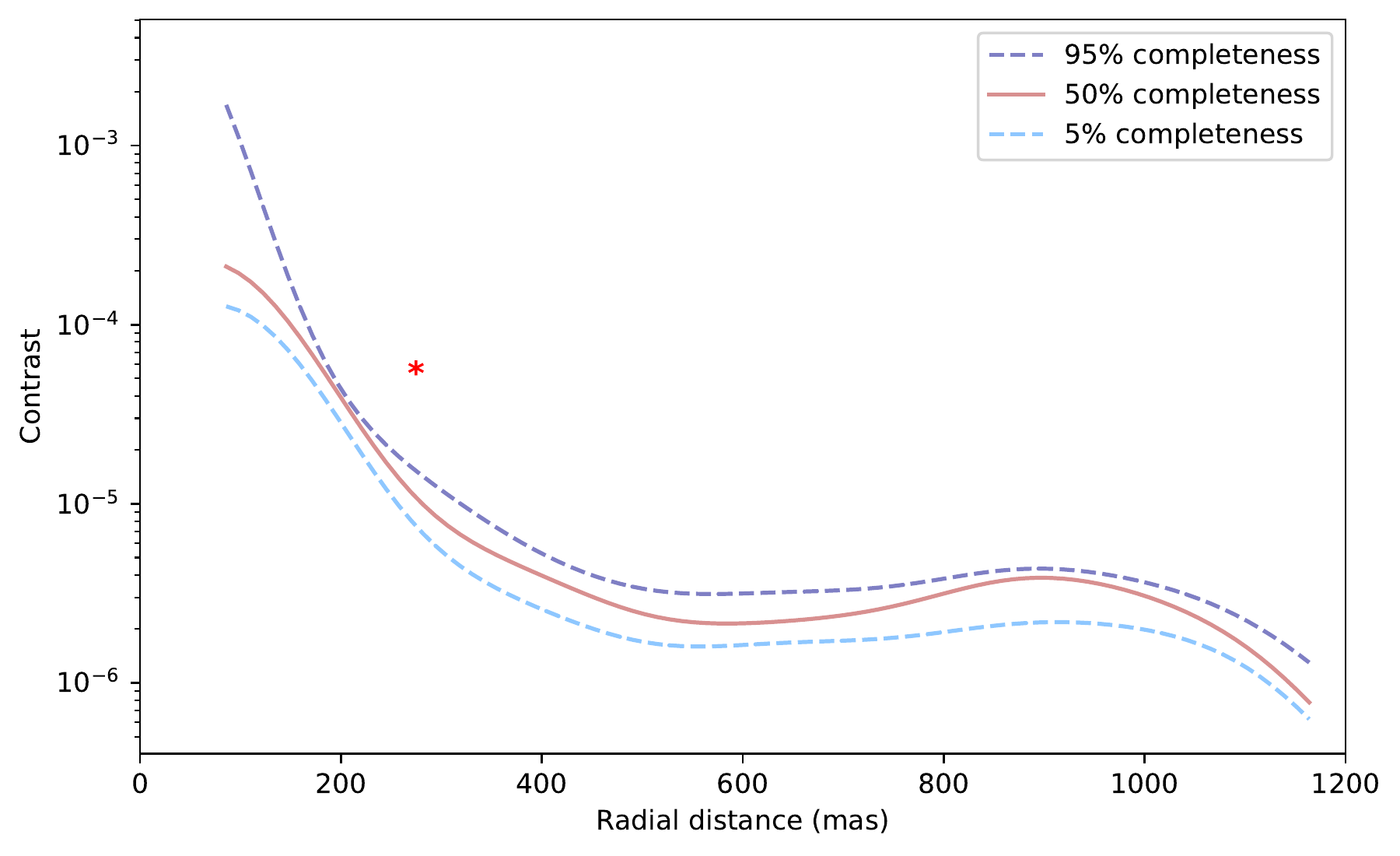}}
  \caption{\label{Target_cc} Contrast curve of the dataset for which one or multiple planetary candidates have been detected. The contrast curves have been computed at a 95\%, 50\%, and 5\% completeness level. The planetary candidate is indicated by a red star at its respective contrast and angular separation from the host star.}
\end{figure}

We applied the RSM-based planetary signal characterisation algorithm on the two targets for which signals were detected. The results are presented in Table \ref{Parameters}. Besides the astrometry and photometry, we estimated contrast curves for HD\,206893 at two additional completeness levels, 50\% and 5\%. This could further help us to classify the detected signal between planetary candidates and bright speckle, by considering its relative distance to these contrast curves. In contrast with S/N-based analysis, which relies on Gaussian assumption, there is no linear relationship between companion brightness and completeness level in RSM detection map. The distance between a companion and contrast curves estimated at different completeness levels should therefore give information about the uncertainty associated with the detection. The contrast curves were computed after removing the detected signal via the negative fake companion subtraction technique, using the parameters from Table \ref{Parameters}. Figure \ref{Target_cc} presents the contrast curves along with the detected signal positioned at its estimated contrast and angular separation. No contrast curves were computed for HD\,114082 pertaining to the difficulty of removing the disk via fake companion injections.

\begin{table*}[t]
                        \caption{Detected targets photometry and astrometry.}
                        \label{Parameters}
\centering

                        \begin{tabular}{lcccc}
                        
                        \hline
Target &Radial distance (mas)  & Position Angle ($^{\circ}$)& Contrast & Epoch\\                           
 \hline
\textbf{Confirmed detections}&&\\
  \hline
HD\,206893 b & $266.58 \pm 3.25$ &$ 159.76 \pm 0.65$ & $ 4.59 \pm 0.37 \times 10^{-5}$&2015-10-05\\
HD\,114082 BKG (1) & $803.93 \pm 1.06$ & $332.10 \pm 0.08$ & $ 7.49 \pm 0.11 \times 10^{-6}$&2016-02-14\\
HD\,114082 BKG (2) & $1082.67 \pm 0.93$ & $56.75 \pm 0.05$ & $ 1.69 \pm 0.01 \times 10^{-5}$&2016-02-14\\
                        \end{tabular}
                                \end{table*}

%Although these two targets contained already known astrophysical signals, we considered several additional measures to confirm these detections, aiming at developing a detection and characterisation pipeline that could be used in other surveys. The first measure is an intra-annulus detection rate. We started by translating, for each target, the obtained contrast into an analog to digital units (ADU) flux. We then injected fake companions at the same angular separation as the target in the emptied RSM detection map, at intervals of two FWHMs. We generated for each injection a new RSM detection map and computed based on these detection maps the probability of observing the signal above the background residual noise within the annulus. The obtained detection rate should indicate if the detected companions are only the top of a local bright structures of if they can be observed at all azimuth. These intra-annulus detection rates are provided in the last column of Table \ref{Parameters}.

We finally computed additional detection maps. We ran the Auto-RSM framework replacing the bottom-up approach by a top-down selection method to define the set of likelihoods cubes used to generate the final RSM detection maps. We also relied on the Auto-SNR framework \citep{Dahlqvist21b} to generate optimised S/N maps. This framework uses the optimised parameters of the Auto-RSM framework for the PSF-subtraction techniques, but relies on a dedicated function to select and combine the optimal set of S/N maps. We eventually computed S/N maps with APCA, NMF, LLSG, and LOCI and simply mean combined them to generate an averaged S/N map. All these detection maps are presented in Figures \ref{Target_map1}, with a yellow circle indicating the position of the detected signals.\\

\noindent\textit{HD\,206893}\\

The first detection of HD\,206893 B dates back to 2015 \citep{Milli16b}, with numerous papers devoted to its characterisation published since \citep[e.g.][]{Grandjean19,Kammerer21}. We see from Figures \ref{Target_cc} and \ref{Target_map1} that HD\,206893 B is a very bright companion, located well above the 95 \% completeness contrast curve, and visible in all detection maps. We estimate a contrast of $ 4.59 \pm 0.37 \times 10^{-5}$, which translates into a mass of  $24.76^{+ 0.67}_{-0.62}$ $M_{Jup}$  and $33.22^{+ 0.37}_{-0.34}$ $M_{Jup}$  for respectively the AMES-COND and AMES-DUSTY evolutionary models, using the estimated stellar age of 0.25 Gy taken from Table 1. These estimated masses lie inside or close to the $[5 -30]$ $M_{Jup}$ range defined in\citep[][]{Kammerer21}, while the estimated angular separation of $266.58 \pm 3.25$ mas (10.88 au) is very close to the one determined for the same epoch in \cite{Milli16b}.\\

\noindent\textit{HD\,114082}\\

Although the RSM approach is not designed to unveil large structures, the debris disk around HD\,114082, first detected by \cite{Wahhaj16}, is clearly visible. Two point-like sources are also visible. They are situated at an estimated distance of $803.93 \pm 1.06$ mas and $1082.67 \pm 0.93$ mas from HD\,114082.  These signals are visible in all detection maps from Fig.~\ref{Target_map1}. HD\,114082 being in a dense field, we rely on TRILEGAL stellar population model \citep{TRILEGAL} to infer the density of background stars around HD\,114082. This density is then used to estimate the probability of observing two or more background stars at a distance below $1082.67$ mas from HD\,114082, using a spatial Poisson point process. This probability is equal to $63.5$ \%, and increase to $88.5$ \% when considering the probability of observing one or more background stars. Considering these high probabilities and the high inclination of these objects compared to the debris disk, these detections are most likely background stars. A second-epoch follow-up and an astrometric analysis is presented in \citet{Engler2022} and confirmed that those two sources are background sources without proper motion.\\

\section{Contrast curves analysis}
\label{sec:CCanalysis}

The contrast curves computed in Section \ref{subsec:Contrastcurve} are used throughout this section as a measure of the ADI sequences quality, as well as a metric for the RSM map algorithm performance. 

\subsection{Influence of clustering}

\begin{figure}[t]
\footnotesize
  \centering

  \includegraphics[width=240pt]{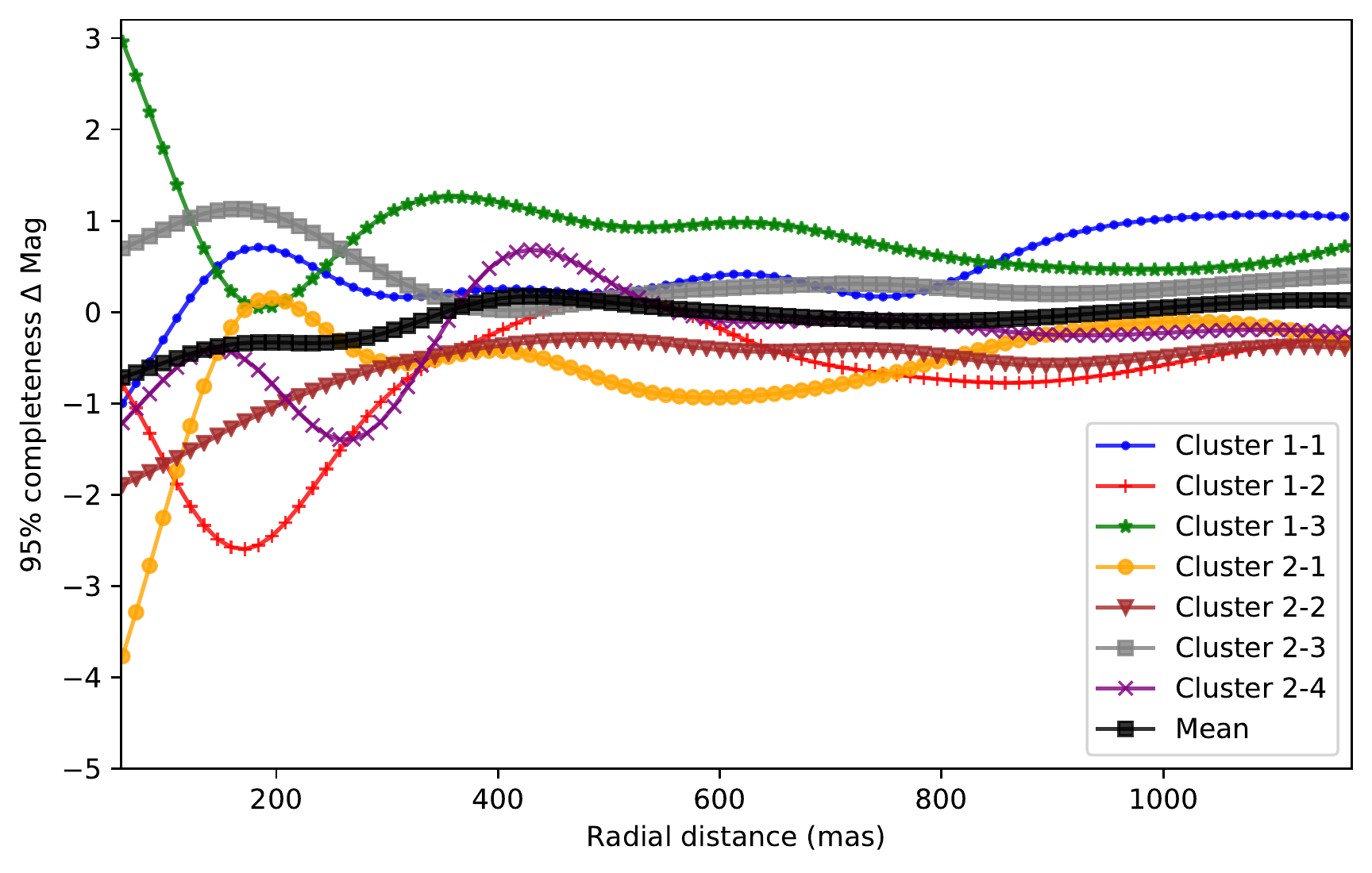}
 
  \caption{\label{Perfmembervscenter} Angular evolution of the average contrast difference between cluster members and their cluster center expressed in terms of $\Delta$ mag. A positive $\Delta$ mag indicates a poorer contrast for the cluster members.}
\end{figure}

We start by comparing the contrast curves obtained for the cluster centroids and the ones obtained by applying the centroids optimal parameters on the remaining targets of the cluster. The comparison aims to determine if the cluster centroids, for which the optimal parametrisations were computed, do perform better than the other members of the cluster in terms of achievable contrast. This should provide an idea of how far from the optimum we are, the optimum being the case where Auto-RSM is applied on every target. We have estimated the difference between each of the members and their cluster centroid in terms of $\Delta$ mag\footnote{We expressed both contrast curves in terms of magnitude and then subtracted the magnitudes of the members from the one of the cluster center.}, and report in Figure \ref{Perfmembervscenter} the radial evolution of this measure averaged, for each cluster, over their set of members. 

Looking at the seven curves\footnote{The cluster composed of HD\,14082B, HD\,82943, HD\,107649 was not included in the analysis as two of the three cluster members were rejected, due to the presence of multiple extended speckle-like bright structures in the HD\,107649 detection map and the existence of better epochs for HD\,82943.}, the center seems to perform better for some clusters (see clusters 1-1, 1-3, and 2-3), while for others the cluster members show a higher performance (see clusters 1-2, 2-1, 2-2, and 2-4). Surprisingly, we observe on average a small increase of the performance in terms of contrast for the cluster members at close angular separations. The average performance gain is close to zero at larger separations. This seems to support the use of a reduced number of optimal parameters, as it does not seem to negatively impact the performance within the different clusters.

\begin{figure}[t]
\footnotesize
  \centering

  \subfloat[]{\includegraphics[width=250pt]{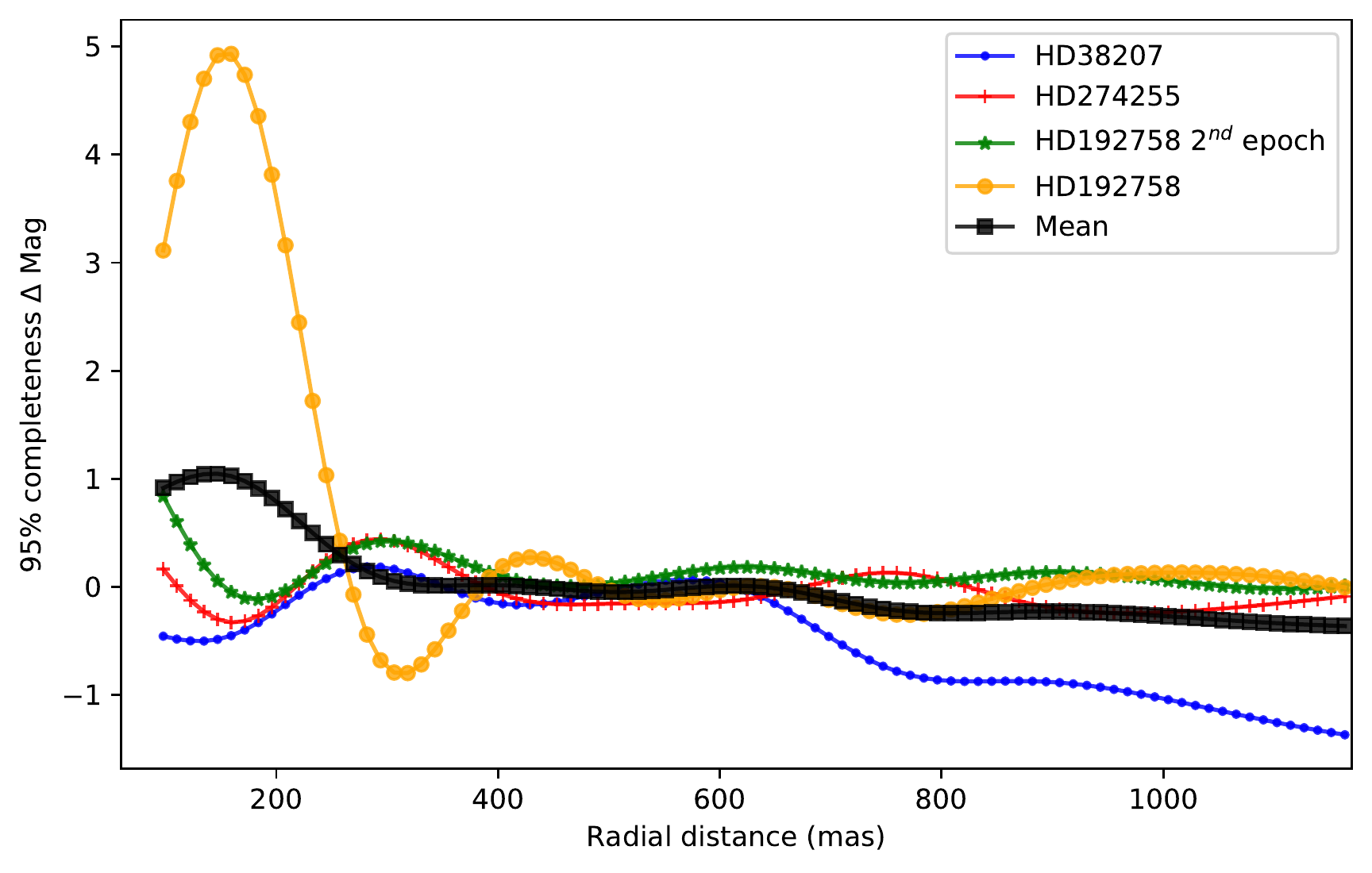}}\\ 
  \subfloat[]{\includegraphics[width=250pt]{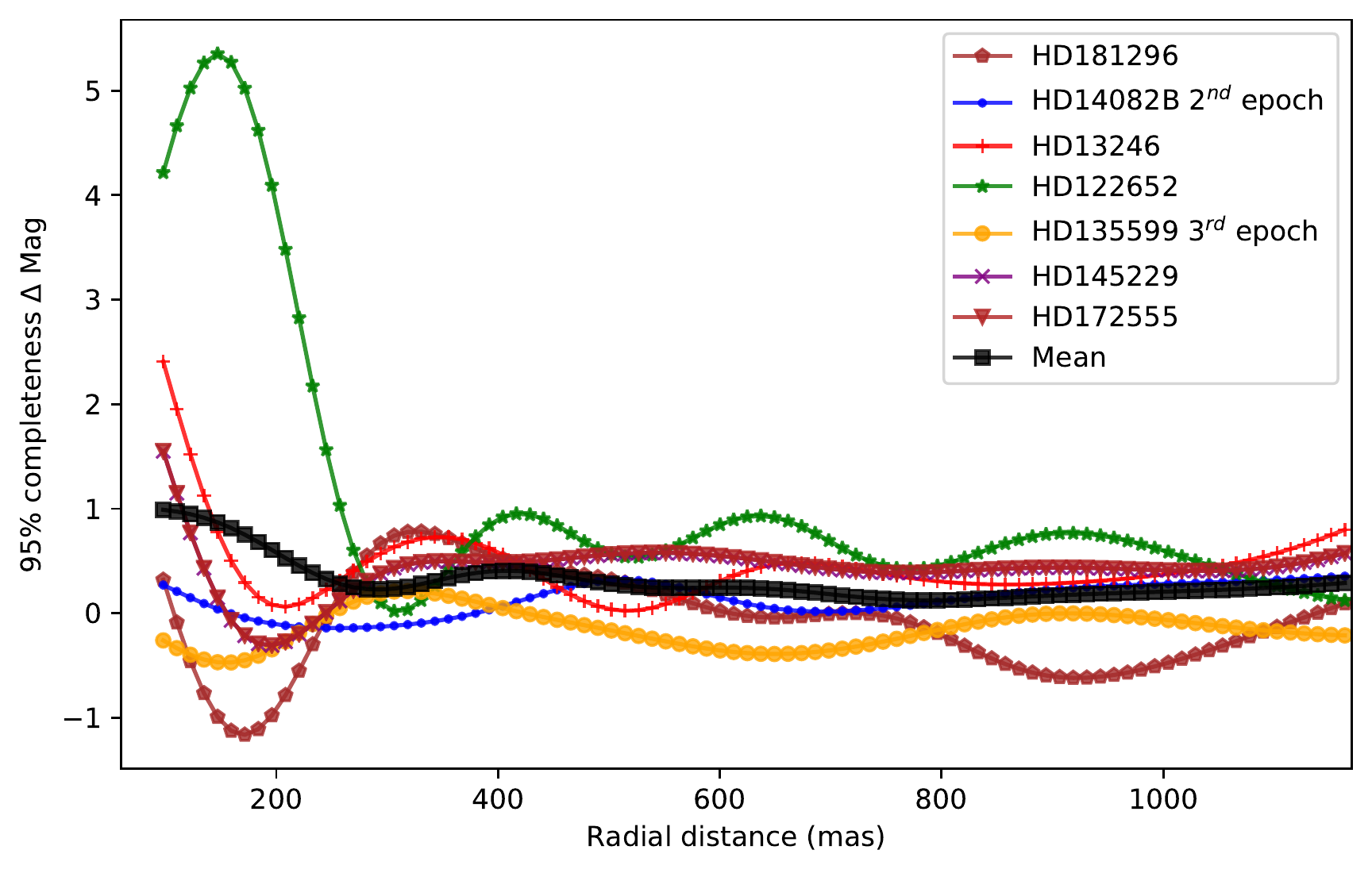}}\\        
                
  \caption{\label{clusterdeltacomp} Angular evolution of the contrast difference, for the cluster 1-1 (top) and 2-3 (bottom), between the contrast obtained with their optimal parametrisation (corresponding to the optimal parametrisation of their respective cluster center, HD\,192758 and HD\,181296) and the contrast obtained with the optimal parametrisation of another cluster center (resp. HD\,3670 and HD\,3003, i.e. the center of the cluster 1-2 and 2-4), expressed in terms of $\Delta$ mag. A positive $\Delta$ mag indicates a poorer contrast achieved with the optimal parametrisation of the other cluster centres compared to their own cluster center. The black curve provide the $\Delta$ mag averaged over the set of considered targets.}
\end{figure}

We used the same approach to assess the necessity to rely on multiple optimal parametrisations instead of a single one for the entire survey. This allows us to investigate also the impact of the degree of dissimilarity between optimal parametrisations on the performance, measured in terms of contrast. We considered two sets of clusters, one set of clusters close in terms of parametrisation, cluster 1-1 and 1-2, and one set of clusters presenting a larger level of dissimilarity, cluster 2-3 and 2-4 (see Figure \ref{optiparamcompcluster}). We computed for cluster 1-1 and 2-3, a new set of contrast curves using respectively the optimal parameters of cluster 1-2 and 2-4 (obtained for respectively HD\,3670 and HD\,3003). We then estimated the difference between these new contrast curves and the contrast curves obtained with the optimal parametrisation of their own cluster centroid (respectively HD\,192758 and HD\,181296). These contrast differences, expressed in terms of $\Delta$ mag, are shown in Fig.~\ref{clusterdeltacomp}\footnote{All targets on which the auto-RSM framework had to be applied due to the presence of bright speckles in the first detection maps, were not included in this analysis.}. As can be seen from the mean curves, using the optimal parameters estimated for their own cluster centroid leads on average to a better performance, especially at small angular separation. We see also that the mean distance is larger for the cluster 2-3, which showed a higher degree of dissimilarity in Figure \ref{optiparamcompcluster}. These results highlight the added value, at close separation, of the definition of local optimal parametrisation via Auto-RSM. The reasons for this higher performance are twofold. First, regions with a high level of background residual noise are more difficult to treat and are therefore more sensitive to parametrisation.  Secondly, Auto-RSM focuses mainly on close separations to optimise the model parameters, which explain its better performance at these distances compared to other approaches. This confirms the interest of computing several sets of optimal parameters for a large survey to account for dissimilarities in the ADI sequences' characteristics.

\subsection{Influence of environmental parameters}

\begin{figure*}[!htbp]
\footnotesize
  \centering

  \includegraphics[width=400pt]{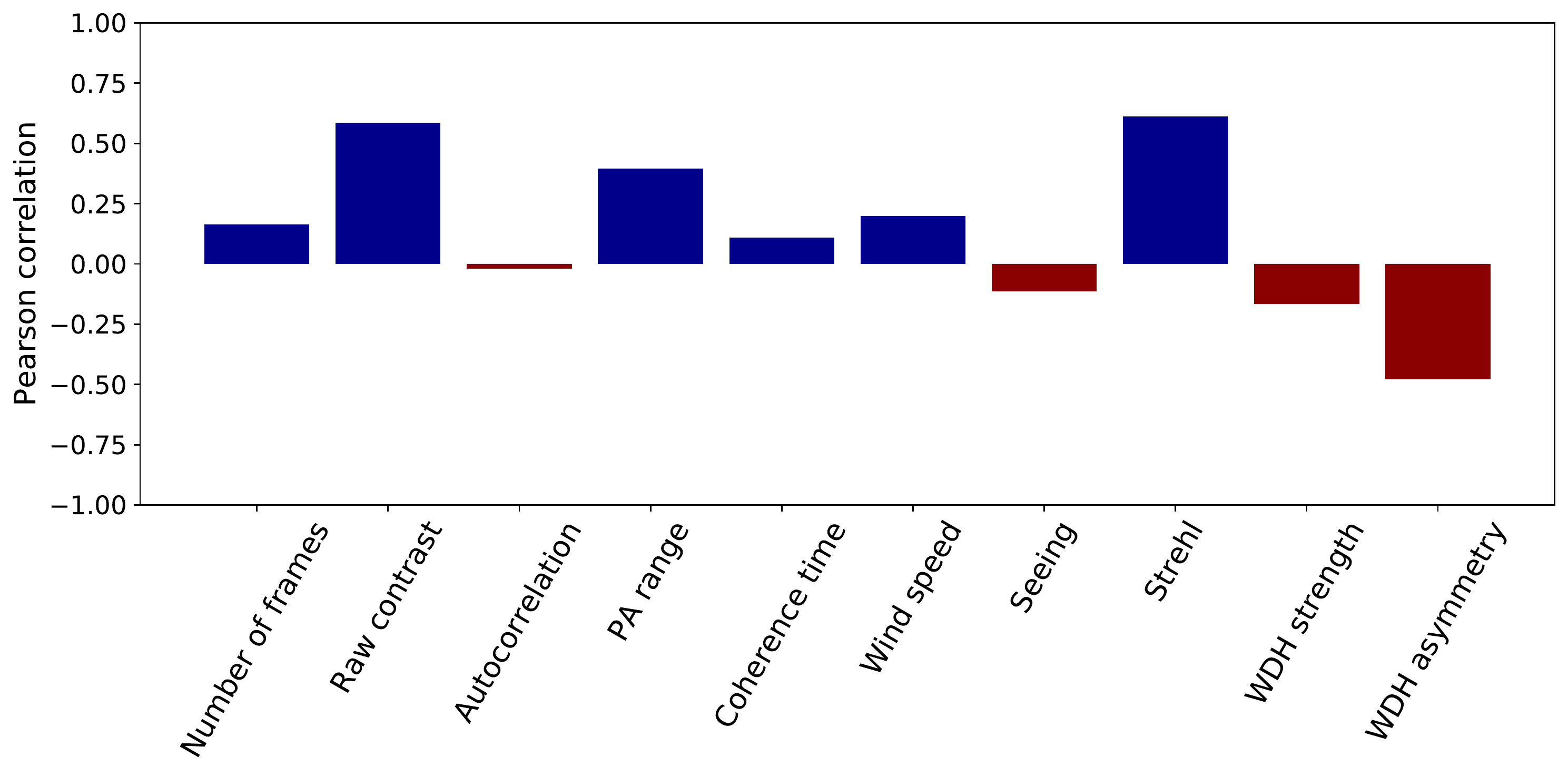}
 
  \caption{\label{PerfParam} Pearson correlations between the contrast curve median values expressed in $\Delta$ mag and the parameters characterising the ADI sequences.}
\end{figure*}

We perform a similar correlation analysis as the one made in Appendix \ref{common}, but focusing here on the relationships existing between the parameters characterising the ADI sequences and the performance in terms of achievable contrast. We start by re-expressing every contrast curve in terms of magnitude and average these magnitudes over the set of considered angular distances. We then compute the Pearson correlations between the parameters characterising the ADI sequences and the median contrast, considering the entire SHARDDS dataset. As can be seen from Figure \ref{PerfParam}, the raw contrast at 500 mas, the Strehl and the WDH asymmetry show relatively high correlations and have the expected sign. A higher asymmetry of the WDH is indeed more difficult to treat by the PSF-subtraction techniques, which do not cope well with anisotropy in the speckle field. Despite their lower correlation, the other parameters show also the expected sign. As in Table \ref{paramcor}, the lowest correlation is associated to the autocorrelation measure, indicating that the decay rate of the autocorrelation function is not the best measure of the temporal relationships between the frames.

In order to further investigate the relationship between the achievable contrast and parameters characterising the ADI sequences, we propose to rely on linear regression to highlight the parameters contributing the most to the quality of the ADI sequences. Considering the relatively low number of data points with only 60 fully treated observing sequences, and the potential co-linearities existing between the parameters, we rely on a bottom-up approach based on the Akaike information criterion \citep[AIC, ][]{Akaike74} to select one by one the parameters to be included in our model. The AIC provides a measure of the amount of information lost by a model. This measure includes a penalty term increasing with the number of parameters, providing a good trade-off between the model complexity and its goodness of fit. We start by computing the AIC for every parameter and select the parameter having the lowest AIC. We include this parameter to the model and compute again the AIC of this model after adding one at a time each of the remaining parameters. The parameter leading to the highest reduction of the AIC is then included in the model. This procedure is repeated until no more reduction of the AIC is observed. 

Table \ref{ParametersReg} gives the set of parameters that were selected using this method, along with the parameter values in the linear regression, their standard error, and p-value. We retrieve all three parameters that were already identified as highly correlated to the contrast in Figure \ref{PerfParam}, with in addition the wind speed showing a positive coefficient most probably attributable to the low wind effect. All the selected parameters show a high significance, especially the raw contrast at 500 mas and the WDH asymmetry. This highlights the importance of finding mitigation strategies to tackle the WDH to increase the quality of the ADI sequences \citep[see][]{Cantalloube20a}. With a $R^2$ adjusted for the number of parameters equal to 0.699, this simple model provides already a good indication of the expected contrast, relying on only four parameters that can be quickly computed or are already available in the metadata. 

\begin{table}[t]
                        \caption{Linear regression coefficients, standard error, and p-value for the five parameters selected via the minimisation of the AIC with as dependent variable the contrast curve median values expressed in $\Delta$ mag.  }
                        \label{ParametersReg}
\centering
                        \begin{tabular}{lcccc}
                        
                        \hline
Parameters &Coefficient&Standard error&p-value\\ 
 \hline
Contrast at 500 mas   &          0.5863   &   0.082      &   0.000 \\
WDH asymmetry   &          -0.0468   &   0.013   &    0.001 \\
Strehl           &  2.2234   &   1.020   &      0.034      \\
Wind speed     &        0.0192   &   0.010       &  0.063 \\
\hline
                        \end{tabular}
                        \tablefoot{
The minimum AIC and the adjusted R$^2$ are respectively equal to 60.04 and 0.699.}
                                \end{table}
                                
\begin{figure}[t]
\footnotesize
  \centering

\includegraphics[width=250pt]{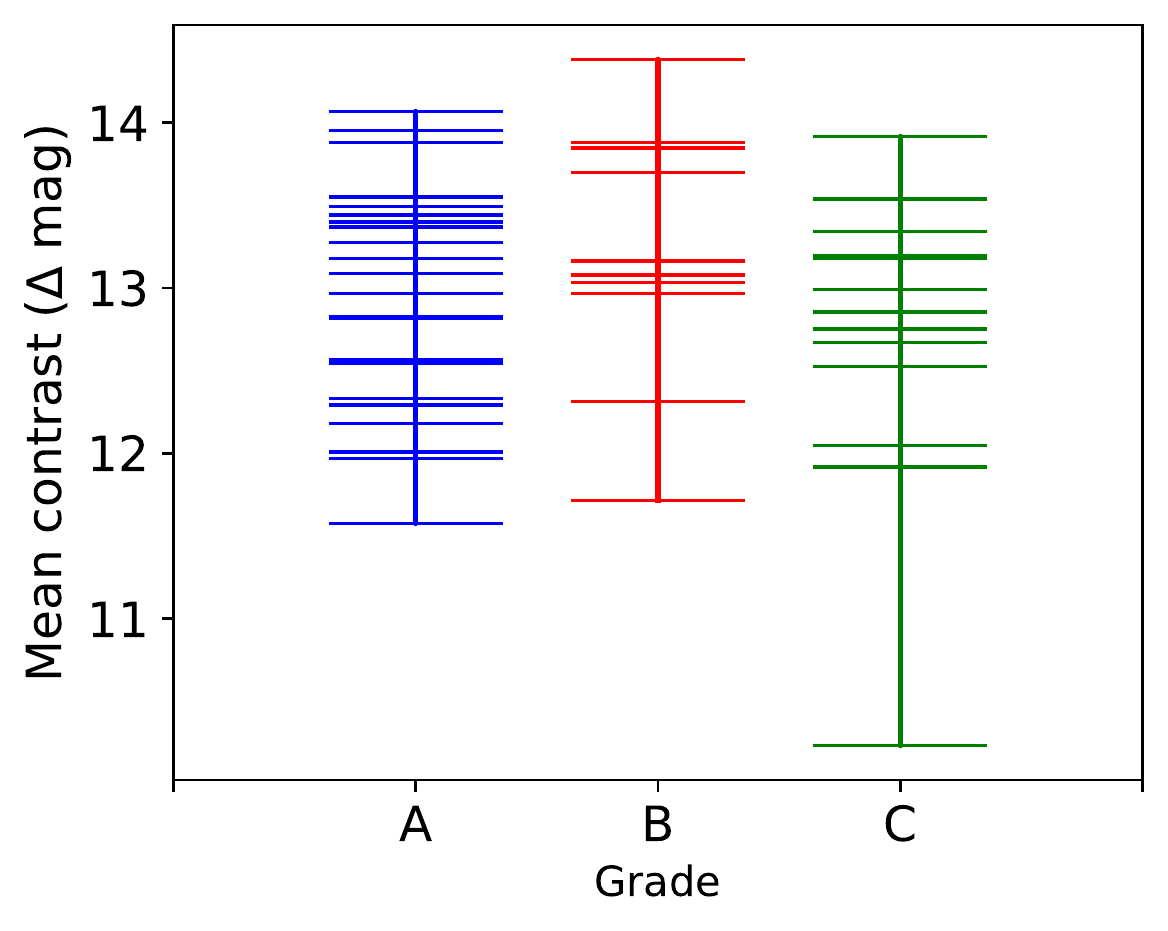}
            
  \caption{\label{Grading} Distribution of the targets mean contrast (expressed in $\Delta$ mag) in the different grading categories.}
\end{figure}

Following this analysis of the parameters driving the most the quality of the ADI sequences in terms of achievable contrast, we propose to look at existing observation quality ratings. In Figure \ref{Grading}, we report the different ADI sequences of the SHARDDS survey classified in terms of ESO observation quality grading and their respective mean contrast. As can be seen from this graph, apart from a single ADI sequence graded C showing a very low mean contrast, there are no major differences between the contrast distribution among the three grades. The ESO grading system used for this survey was mainly based on the seeing. A more robust multi-factor grading system was introduced in April 2018 \citep{Milli19}. However, a more HCI-oriented grading system based on a multi-factor linear regression, such as the one presented in Table \ref{ParametersReg}, could be an interesting tool to grade HCI observations at the telescope, and/or inform the post-processing of large surveys.

\section{Survey sensitivity}
\label{sec:Sensitivity}

\subsection{Target detection probability}
\label{subsec:Detectprob}

The median contrast curves provide a good metric for the quality of the ADI sequences of the SHARDDS survey, and its relationship with the observing conditions. However, this analysis did not provide information about the global sensitivity of the SHARDDS survey to planets. In this section, we translate these contrast curves into upper limits on the detectability of planets depending on their semi-major axis and their mass, using respectively an astrodynamic and an evolutionary model. The astrodynamic model relies on Keplerian motion to determine the range of angular separations covered by a planet depending on its orbital elements. The evolutionary model describes how planets cool down over time depending on their mass.

Different evolutionary models were developed and refined in the past decades. For the sake of continuity with previous studies, we choose two well-known models, namely the AMES-DUSTY \citep{Chabrier00} and AMES-COND \citep{Baraffe03} models. Both models assume planet formation via direct collapse of part of the disk due to gravitational instabilities. Disk instabilities are assumed to be the main scenario for the formation of giant planets and brown dwarfs at large distance from their host star (>10 au). The tables of cloud-free atmosphere AMES-COND, and dusty atmosphere AMES-DUSTY models for SPHERE were used to convert the contrast curves ($\Delta mag$) into planetary mass curves, knowing the age and the magnitude in H-band of the host star.

Having computed the planetary mass sensitivity curves for all targets, we have now to determine the accessible range of angular separations corresponding to a given semi-major axis. This range of angular separations is used alongside the planetary mass curves to compute the detection probabilities for the set of masses and semi-major axis that form the grid points of the planetary detection probability map. We define the range of angular separations for a given semi-major axis, by computing the projected distance between the planet and the host star, as seen from the Earth, for multiple sets of randomly generated orbital elements (eccentricity, inclination, argument of the periapsis, longitude of the ascending node, and mean anomaly). The detailed computation of the projected angular separations is provided in Appendix \ref{Orbit}. 

For each target of the survey, 150 semi-major axes, ranging from 0.1 to 1 000 au and 100 planetary masses, ranging from 0.1 to 100 $M_{Jup}$ are uniformly distributed in log space to form our grid. For each point in the grid, 5000 sets of orbital elements are defined, using a uniform distribution for the inclination, the argument of the periapsis, the longitude of the ascending node, and the mean anomaly. For the inclination, we rely on a uniform distribution in sine to take into account the higher number of configurations for near edge-on orientations compared to face-on orientations, and ensure isotropy. The eccentricity follows a Beta distribution with parameters $\alpha=0.95$ and $\beta=1.30$, corresponding to the best fit to the full sample of wide substellar companions obtained by \cite{Bowler20}. The planetary detection rate is then computed for each target and each grid point, as the fraction of the 5000 drawn angular distances for which the considered mass lies above the planetary mass sensitivity curves. The obtained values are then averaged over the entire set of targets and multiplied by 0.95 to account for the selected completeness of the contrast curves. 

\begin{figure*}[!htbp]
\footnotesize
  \centering

  \subfloat[]{\includegraphics[width=500pt]{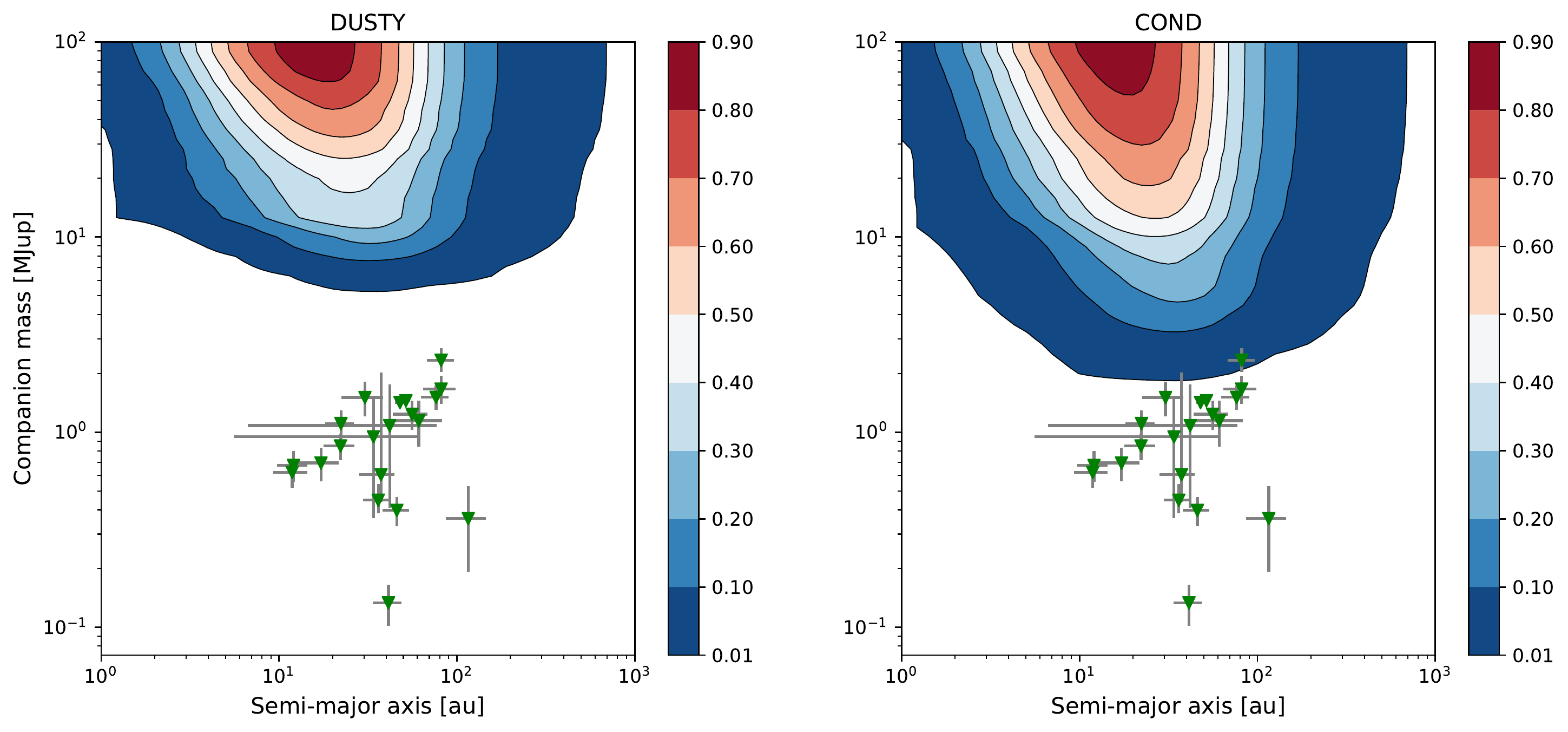}}\\       
                
  \caption{\label{detectprob} Detection probability as function of companion mass and semi-major axis. The contour plots have been calculated using the AMES-COND and AMES-DUSTY evolutionary models, relying on the contrast curves generated for the 53 targets of the SHARDDS survey (Fomalhaut C and HD\,107649 have been rejected due to respectively adverse observing conditions and the presence of extended bright structures). The estimated mass and semi-major axis estimated for 21 targets of the SHARDDS survey by \cite{Pearce22} are injected in the probability map along with the associated uncertainties.}
\end{figure*}

Figure \ref{detectprob} shows the resulting planet detection probability maps as a function of companion mass and semi-major axis. We see that higher detection rates are obtained for a semi-major axis range of $[10,100]$ au with masses above $10$ $M_{Jup}$. We have superimposed on this plot, the predicted planets derived from the dynamical constraints presented in \citet{Pearce22}. This study inferred the planet properties (mass, semi-major axis and eccentricity) if the inner edge of the disk is sculpted by one or several planets, and modelled the disk morphology based on ALMA, Herschel or the star spectral energy distribution (SED). We have plotted in Figure \ref{detectprob} the minimum masses and maximum semi-major axes of the planets predicted to be sculpting the inner edges of the disks if one planet is responsible in each of the 21 systems that are common between the SHARDDS sample and that of \citet{Pearce22}. These 21 targets are presented in Appendix \ref{app_disks}. These are the minimum masses and maximum semi-major axes that a single planet would need to sculpt the inner edge of the disk. Alternatively, a more massive planet located further inwards could also have the same effect. The planet masses could also theoretically be lower if multiple planets sculpt each disc, rather than just one planet, or if the inner edge of the disk is smaller than estimated. The disk inner edge was estimated from either a blackbody fit to the Spectral Energy Distribution (SED), or if available, from resolved observation with Herschel or ALMA \cite[see Fig.~9 left in][the data being reproduced here in Appendix \ref{app_disks}]{Pearce22}. Considering the conservative limits we computed for the detection probabilities (95 \% completeness), these planets are relatively close to the detection limit when considering the AMES-COND evolutionary model.

\begin{figure}[t]
\footnotesize
  \centering
    \subfloat[]{\includegraphics[width=255pt]{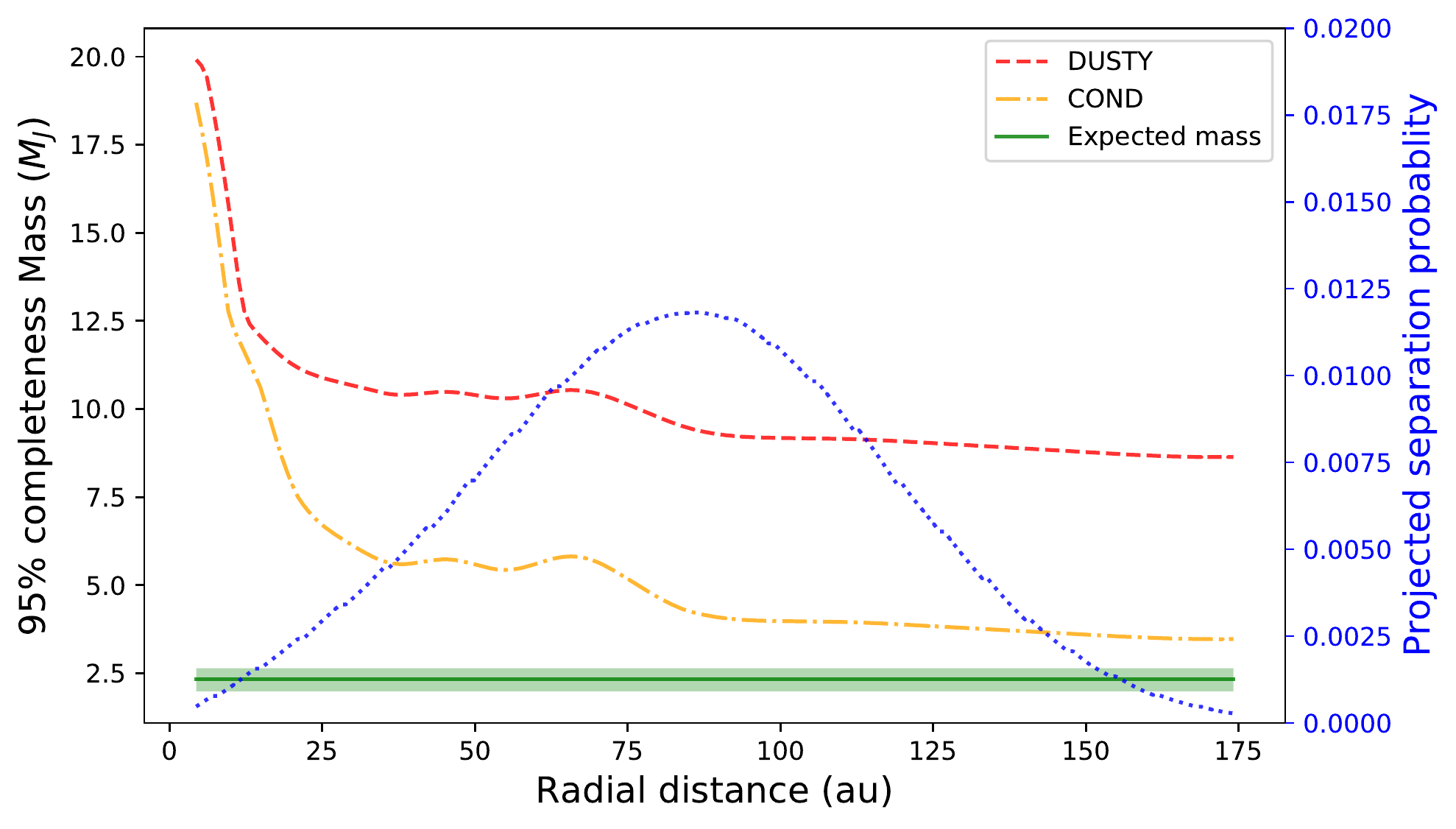}}
  \caption{\label{HD38206} Translation of the HD38206 95 \% completeness contrast curve into mass curves, using the AMES-COND (yellow dashed-dotted line) and AMES-DUSTY (red dashed line) evolutionary models. The green line provides the expected mass of the companion \citep{Pearce22} and the blue line the probabilities associated with the considered range of projected separations.}
\end{figure}

Figure \ref{HD38206} shows the contrast curve of HD38206, the most favourable target in terms of mass and semi-major axis, translated into mass curves using the AMES-COND and AMES-DUSTY evolutionary models. We computed the probability distribution of the companion's expected projected separation, using the orbital elements provided in \citep{Pearce22} and assuming a Gaussian distribution for these different orbital elements. As can be seen, the mass curve obtained with AMES-COND is very close to the expected mass of the companion for the region with the highest probability for the projected separation. 

\subsection{Occurrence rate}
\label{subsec:Occurfreq}

The definition of planetary detection probabilities allows us to derive statistical constraints on the planet occurrence rate. We consider the statistical approach proposed by \cite{Lafreniere_2007} who build confidence intervals for the planet occurrence rate relying on a Bayesian approach. We start by defining the likelihood of observing a planet characterised by a mass $m\in[m_{min},m_{max}]$ and a semi-major axis $a\in [a_{min},a_{max}]$ around star $i \in [1,N]$\footnote{For the SHARDDS  survey $N=53$ as we removed two targets from the initial set of 55 stars because of adverse observing conditions, i.e. Fomalhaut C and HD\,107649} as follows:

\begin{eqnarray}
\mathcal{L}([d_j]\vert f)=\prod_{i=0}^N (1-fp_i)^{(1-d_i)}(fp_i)^{d_i}\;,
\end{eqnarray}
where $f$ is the planet occurrence rate we are looking for, $p_i$ the previously derived planet detection probability, and $d_i$ the detections, with $d_i=1$ for the detection of a planet with $m\in[m_{min},m_{max}]$ and $a\in [a_{min},a_{max}]$ around target $i$. The occurrence rates are computed for specific points in the mass-semi-major axis space defined for the estimation of the planet detection probabilities. We replace therefore each of the ranges $m\in[m_{min},m_{max}]$ and $a\in [a_{min},a_{max}]$ by a single mass and semi-major axis point.

Following Bayes' theorem, we estimate the posterior probability distribution from the likelihood and the prior probability distribution, which we set to $p(f)=1$, assuming no prior knowledge about the distribution of the occurrence rate. The posterior probability reads:

\begin{eqnarray}
p(f\vert [d_j])=\frac{\mathcal{L}([d_j]\vert f)p(f)}{\int_0^1 \mathcal{L}([d_j]\vert f)p(f)df}\;,
\end{eqnarray}
from which we derive the minimum and maximum occurrence rate at a given level of confidence $\alpha$ by solving:
\begin{eqnarray}
\frac{1-\alpha}{2}=\int_0^{f_{min}} p(f\vert [d_j])df\;,\;\;\;\;\;\; \frac{1-\alpha}{2}=\int^1_{f_{max}} p(f\vert [d_j])df\;.
\end{eqnarray}
These last expressions simplify for grid points where no detection has been made within the considered set of targets. This is the case for all grid points except the one associated with HD\,206893 B. The simplified expression provides only the maximum occurrence rate, $f_{max}$:
\begin{eqnarray}
\alpha=\int_0^{f_{max}} p(f\vert [d_j])df\;.
\end{eqnarray}
For each considered grid point, the occurrence rates are obtained via simplex minimisation using the Nelder-Mead approach imposing a confidence level $\alpha=0.95$. 

\begin{figure*}[!htbp]
\footnotesize
  \centering

    \subfloat[]{\includegraphics[width=500pt]{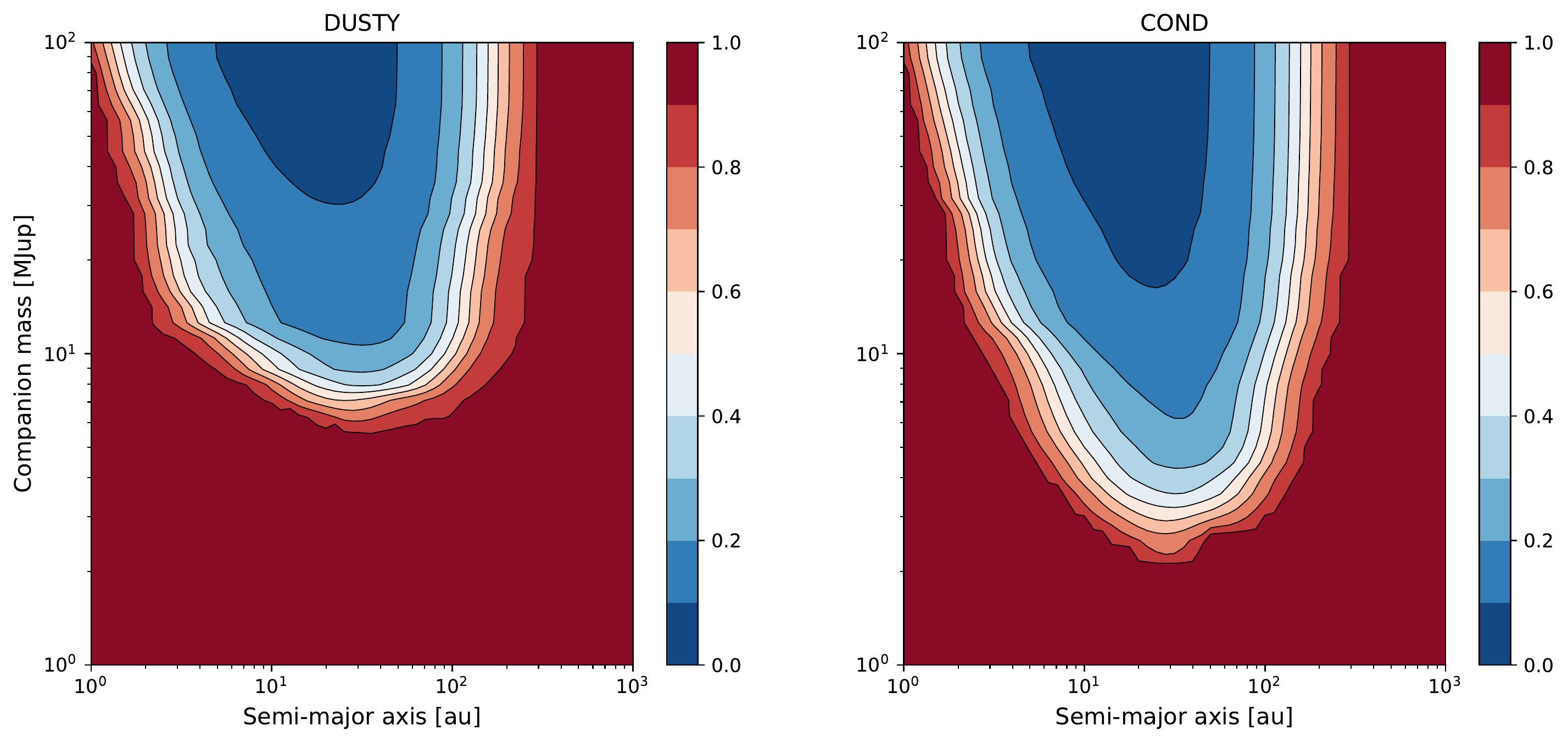}}

  \caption{\label{occfreq} 95\% confidence upper limits on the planet occurrence rate as function of companion mass and semi-major axis. The contour plots are calculated using the AMES-COND and AMES-DUSTY evolutionary models, relying on the contrast curves generated for the 53 targets of the SHARDDS survey.}
\end{figure*}

Figure \ref{occfreq} presents the upper limit of the companion occurrence rate obtained for the two considered evolutionary models, as a function of semi-major axis and mass. We see that the occurrence rate is especially low (below 10\%) for companion with masses above $20$ $M_{Jup}$ with a semi major axis ranging between $10$ and $60$ au, because of the high sensitivity of our survey to this region. The lower sensitivity towards the larger semi-major axis, and the sensitivity peak at $30$ au are explained by the stellar distances limited to 100 pc in the SHARDDS survey, as well as the field of view of 1.25 arcsec used in this study. Having considered a completeness level of 95\%,  we discarded a large fraction of the cumulative probability distribution of the contrast versus the detection probability. This approach is therefore conservative as it considers the lower bound of the planet detection probability, providing an upper limit of the planet occurrence rates.

\section{Conclusion}
\label{sec:Conclusion}

In this paper, we present an in-depth analysis of the SHARDDS survey in terms of point-source detection, based on the Auto-RSM framework. This framework is an automated optimisation algorithm relying on the RSM algorithm and multiple PSF-subtraction techniques to generate detection maps and unveil potential point sources. Although the SHARDDS survey was mainly designed to image bright debris disks in near-infrared scattered light, the detection of point sources may provide a better understanding of the interaction between planets and debris disks, and give information about the formation and evolution of circumstellar systems. 

Considering the computational cost of the Auto-RSM framework, as well as the high degree of similarity observed between the optimal parametrisations of different ADI sequences \citep[see ][]{Dahlqvist21b}, we decided to rely on clustering to reduce the number of required optimisations. We divided our dataset into eight clusters using K-means clustering algorithm, based on parameters characterising the ADI sequence itself and the related observing conditions. For each cluster, the most representative ADI sequence was selected and the Auto-RSM framework was applied on it. The generated set of optimal parameters for both the PSF-subtraction techniques and the RSM algorithm was then used to generate detection maps for all the ADI sequences contained in the cluster. The analysis of the obtained detection maps showed the presence of a higher number of bright speckles when reversing the parallactic angles, providing an important reminder that care should be taken when computing detection thresholds based on reversed parallactic angles.  

Based on the detection maps, we identified high-probability signals in only two ADI sequences: HD\,206893 B which had already been previously detected, and the bright debris disk around HD\,114082.  Although these astrophysical objects had already been identified, we proposed a multi-factor detection and characterisation pipeline to confirm the detections and characterise the signals in terms of astrometry and photometry. 

Following the analysis of the detection maps, we computed for each target a contrast curve at a 95\% completeness level, subtracting the detected signal via the negative fake companion approach when necessary. The median contrast curve demonstrated the high performance of the Auto-RSM framework, reaching a contrast of $10^{-5}$ at 300 mas and $3\times 10^{-6}$ at 600 mas. These contrast curves were then used to assess the performance of the proposed clustering approach. Using the contrast as a performance metric, we found that on average the optimal parametrisation led to slightly higher performance for cluster members compared to cluster centroids. Shifting the optimal parametrisation between clusters led to lower performance in term of contrast, especially at close separation, highlighting the interest of a clustering approach to account for dissimilarities in the ADI sequences characteristics. The quality of an ADI sequence is also shown to be driven by some key observing condition metrics such as the WDH, the Strehl, the wind speed, or the raw contrast, which could allow to one develop a simple and efficient HCI-oriented grading measure.

A planet detection probability map was then generated based on these contrast curves and on two different evolutionary models, AMES-COND and AMES-DUSTY. The planet detection probability map showed a high detection probability for a semi-major axis range of $[10,100]$ au with mass above $10$ $M_{Jup}$. We finally computed two planet occurrence rate maps based on the estimated detection probabilities, which showed a very small occurrence rate for companions with masses above 20 $M_{Jup}$ having a semi-major axis between $10$ and $60$ au.

The analysis of the SHARDDS survey allowed the development of new tools as well as the improvement of the Auto-RSM framework, allowing it to gain in maturity and become a robust HCI post-processing pipeline, achieving good performance in terms of contrasts.

%Astrometric evolution of HD192758
%https://github.com/agabrown/astrometric-sky-path/blob/master/PythonVersion/AstrometricSkyPaths.ipynb

\begin{acknowledgements}
We dedicate this work to our friend an colleague Matthew Willson who tragically passed away on 18 January 2022. He will be missed by his family, friends, and colleagues. We thank Tim Pearce for sharing his planet mass and orbital parameters dataset. This work was supported by the Fonds de la Recherche Scientifique - FNRS under Grant n$^{\circ}$ F.4504.18 and by the European Research Council (ERC) under the European Union's Horizon 2020 research and innovation program (grant agreement n$^{\circ}$ 819155). This work has made use of the SPHERE Data Centre, jointly operated by OSUG/IPAG (Grenoble), PYTHEAS/LAM/CeSAM (Marseille), OCA/Lagrange (Nice), Observatoire de Paris/LESIA (Paris), and Observatoire de Lyon/CRAL, and supported by a grant from Labex OSUG@2020 (Investissements d’avenir – ANR10 LABX56). Jonathan P. Marshall acknowledges support from the Ministry of Science and Technology of Taiwan under grant MOST109-2112-M-001-036-MY3.

\end{acknowledgements}

\bibliographystyle{aa}
\bibliography{SHARDDS.bib}

\begin{appendix}

\section{Parametrisation commonalities and relationship with ADI sequence characteristics}
\label{common}

\begin{figure}[t]
\footnotesize
  \centering
    \subfloat[]{\includegraphics[width=250pt]{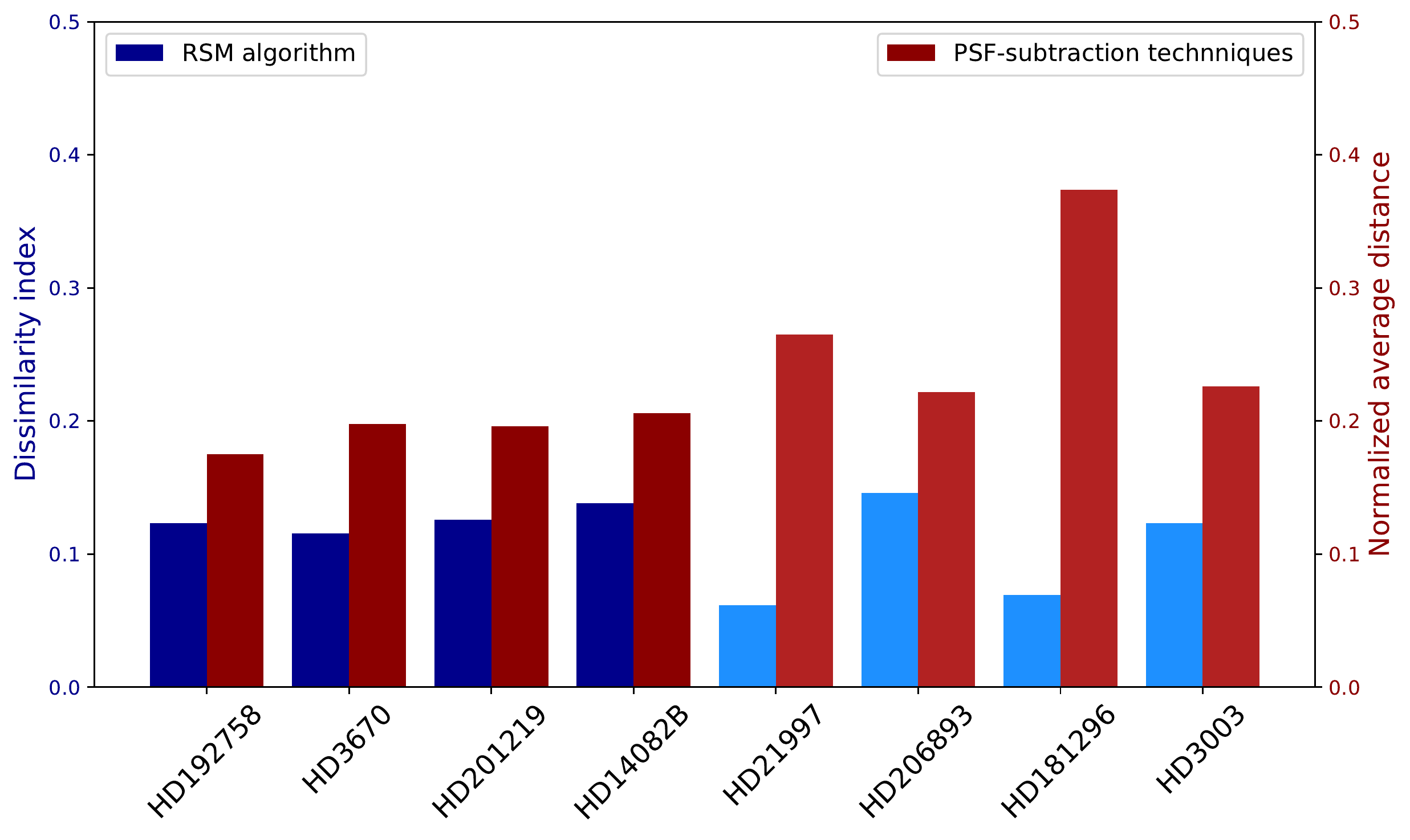}}
  \caption{\label{optiparamcompcluster} Comparison of the optimal parametrisation of the cluster centroids in terms of dissimilarity index of the RSM algorithm parametrisation and normalised average distance for the PSF-subtraction techniques, for the subset containing ADI sequences with less than 151 frames (dark colors) and the one with a number of frames above 151 images (light colors). }
\end{figure}

Following the computation of the optimal set of parameters for the eight cluster centroids, we propose to investigate the similarities existing between these eight optimal parametrisations. We also consider the relationships existing between the centroids optimal parameters and the set of metrics characterising their ADI sequence. We start by comparing in Figure \ref{optiparamcompcluster}, the obtained optimal set of parameters via a normalised distance for the PSF-subtraction techniques and a dissimilarity index for the RSM algorithm. These measures were computed for each pair of cluster centroids and then averaged over the three possible pairs within each size subset (e.g. for HD\,192758, we have HD\,192758-HD\,3670, HD\,192758-HD\,201219, and HD\,192758-HD\,14082B). The normalised distance was computed considering the 19 parameters required by the ten selected PSF subtraction techniques. For each pair of cluster centroids, we computed the absolute value of the distance between their parameters and normalised them with the mean values of these pairs of parameters\footnote{For centroid A with 20 principal components for APCA and centroid B with 24 principal components, the normalised distance is equal to $4/22=0.18$}. We then averaged the resulting distances over the 19 parameters. The normalisation ensures a proper comparison between the different parameters when consolidating the distances. For the RSM algorithm, a dissimilarity metric replaces the normalised distance as most parameters are non-numerical. This dissimilarity index is simply computed as one minus the percentage of common RSM parameters between a pair of centroids, averaged over the five parameters of the RSM algorithm.

Looking at the degree of similarity of the parametrisations within the two size subsets, Figure \ref{optiparamcompcluster} shows an overall higher degree of similarity. We observe a lower degree of dissimilarity for the RSM parametrisation and a lower normalised distance for the PSF subtraction-techniques for the centroids of the subset containing less than 151 observations. For the subset containing more than 151 observations, the slightly higher normalised distance pertain to the high degree of dissimilarity of HD\,181296, which affects strongly the averaged normalised distance. The main drivers of the dissimilarity is the number of segments used for APCA and LLSG\footnote{The number of segments correspond to the number of subdivisions of every annulus during the estimation of the reference PSF when relying on APCA and LLSG.}, the tolerance parameter of LOCI, and the method used to compute the residual speckle noise statistics within the RSM algorithm. These results tend to demonstrate the relatively high stability of the ADI sequence imaged by the VLT/SPHERE instrument and confirm the conclusions drawn in \cite{Dahlqvist21b}. The impact of the dissimilarities in the optimal parametrisations on the performance in terms of achievable contrast is further investigated in Section\ref{sec:CCanalysis}.
                                
We now turn to the analysis of the relationship existing between the parameters that we selected in Section \ref{subsec:Clustering} to describe our dataset and the parametrisations of the PSF-subtraction techniques\footnote{Such an analysis is not possible with the parametrisation of the RSM map algorithm as most parameters are non numerical.}. We computed the Pearson correlation between the ten parameters characterising our sample and the PSF-subtraction techniques parameters, considering the eight cluster centroids as data-points. The raw correlations show a significant correlation between these sets of parameters, with overall, around 25\% of the obtained values over 0.5. Table \ref{paramcor} gives the absolute values of the obtained correlations averaged over five classes of parameters, the number of principal components, the FOV rotation threshold, the number of segments, the rank of LLSG, and the tolerance of LOCI. Looking at these consolidated results, the contrast at 500 mas shows the highest average correlation rate, while the exponent of the autocorrelation function has the lowest one. Once averaged over the five considered classes, the percentage of consolidated correlations above 0.5 reach only 16\%, indicating the existence of some discrepancies between the different PSF-subtraction techniques relying on the same parameter. 

\begin{table*}[h!]
                        \caption{Average absolute Pearson correlations between the PSF-subtraction techniques parameters and the parameters selected to characterised the SHARDDS survey dataset.}
                        \label{paramcor}
\centering
\footnotesize

                        \begin{tabular}{lcccccccccc}
                        
                        \hline
Parameters  & \# frames & Contrast & Auto-corr exp & PA & Coherence &Wind speed&Seeing&Strehl&WDH S&WDH A\\                           
 \hline
Principal components&0.44&0.41&0.32&0.44&0.36&0.28&0.37&0.39&0.33&0.54\\
FOV rotation threshold&0.32&0.54&0.36&0.16&0.35&0.7&0.37&0.13&0.55&0.16\\
Number of segment&0.41&0.34&0.22&0.42&0.33&0.42&0.29&0.45&0.15&0.24\\
Rank&0.36&0.29&0.14&0.33&0.19&0.14&0.19&0.41&0.17&0.51\\
Tolerance&0.49&0.6&0.17&0.71&0.24&0.34&0.64&0.31&0.21&0.13\\
\hline
                        \end{tabular}
                                                \tablefoot{
WDH S and WDH A stand respectively for wind driven halo strength and asymmetry.}
                                \end{table*} 
\FloatBarrier

\onecolumn
\section{RSM detection maps}
\label{detmap}
This section contains the RSM detection maps containing no plausible planetary signals.\\

\begin{figure*}[!htbp]
\footnotesize
  \centering

  \subfloat[HD40540]{\includegraphics[width=120pt]{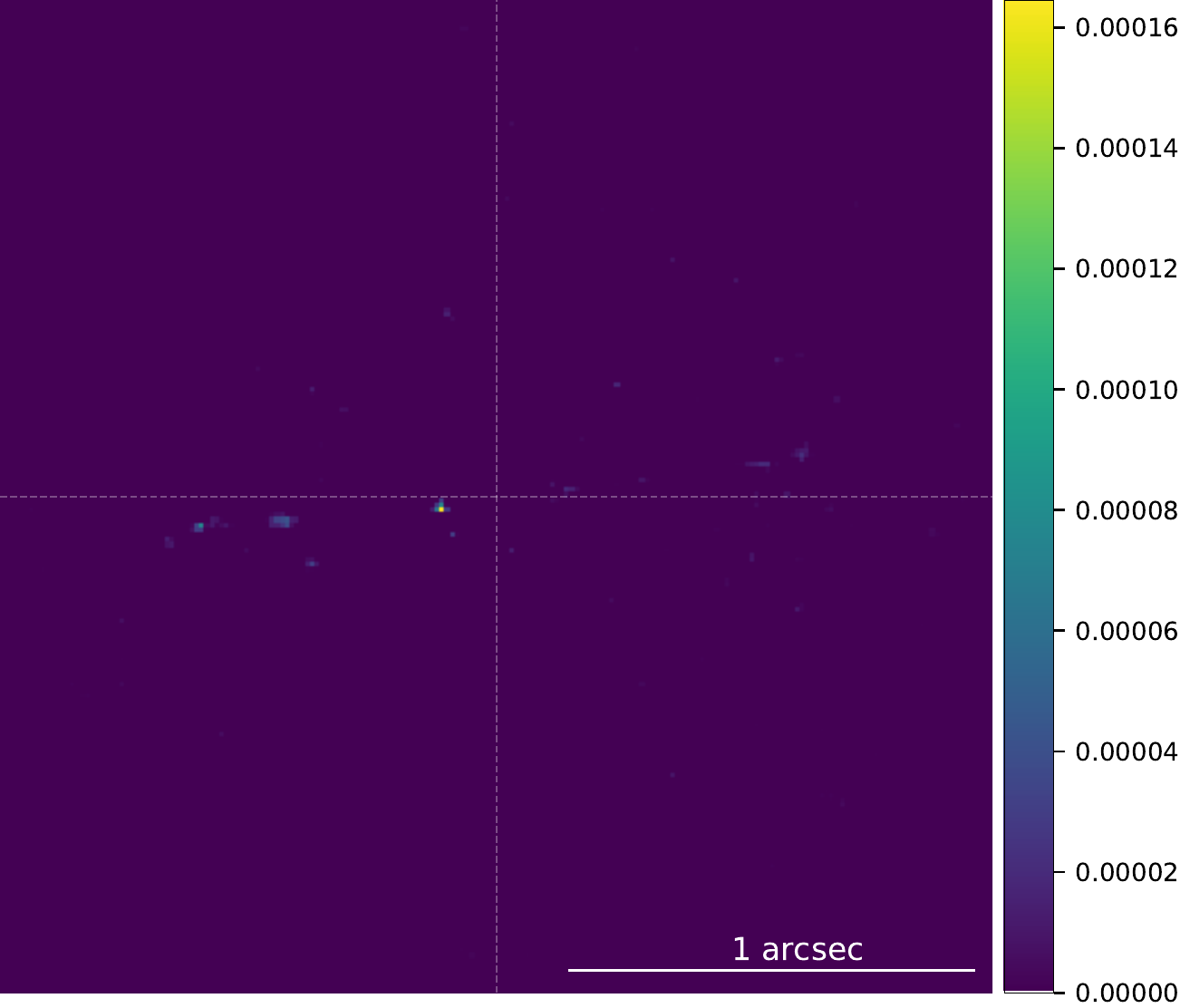}}
    \subfloat[HD38207]{\includegraphics[width=120pt]{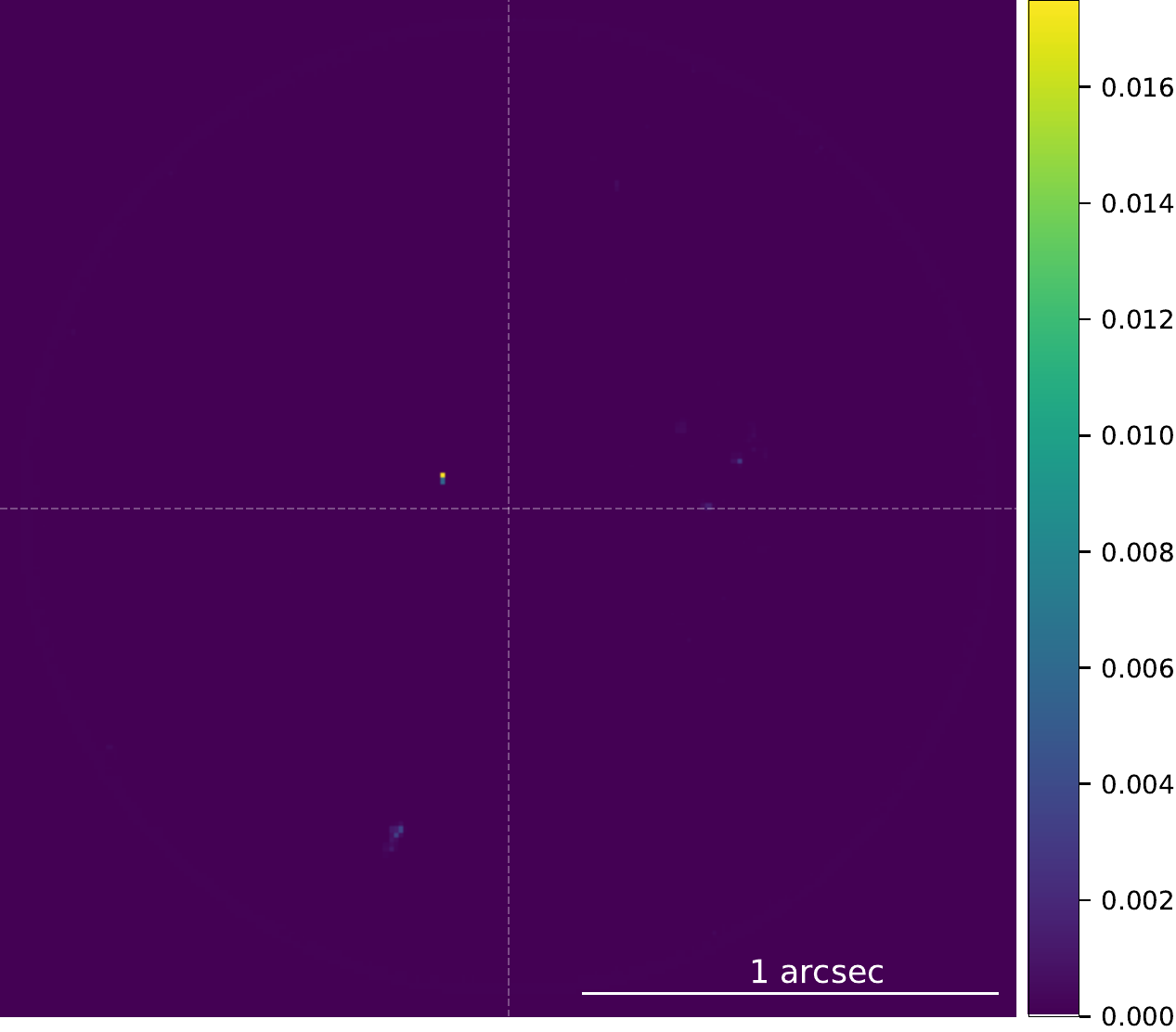}}
        \subfloat[HD38206]{\includegraphics[width=120pt]{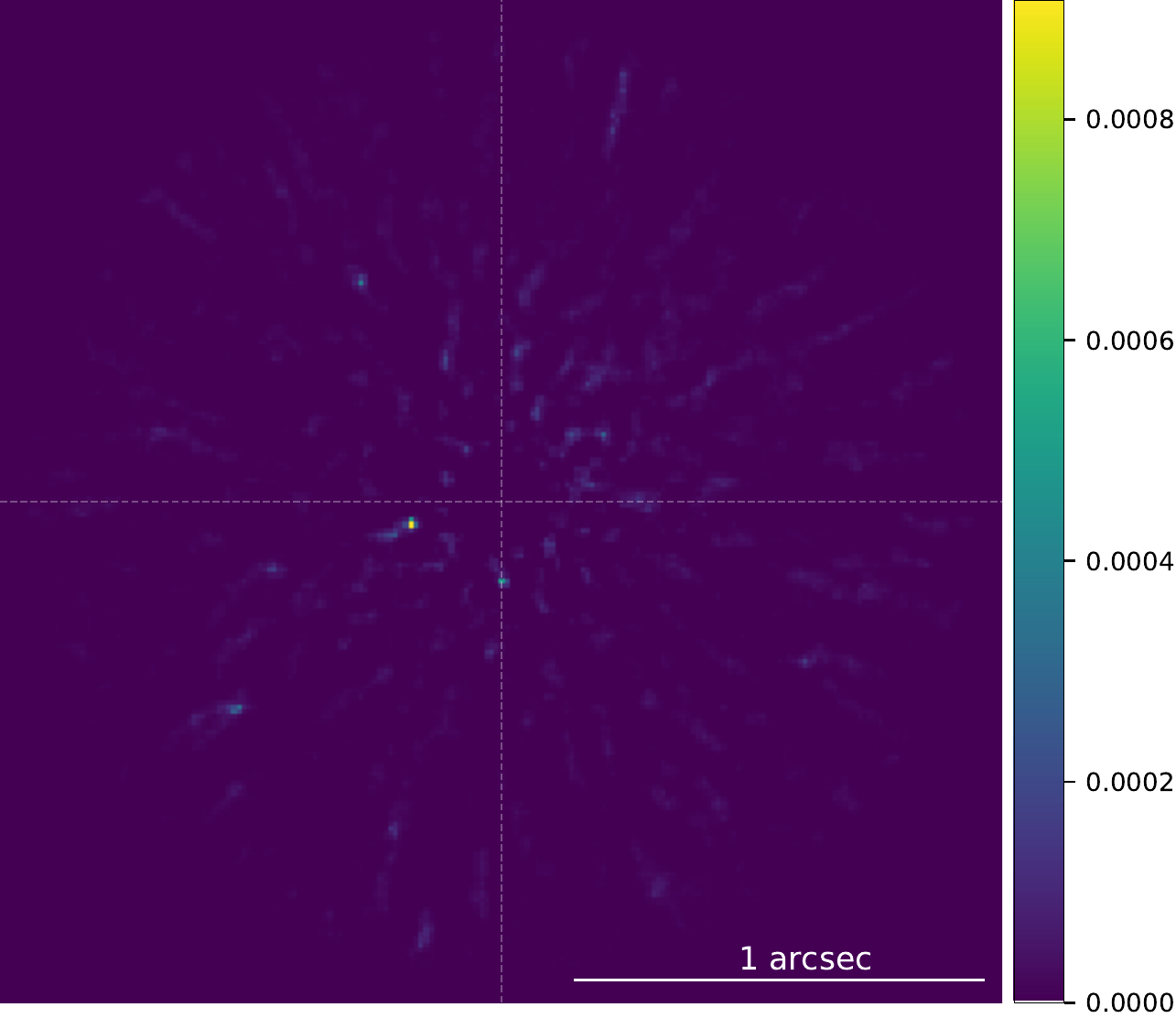}}
        \subfloat[HD37484*]{\includegraphics[width=120pt]{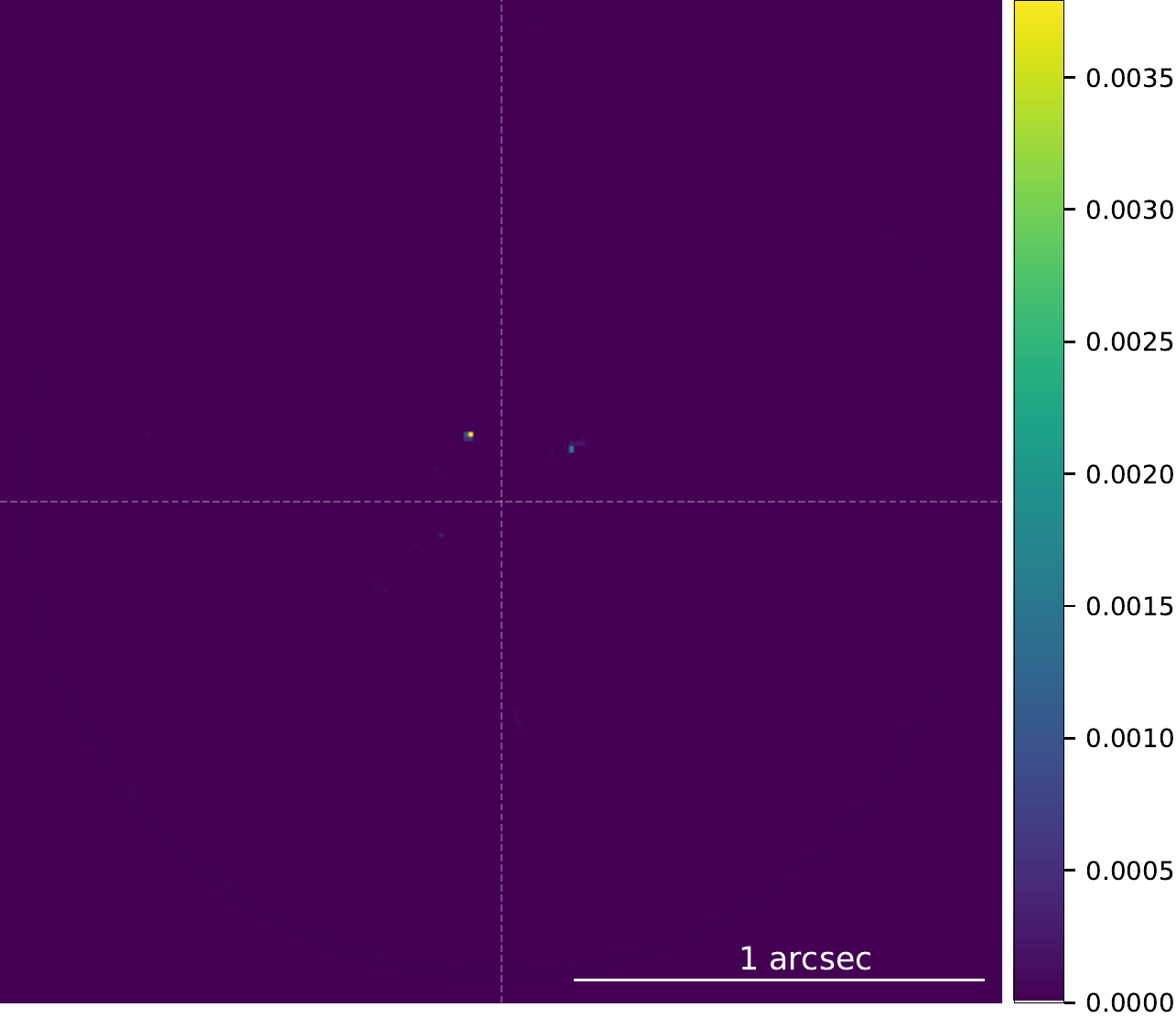}}\\
          \subfloat[HD37484 (2$^{nd}$ epoch)*]{\includegraphics[width=120pt]{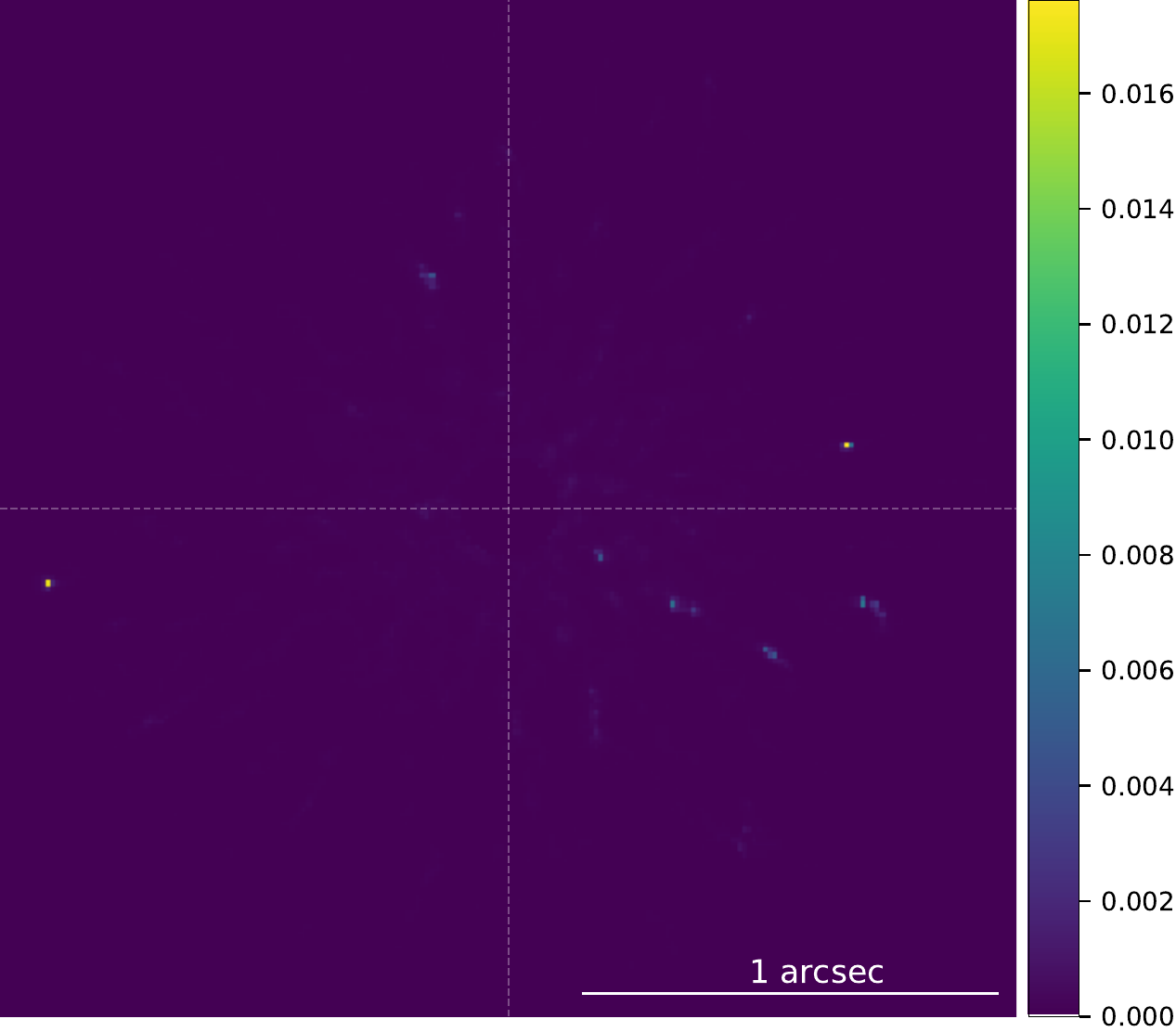}}
    \subfloat[HD35650]{\includegraphics[width=120pt]{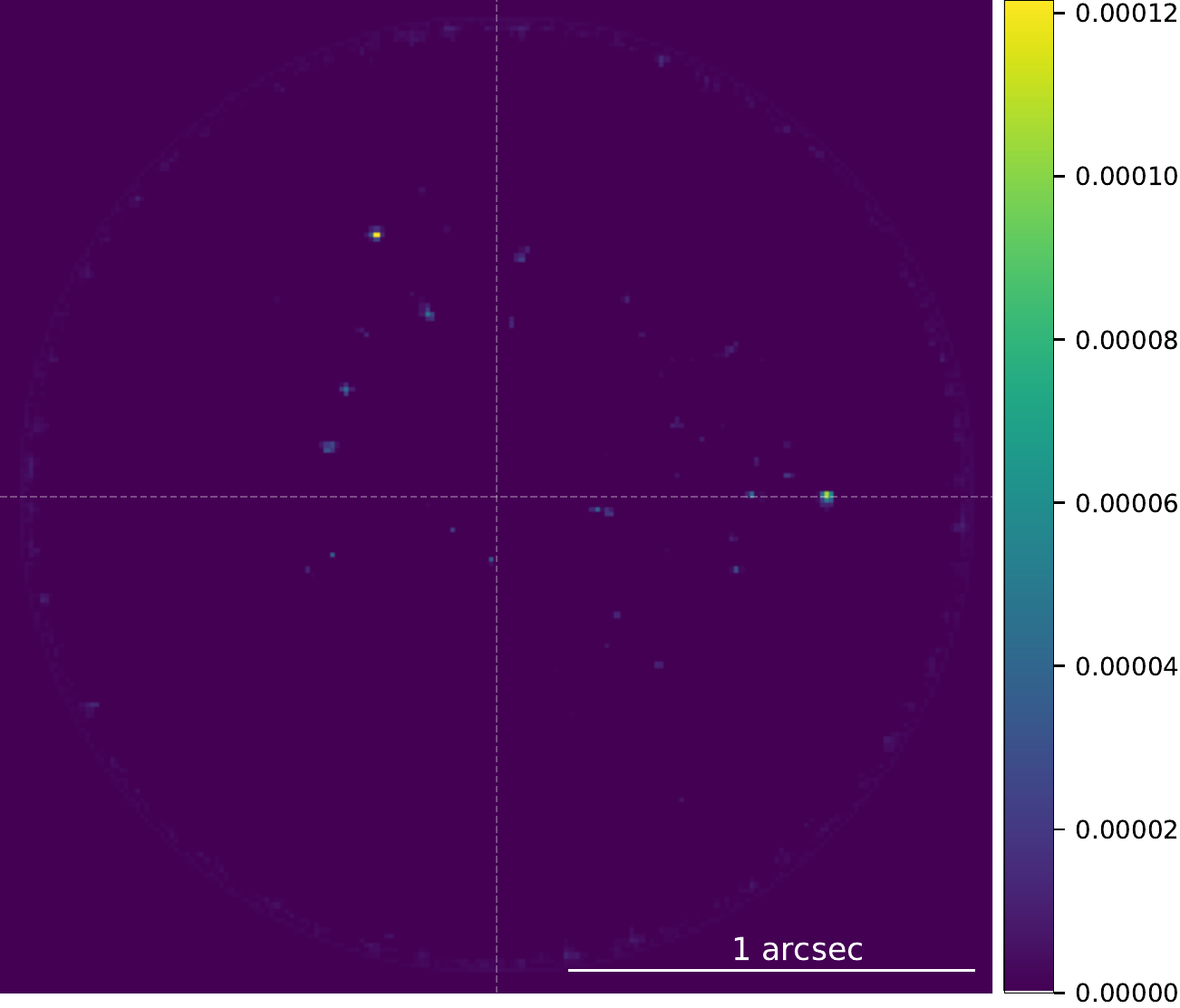}}
        \subfloat[HD31392]{\includegraphics[width=120pt]{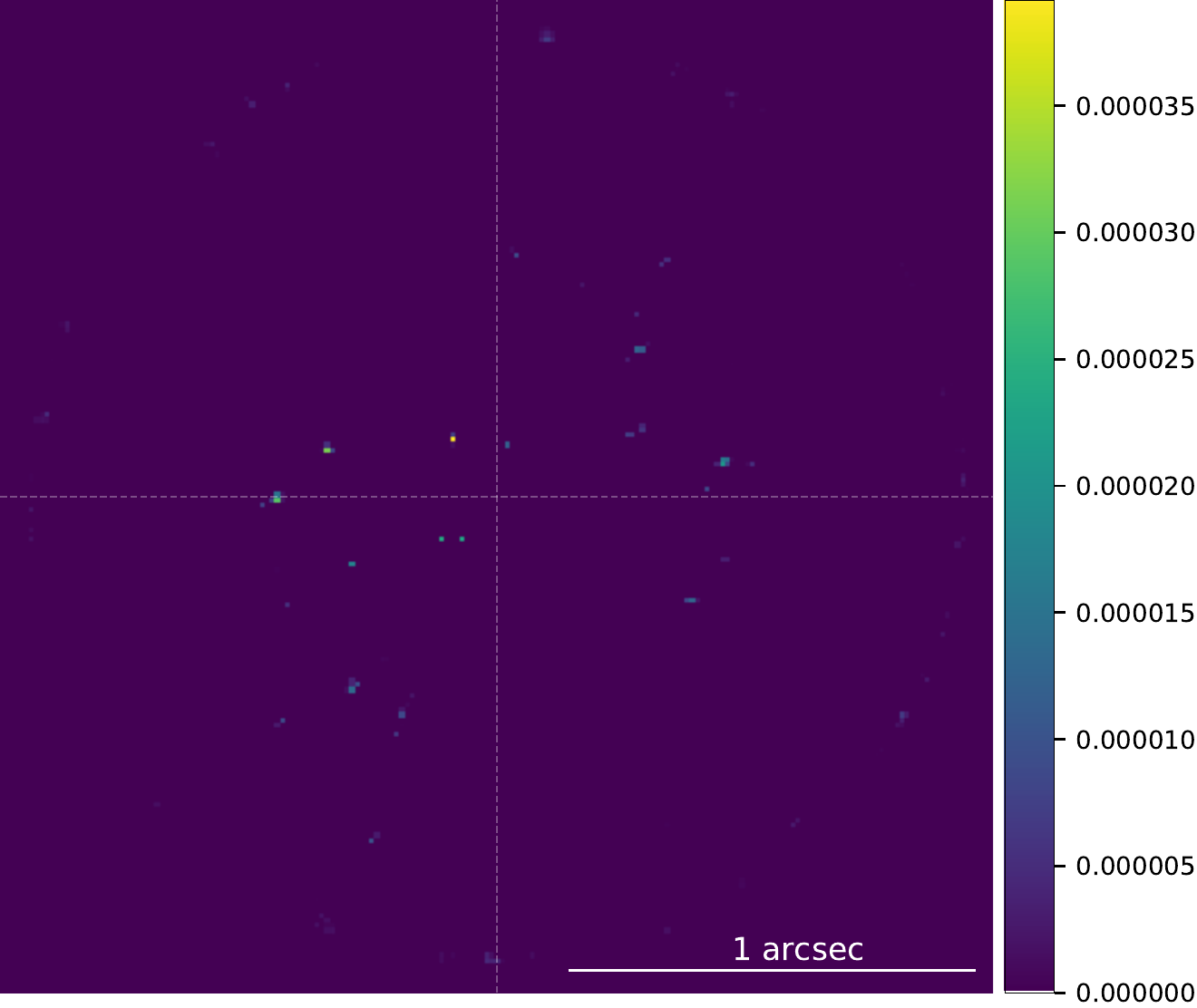}}
        \subfloat[HD25457]{\includegraphics[width=120pt]{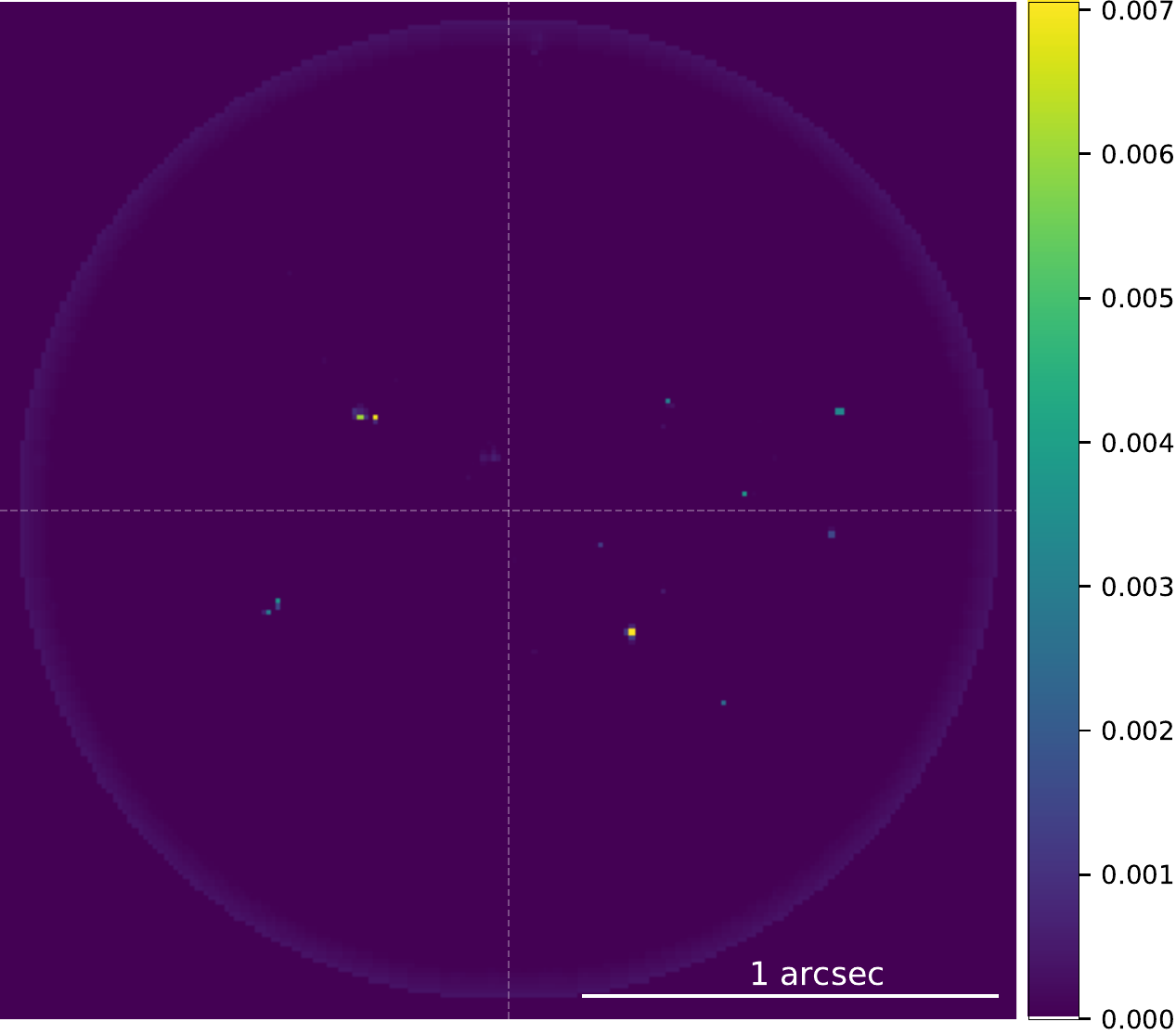}}\\
          \subfloat[HD24636]{\includegraphics[width=120pt]{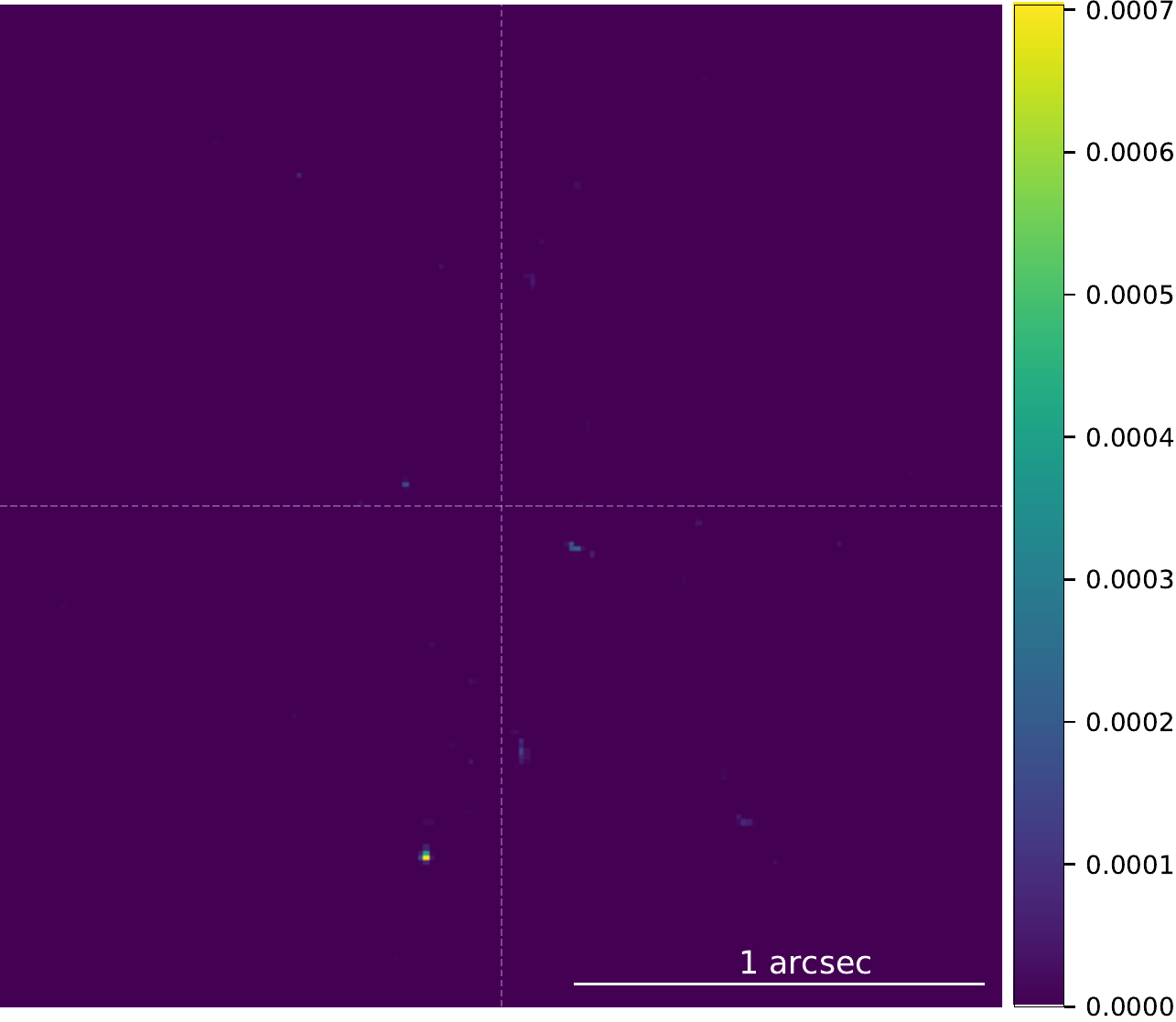}}
    \subfloat[HD22179]{\includegraphics[width=120pt]{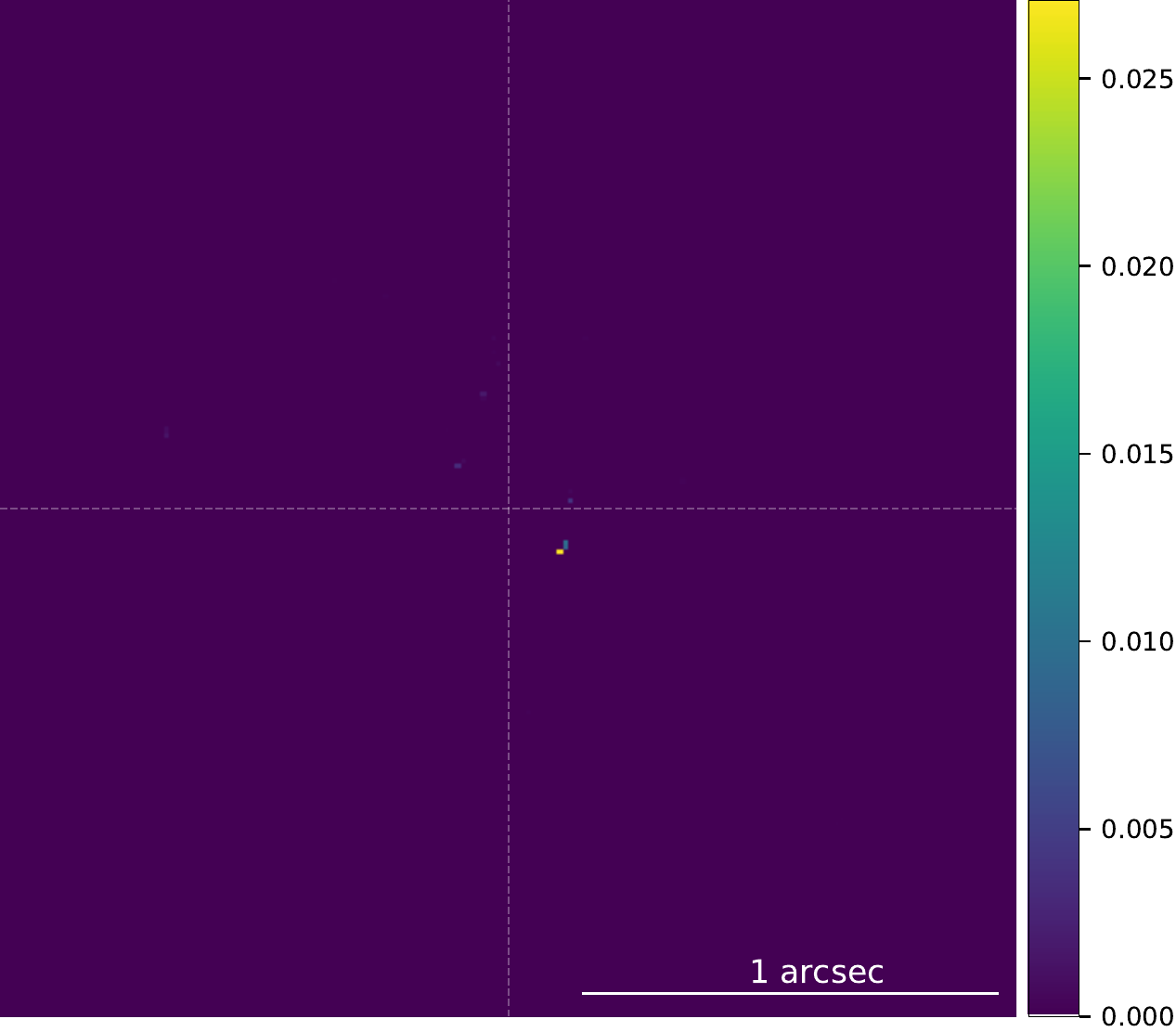}}
        \subfloat[HD17390]{\includegraphics[width=120pt]{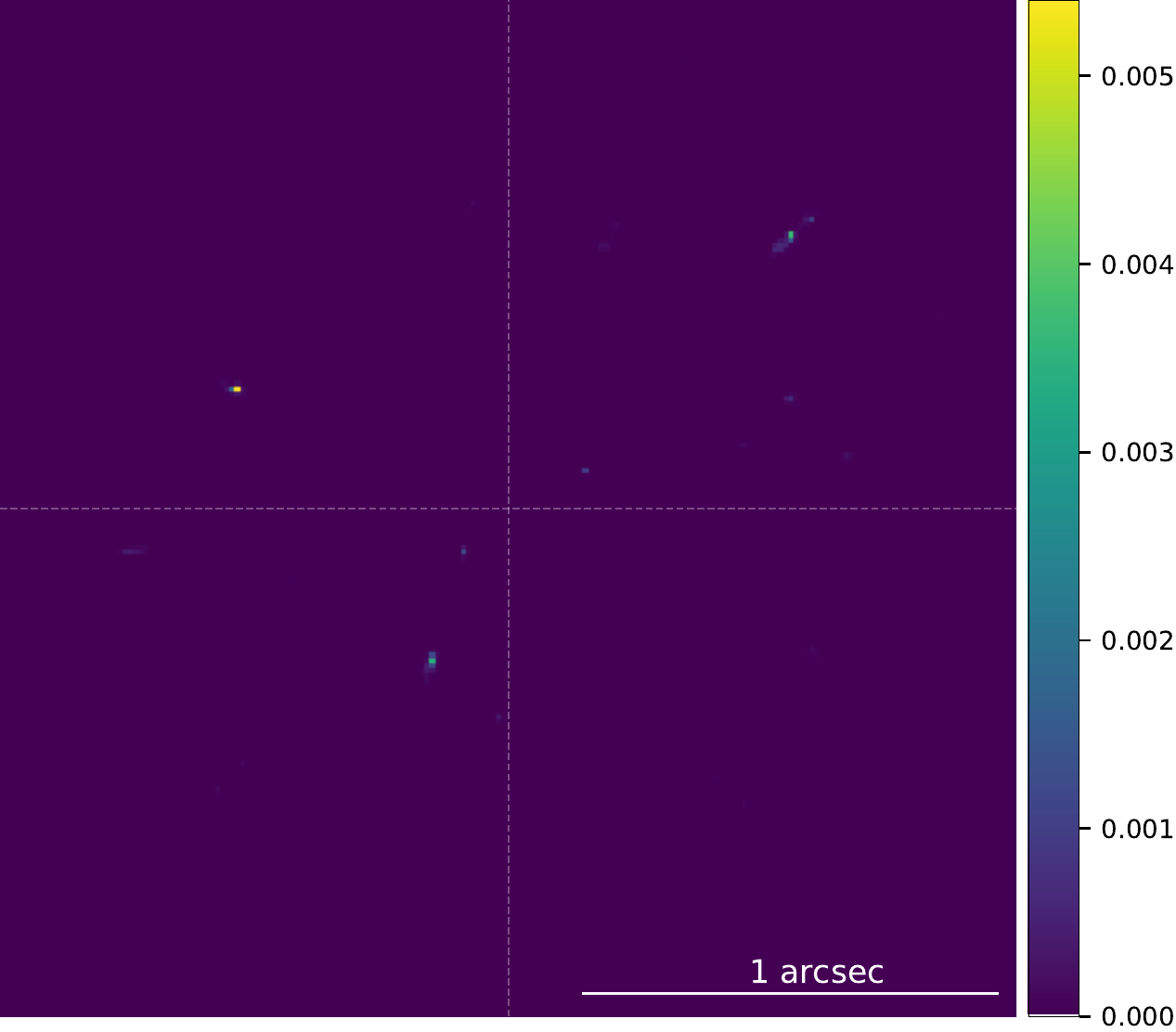}}
          \subfloat[HD16743]{\includegraphics[width=120pt]{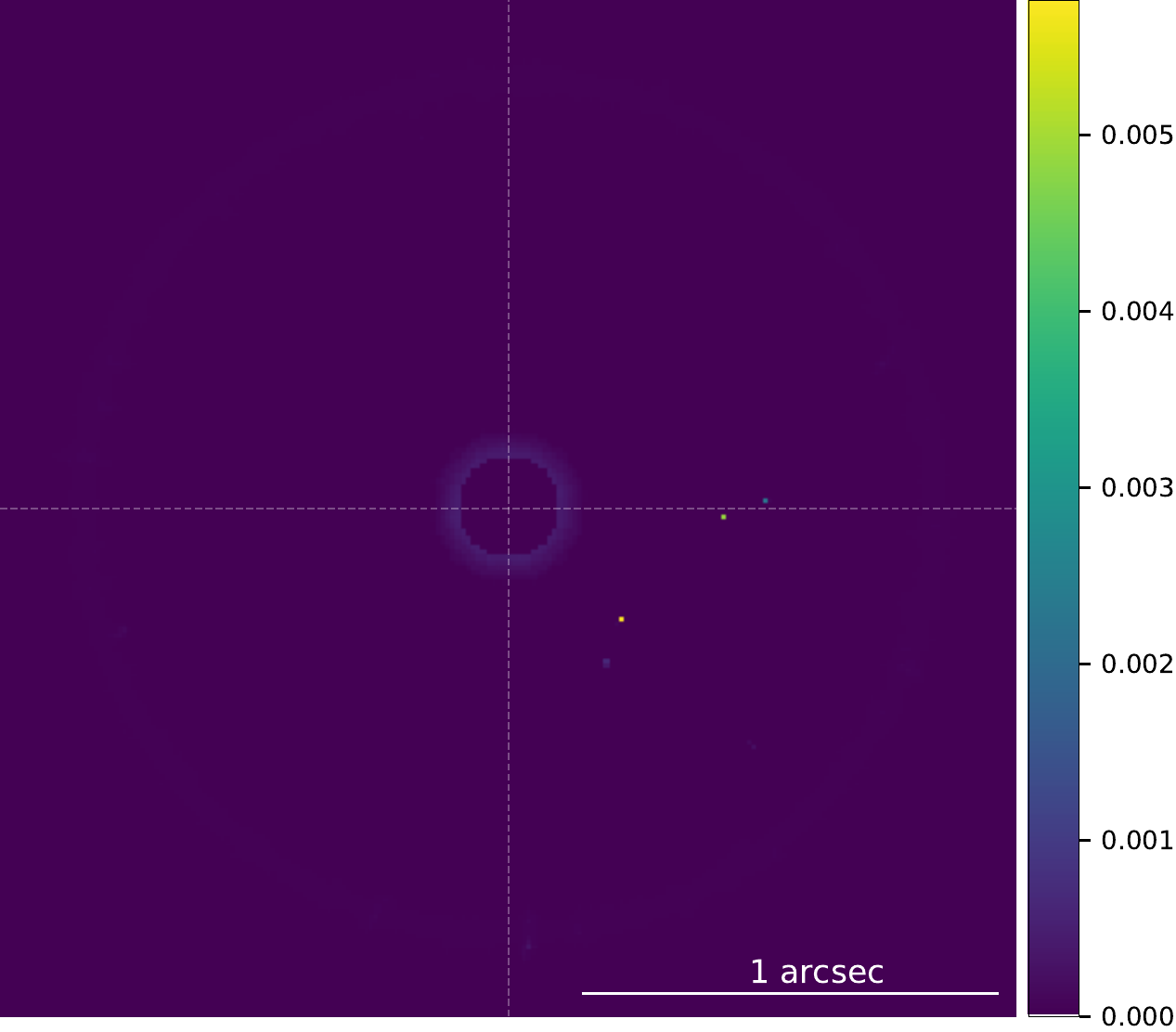}}\\
    \subfloat[HD15257]{\includegraphics[width=120pt]{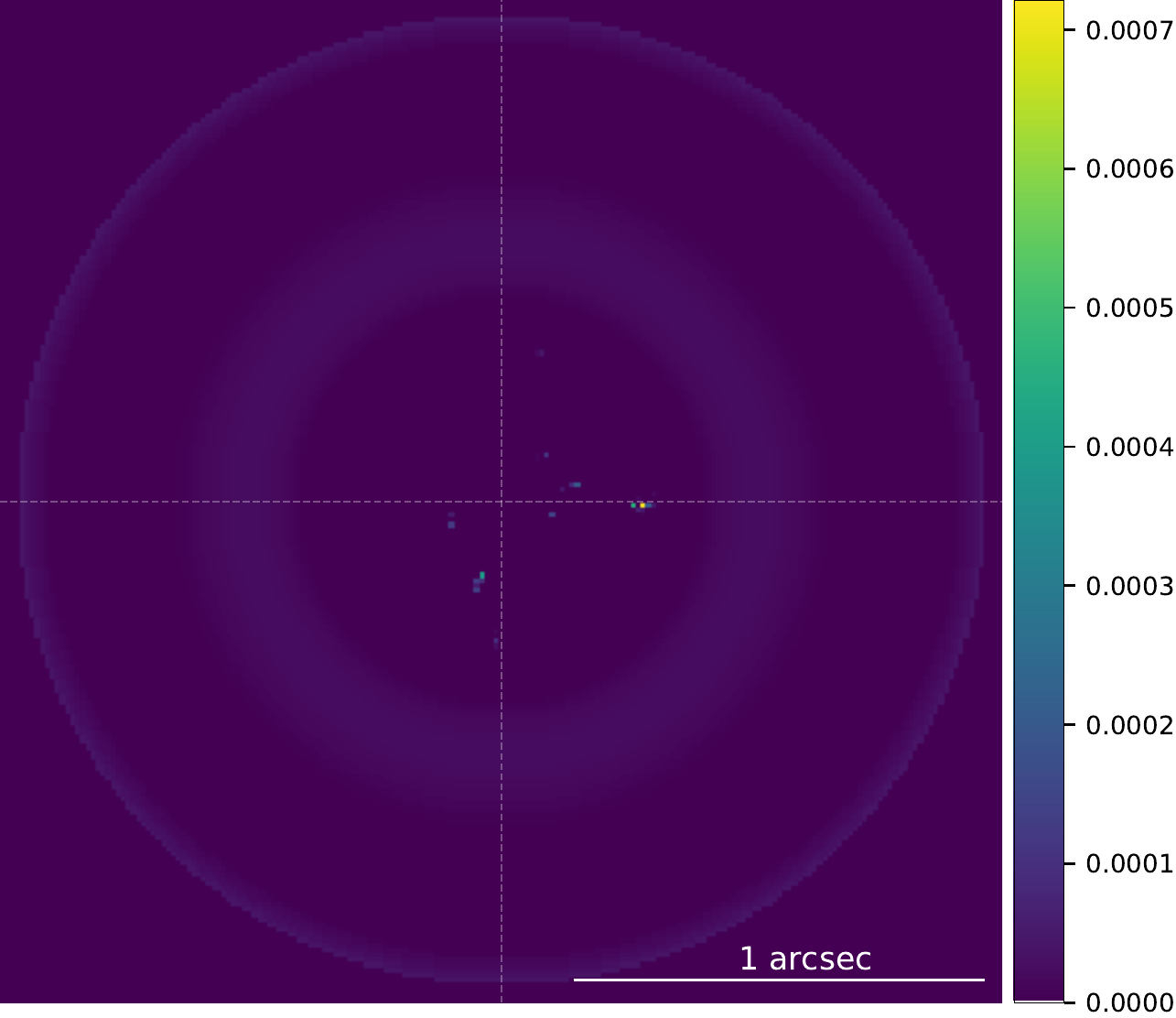}}
        \subfloat[HD14082B*]{\includegraphics[width=120pt]{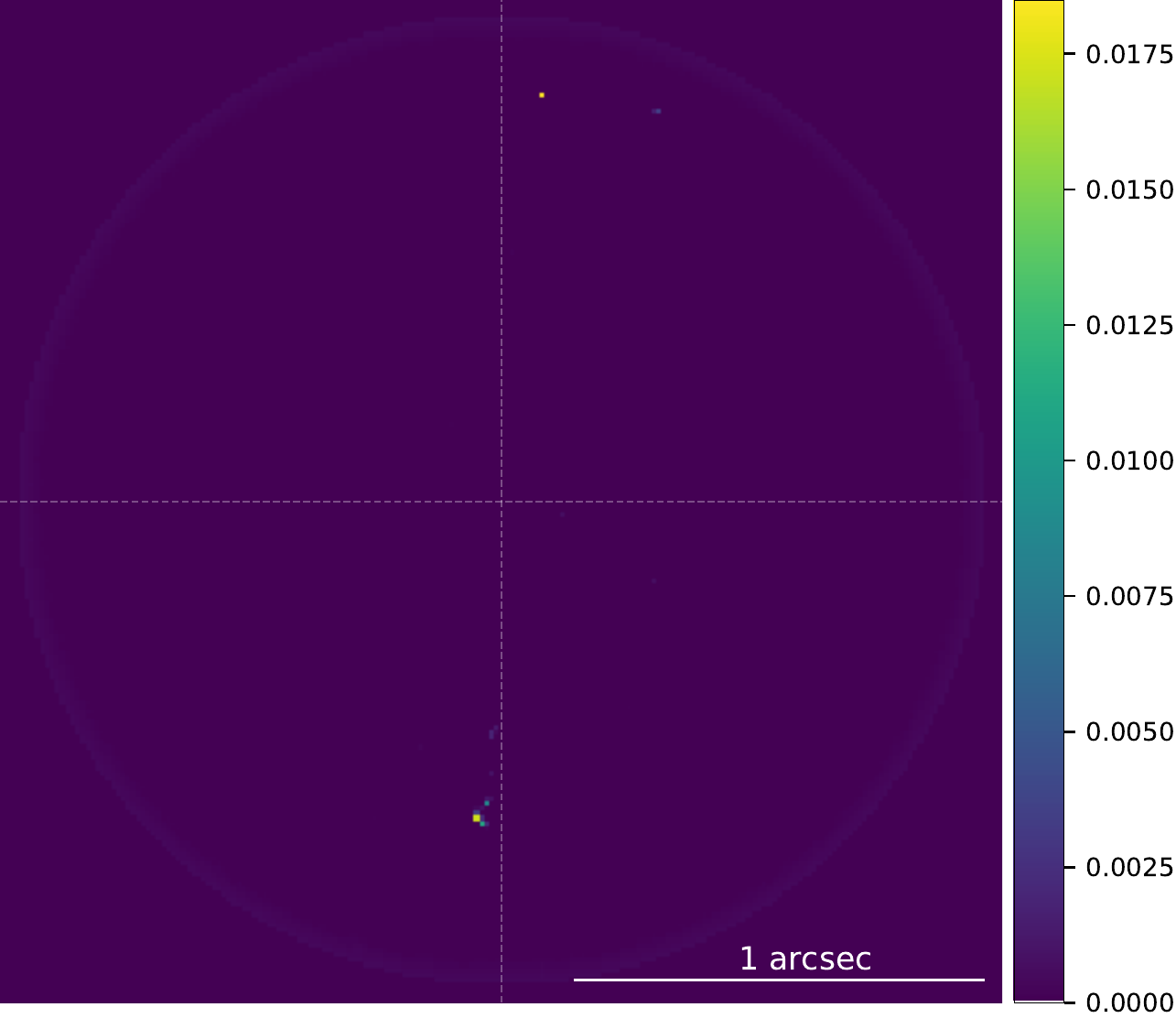}}
        \subfloat[HD14082B (2$^{nd}$ epoch)]{\includegraphics[width=120pt]{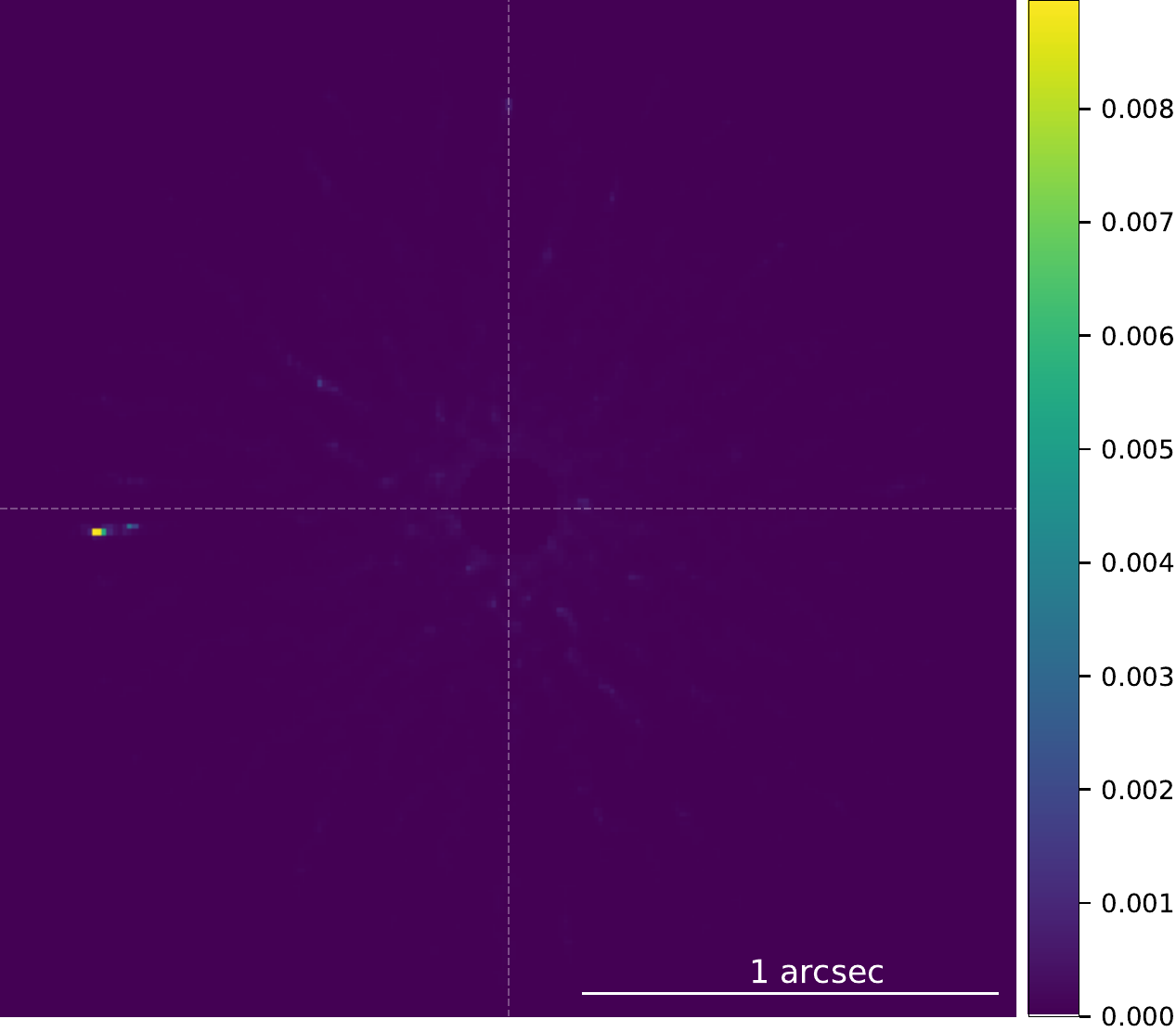}}
          \subfloat[HD53842]{\includegraphics[width=120pt]{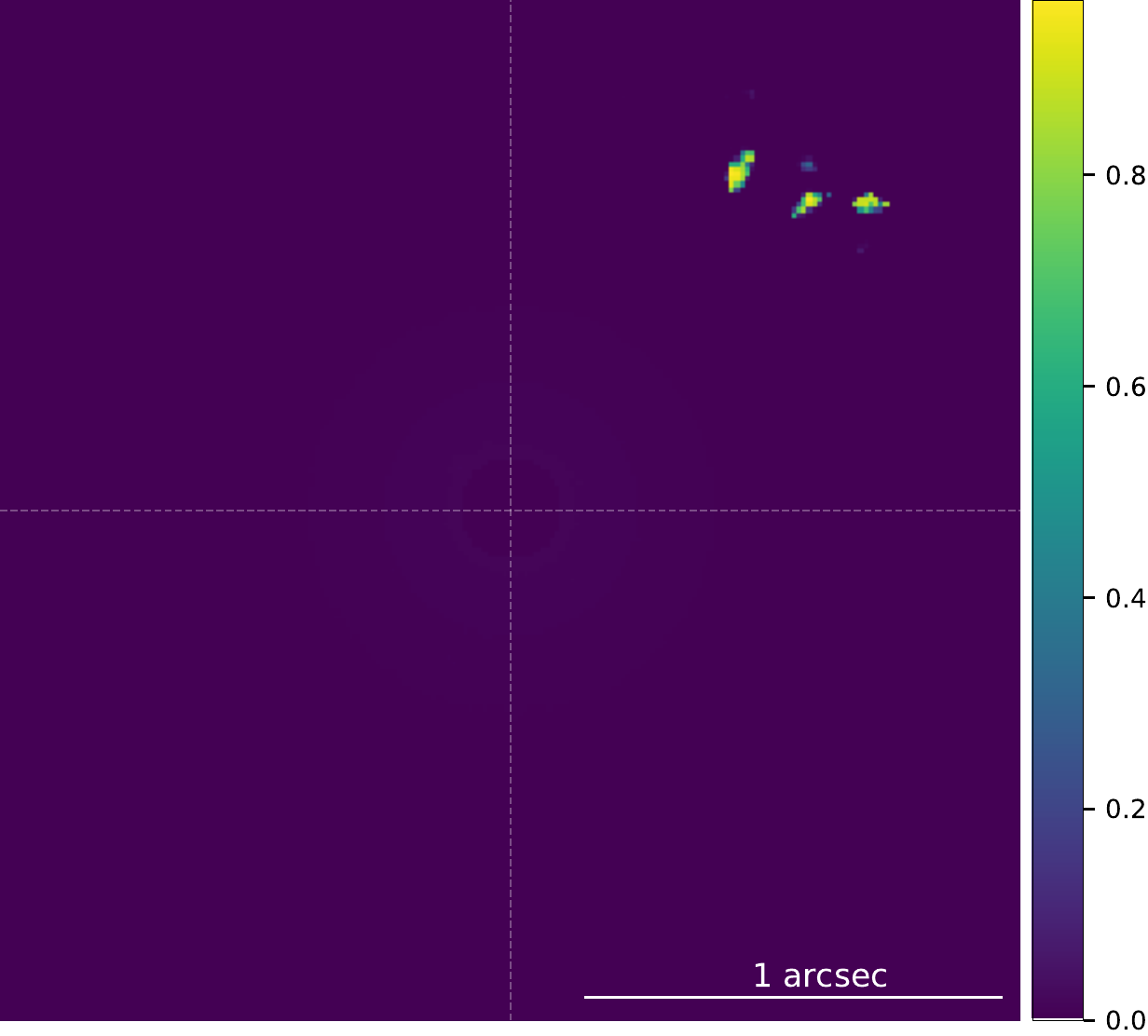}}\\
    \subfloat[HD80950]{\includegraphics[width=120pt]{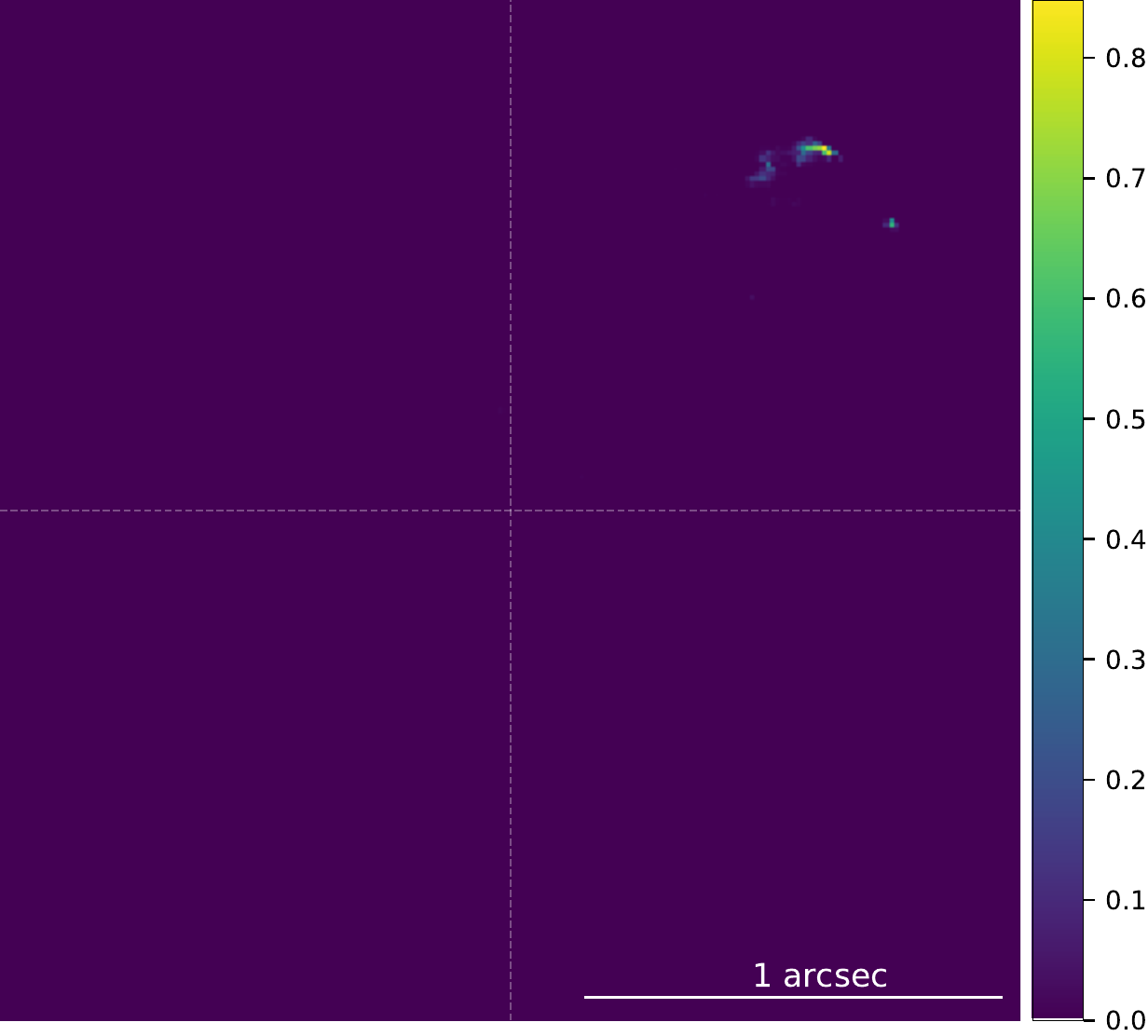}}
        \subfloat[HD203*]{\includegraphics[width=120pt]{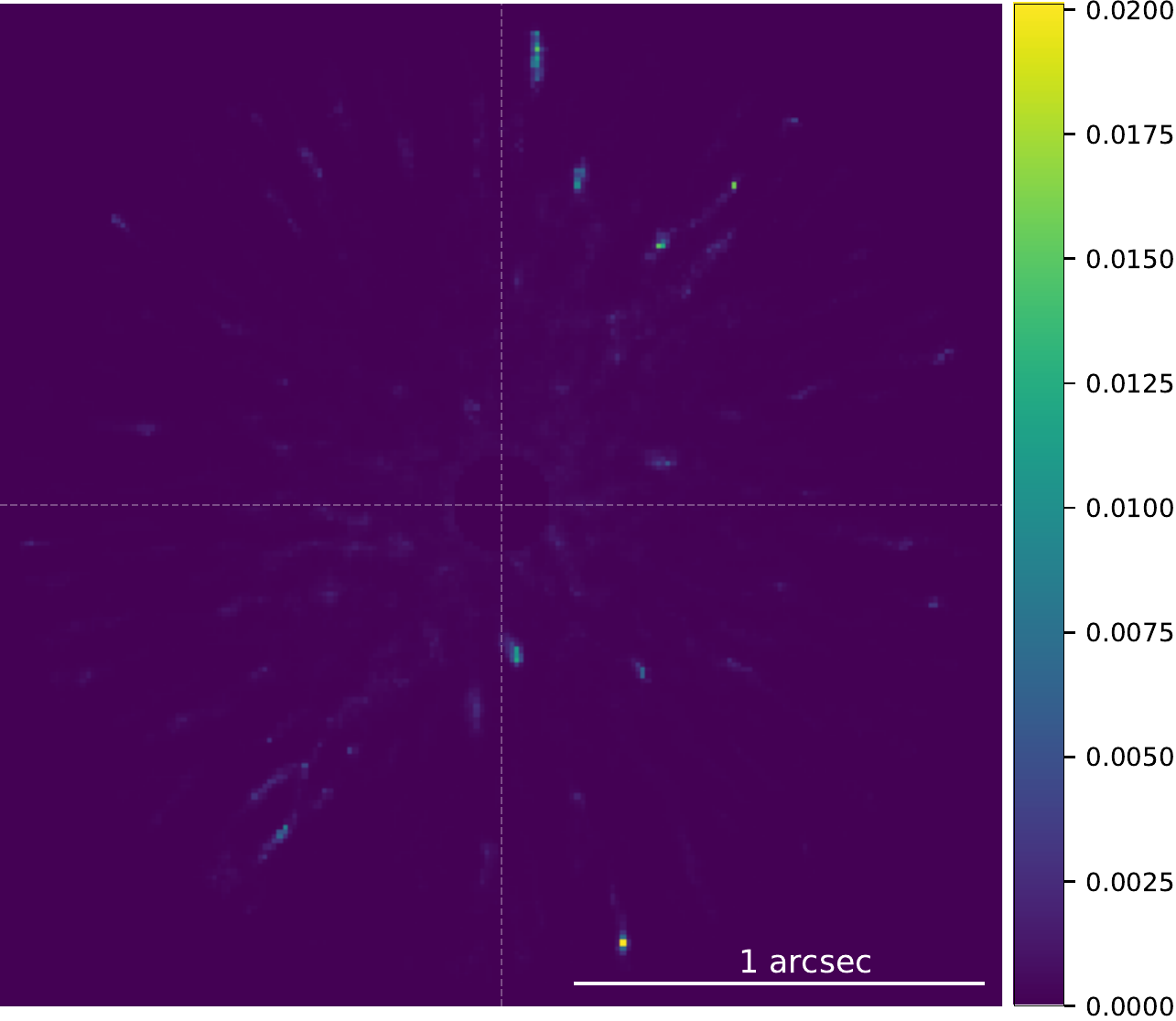}}
  \subfloat[HD60491*]{\includegraphics[width=120pt]{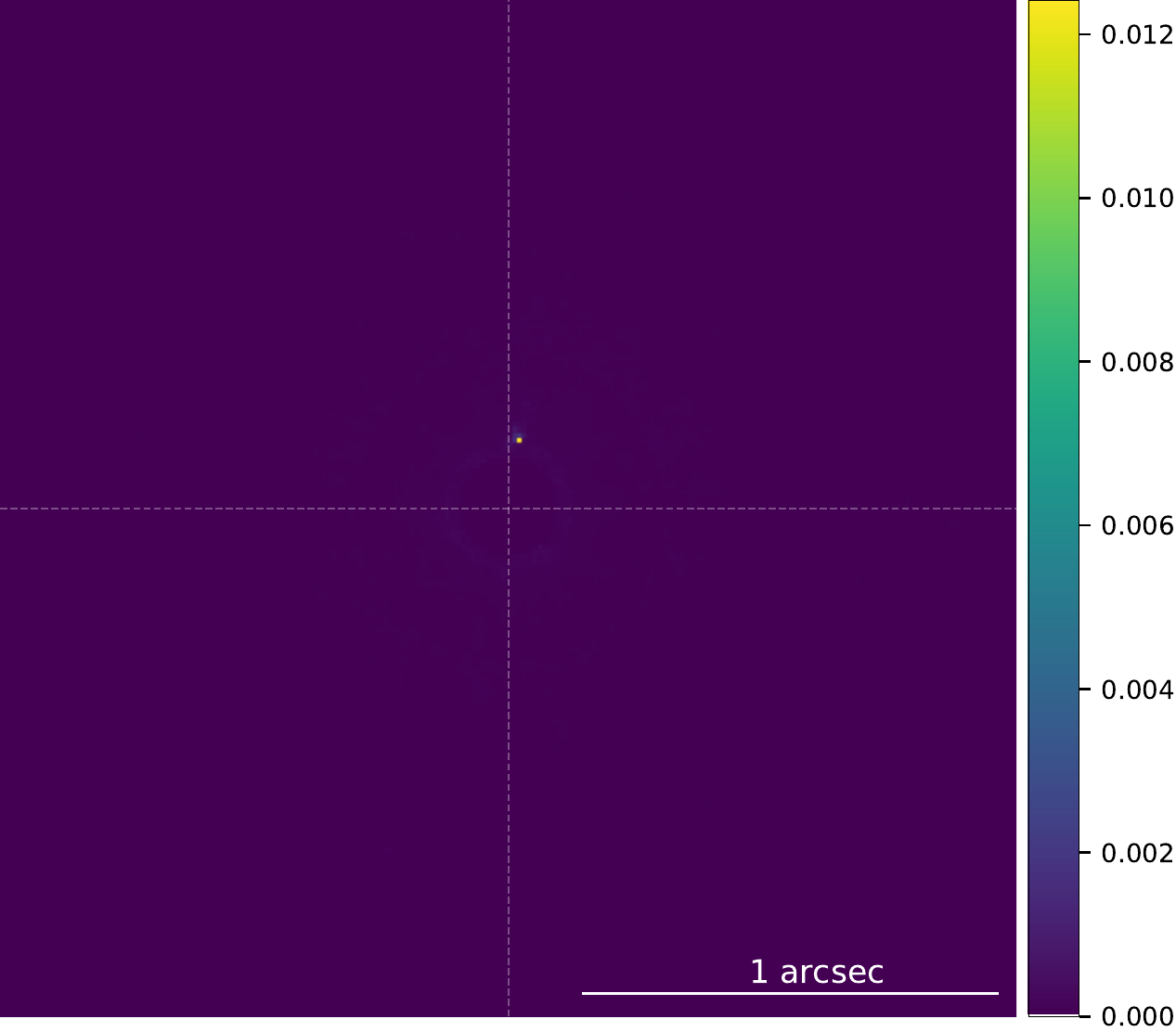}}
    \subfloat[HD76582*]{\includegraphics[width=120pt]{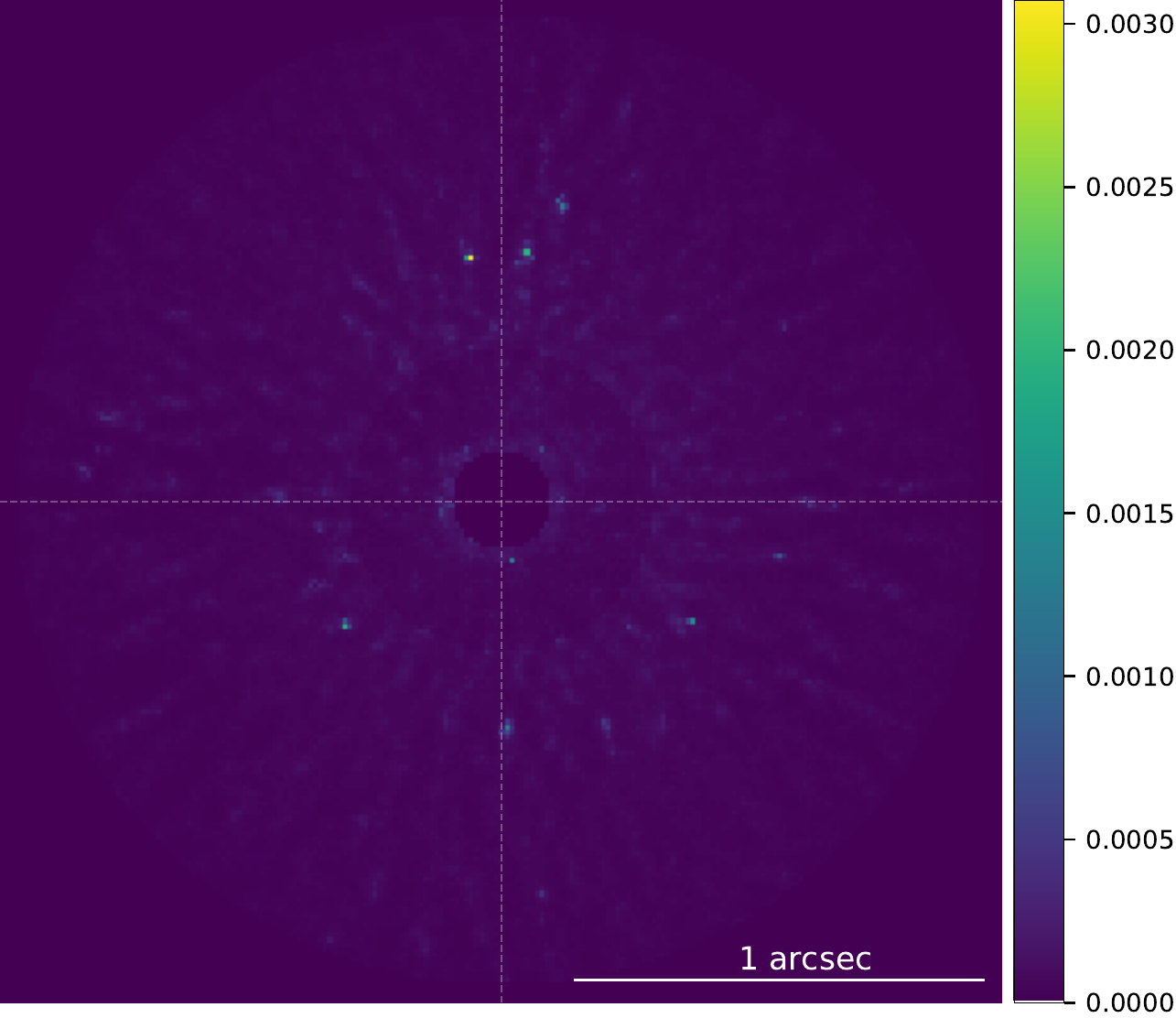}}\\
  \caption{\label{Empty_map1} RSM detection maps generated using auto-RSM or the optimal parameters obtained with auto-RSM for the dataset at the center of the clusters (see Table \ref{Clusters})).  These detection maps did not lead to the detection of a target. The asterisks indicate the targets on which the full Auto-RSM framework was applied. }
\end{figure*}

        \begin{figure*}[!htbp]
\footnotesize
  \centering
          \subfloat[HD13246]{\includegraphics[width=115pt]{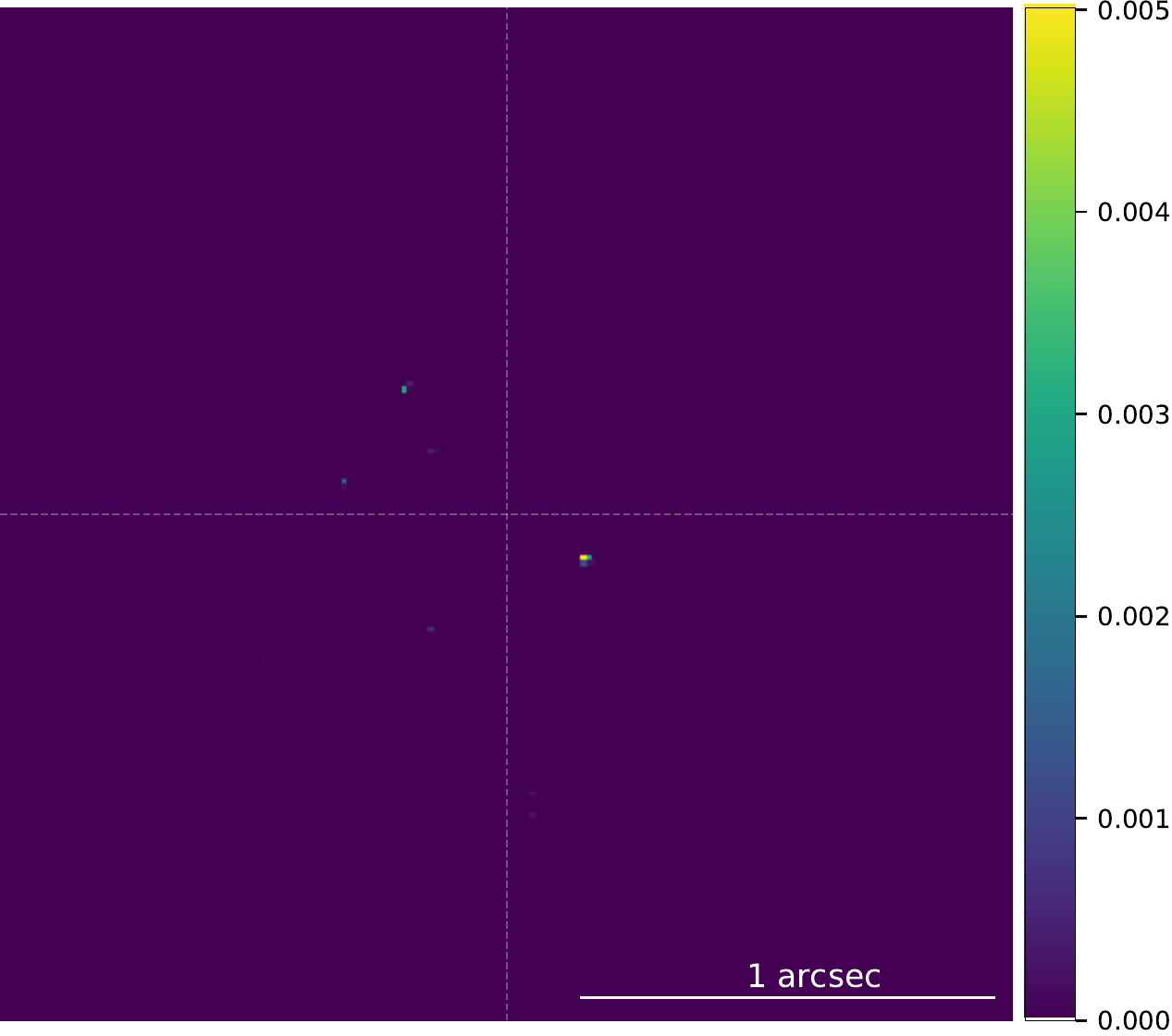}}
    \subfloat[HD10472]{\includegraphics[width=115pt]{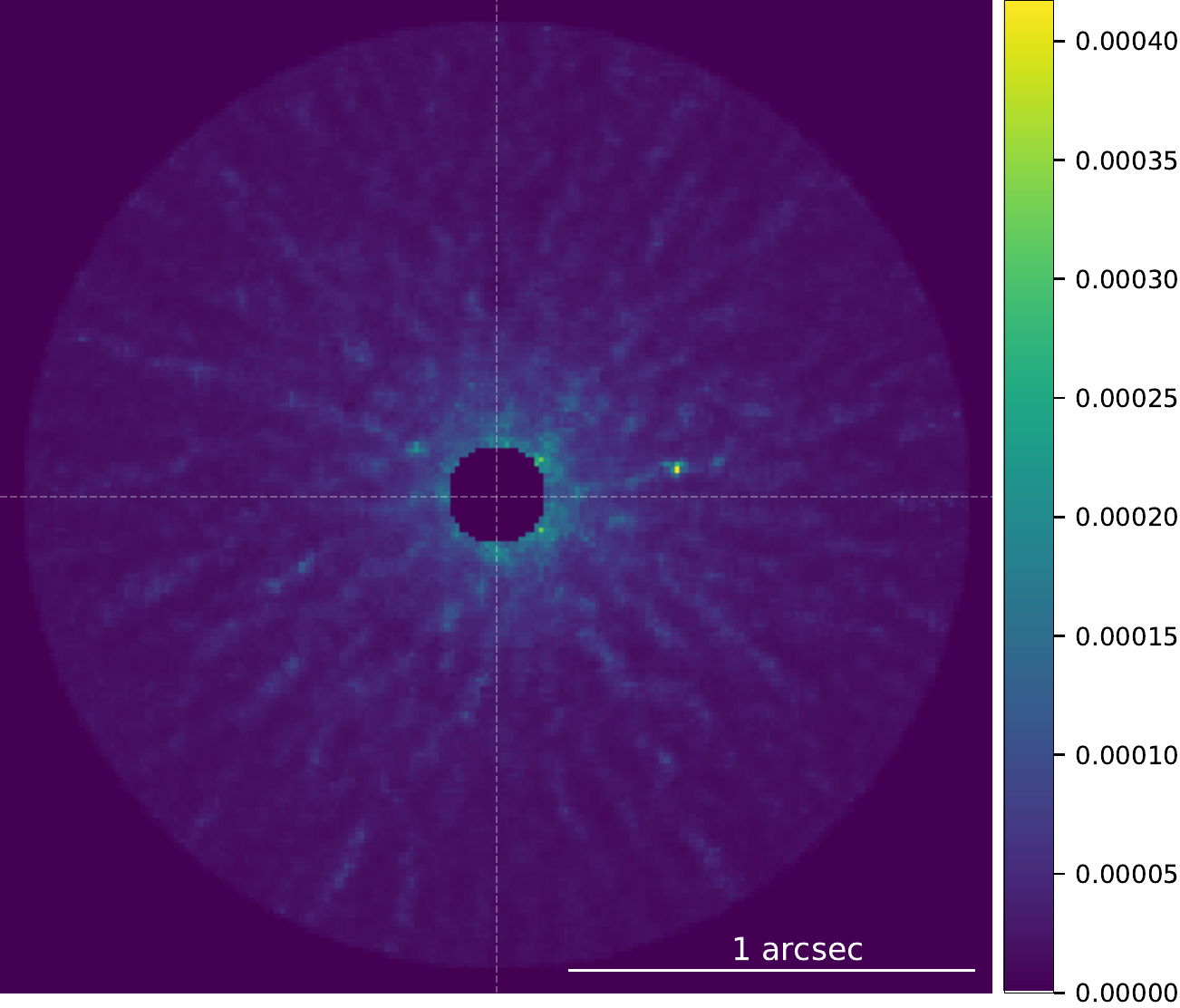}}
        \subfloat[HD10472 (2$^{nd}$ epoch)]{\includegraphics[width=115pt]{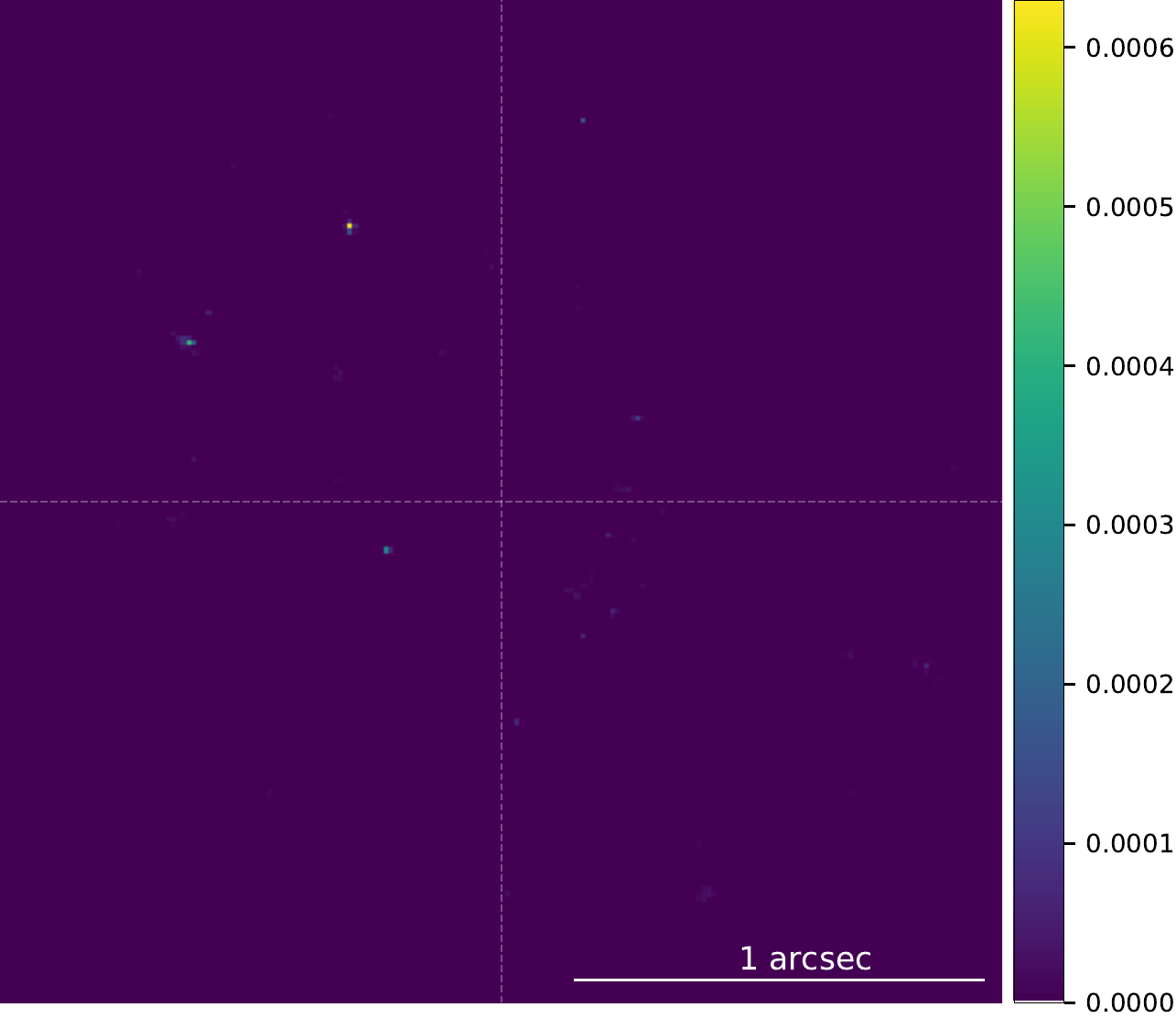}}
        \subfloat[HD9672]{\includegraphics[width=115pt]{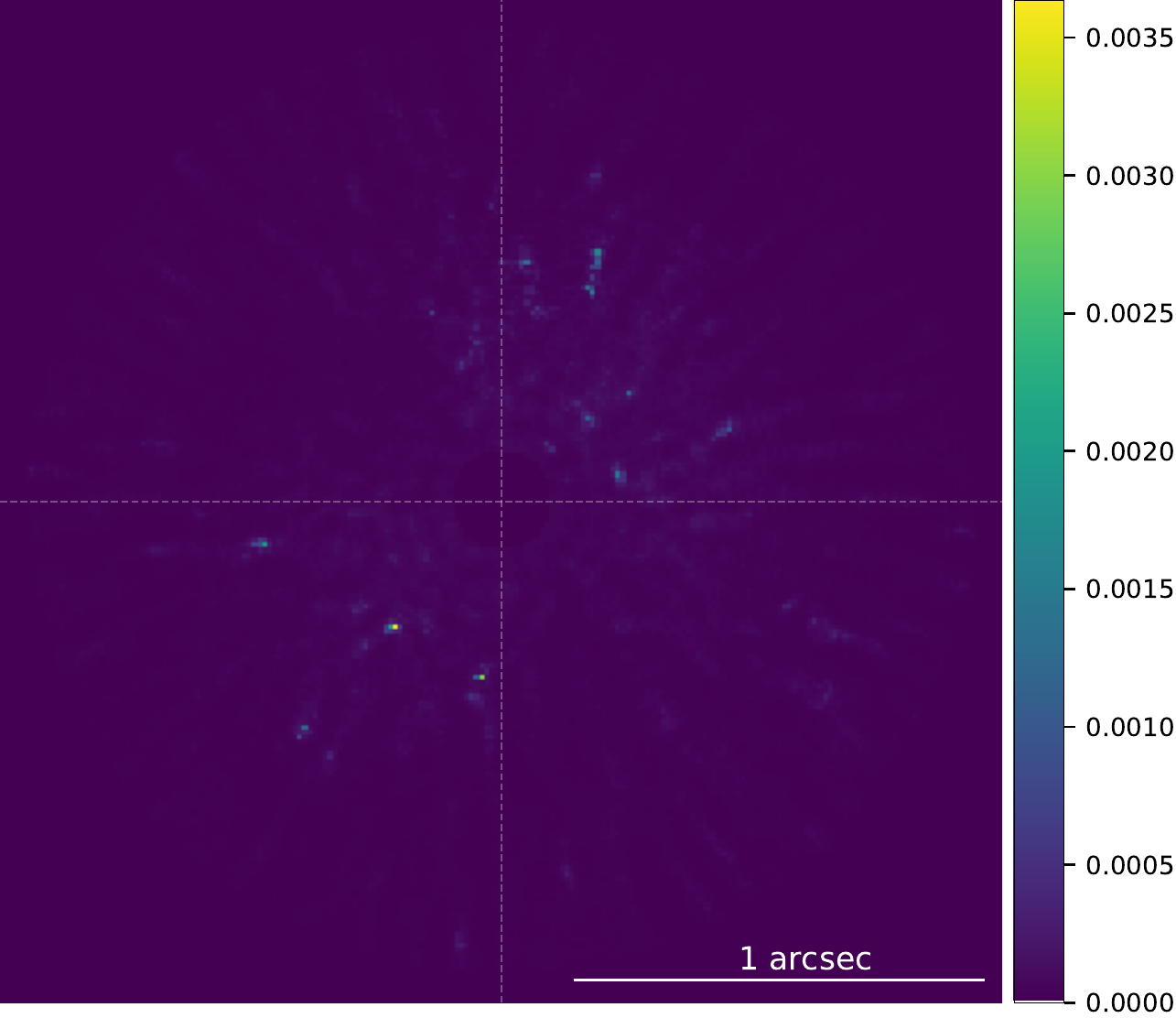}}\\
          \subfloat[HD3670*]{\includegraphics[width=115pt]{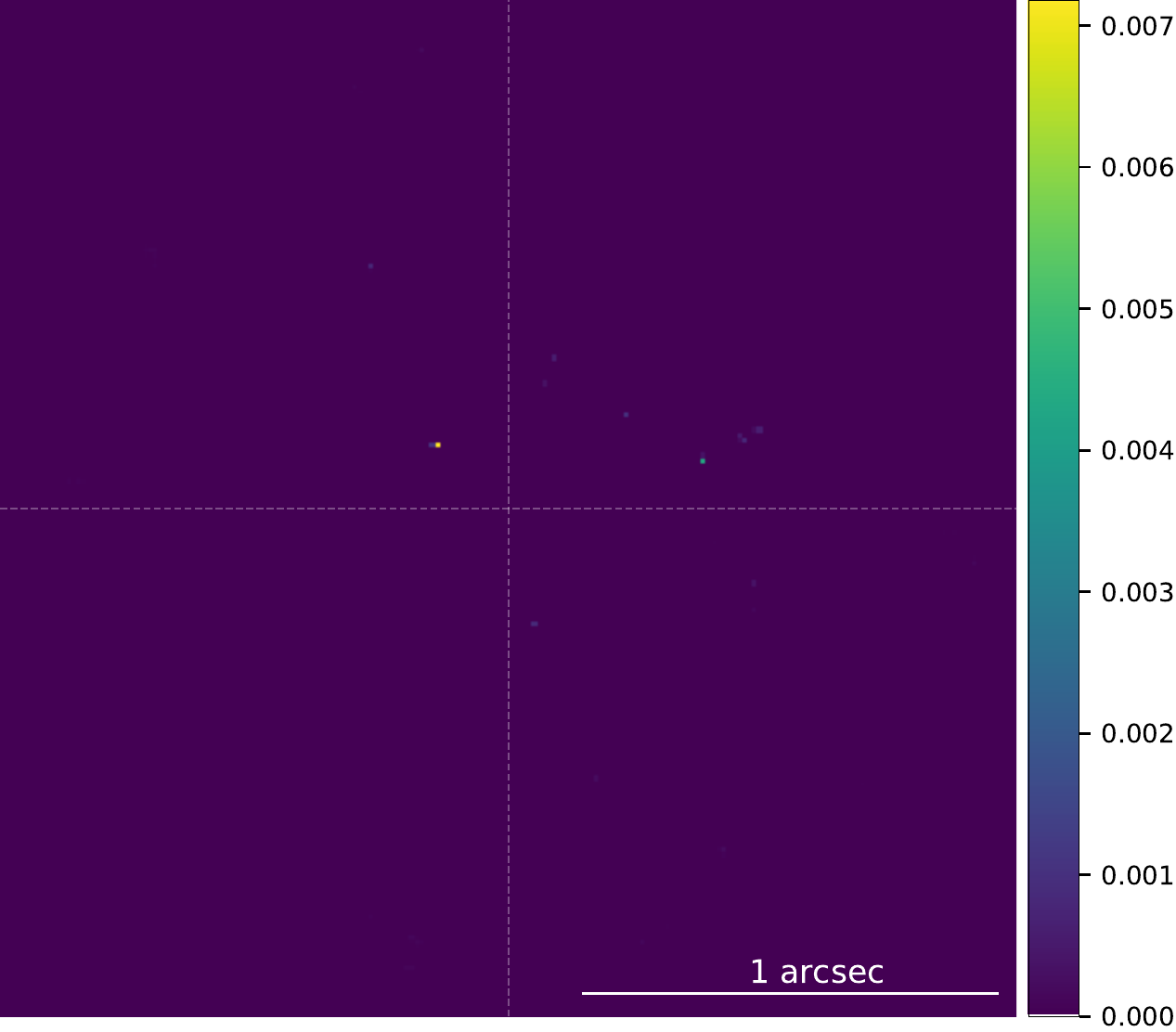}}
    \subfloat[HD3003*]{\includegraphics[width=115pt]{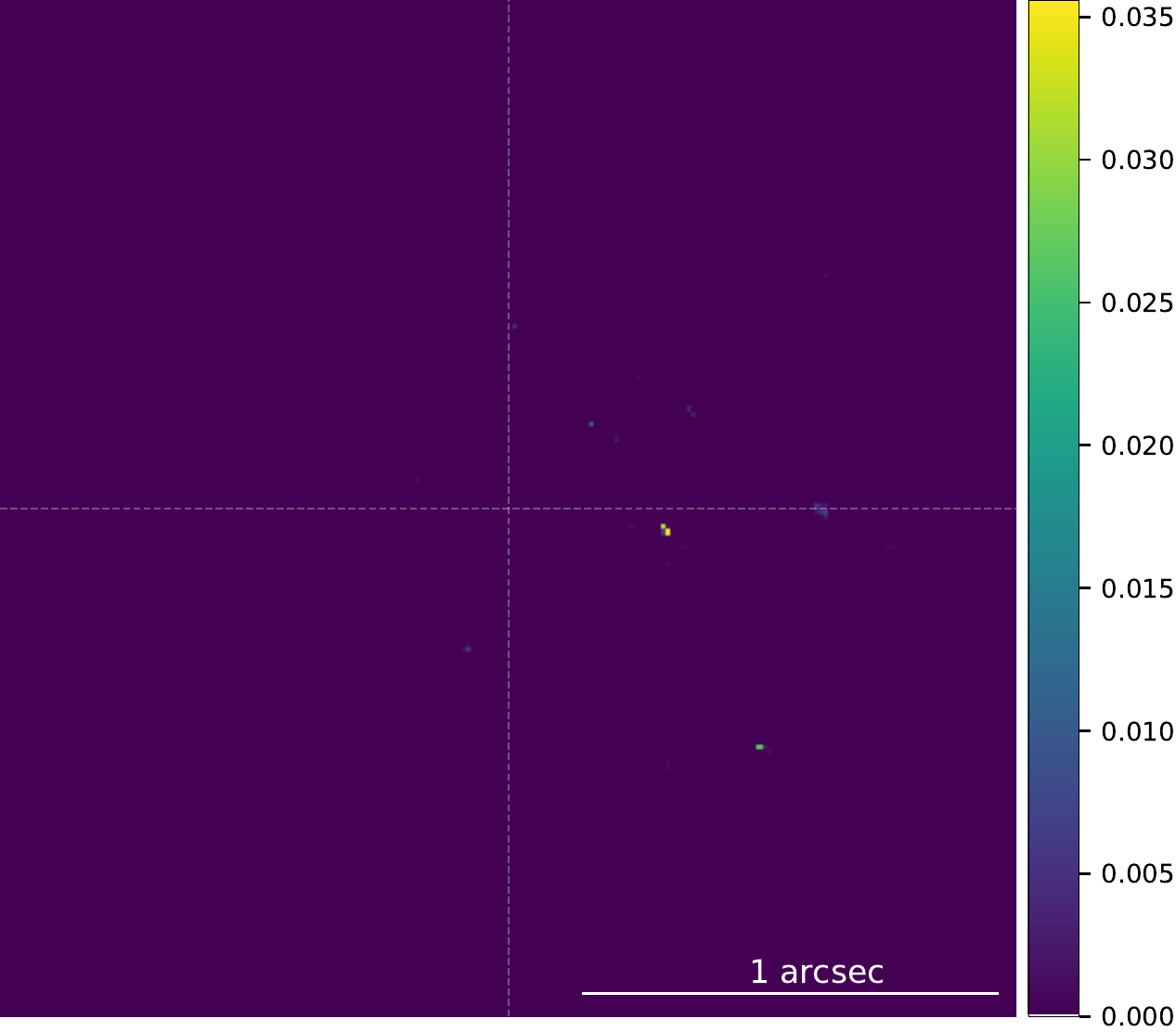}}
        \subfloat[HD377]{\includegraphics[width=115pt]{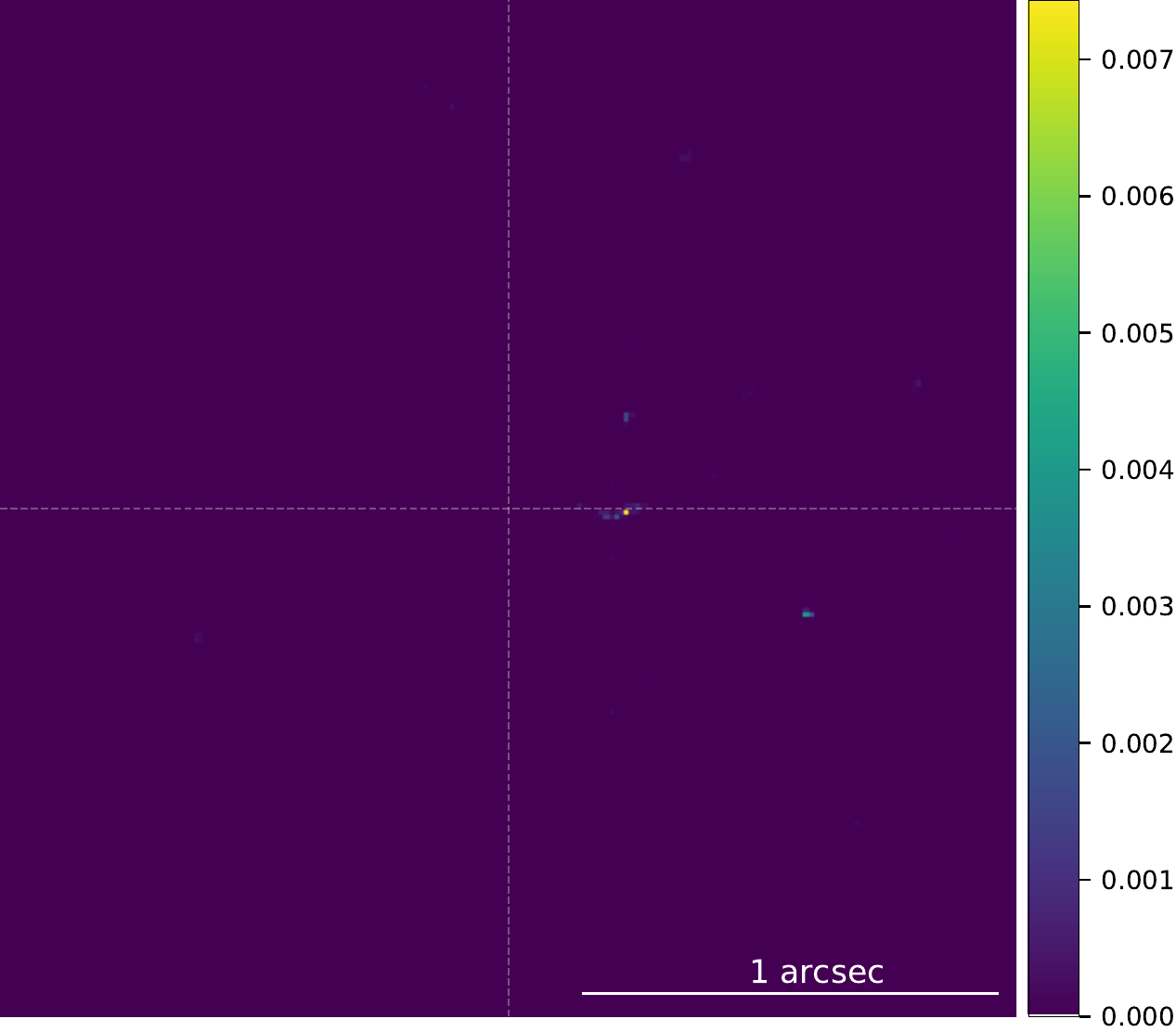}}
        \subfloat[HD105]{\includegraphics[width=115pt]{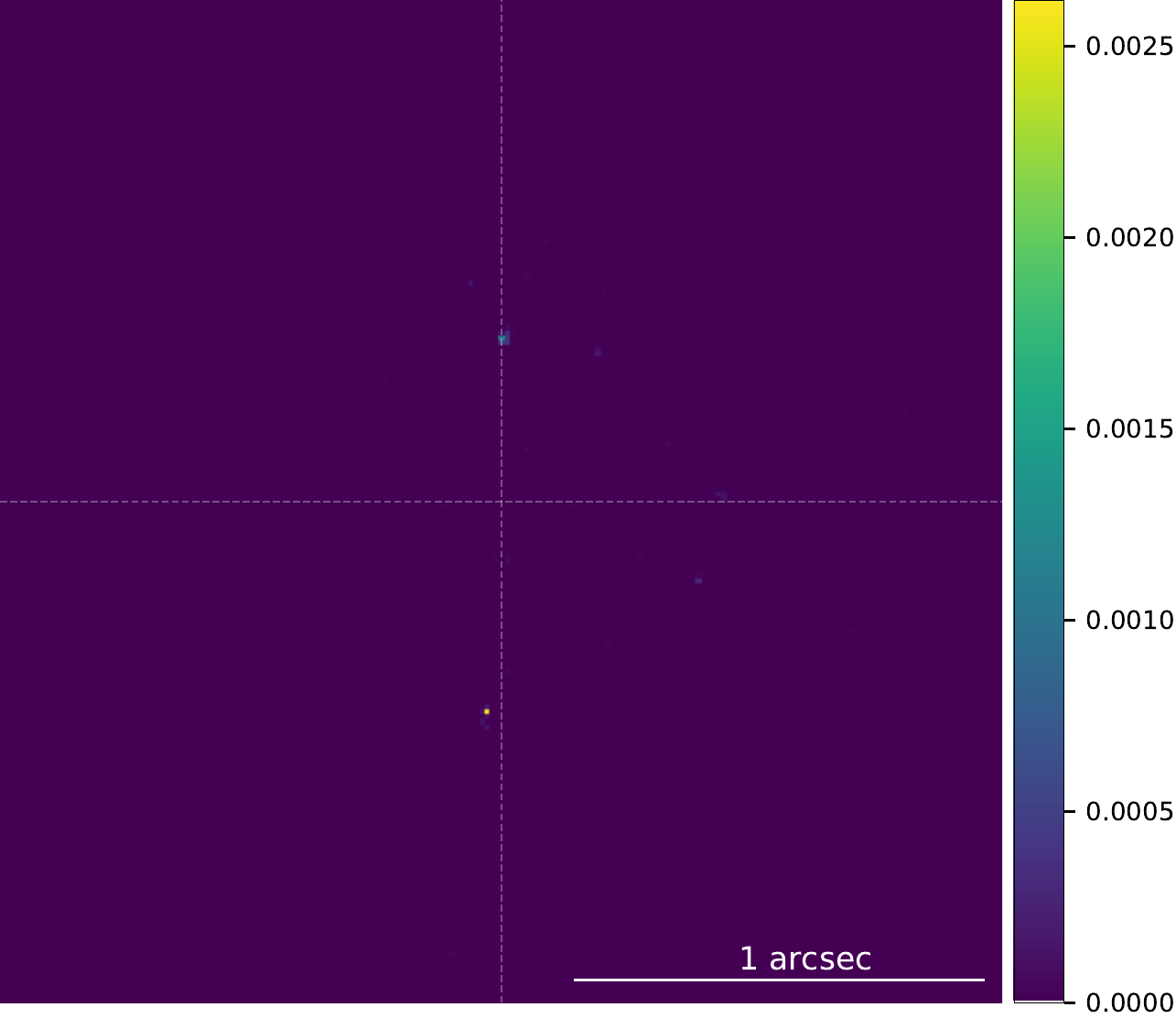}}\\
  \subfloat[AG Tri*]{\includegraphics[width=115pt]{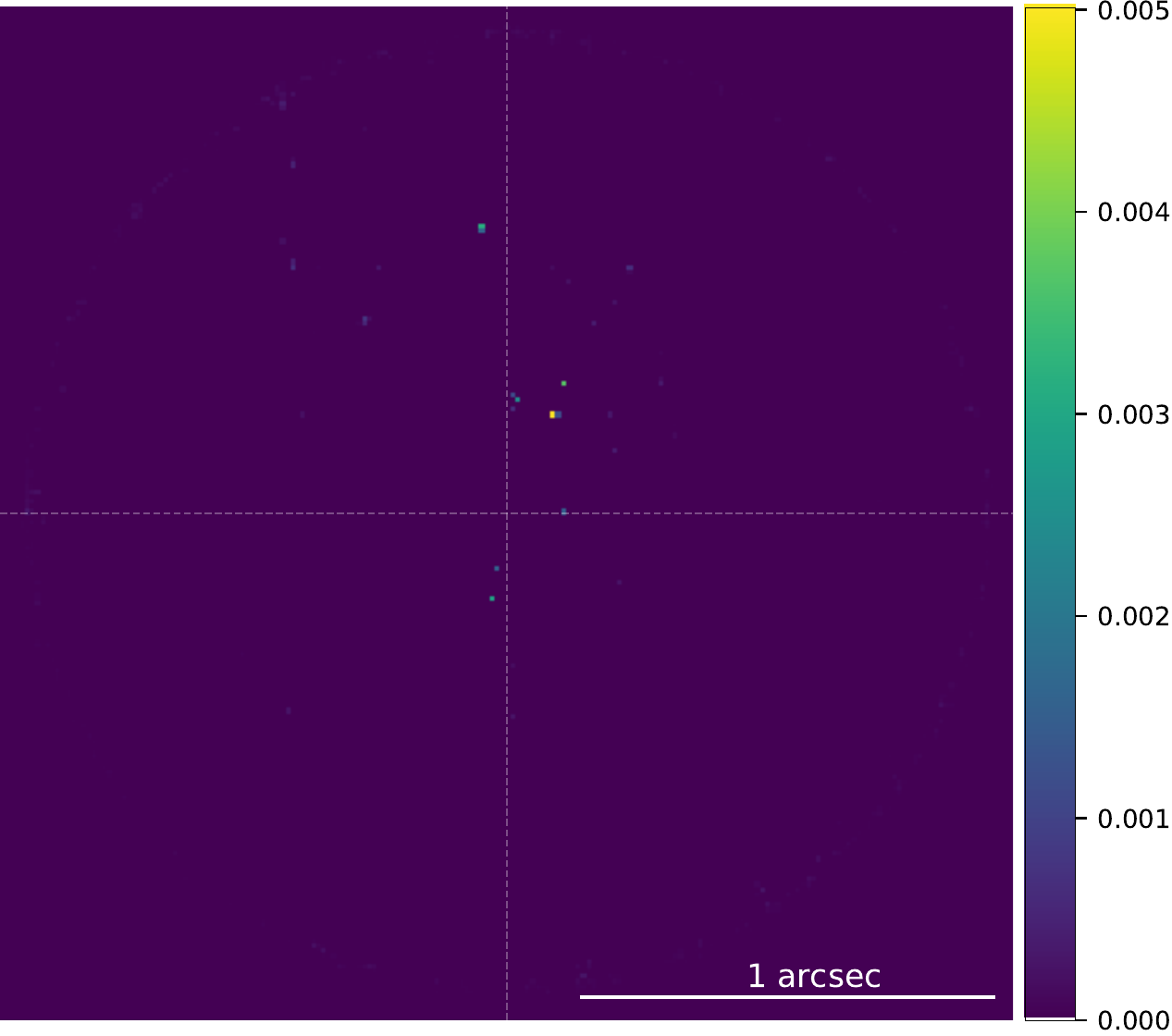}}
    \subfloat[HD69830]{\includegraphics[width=115pt]{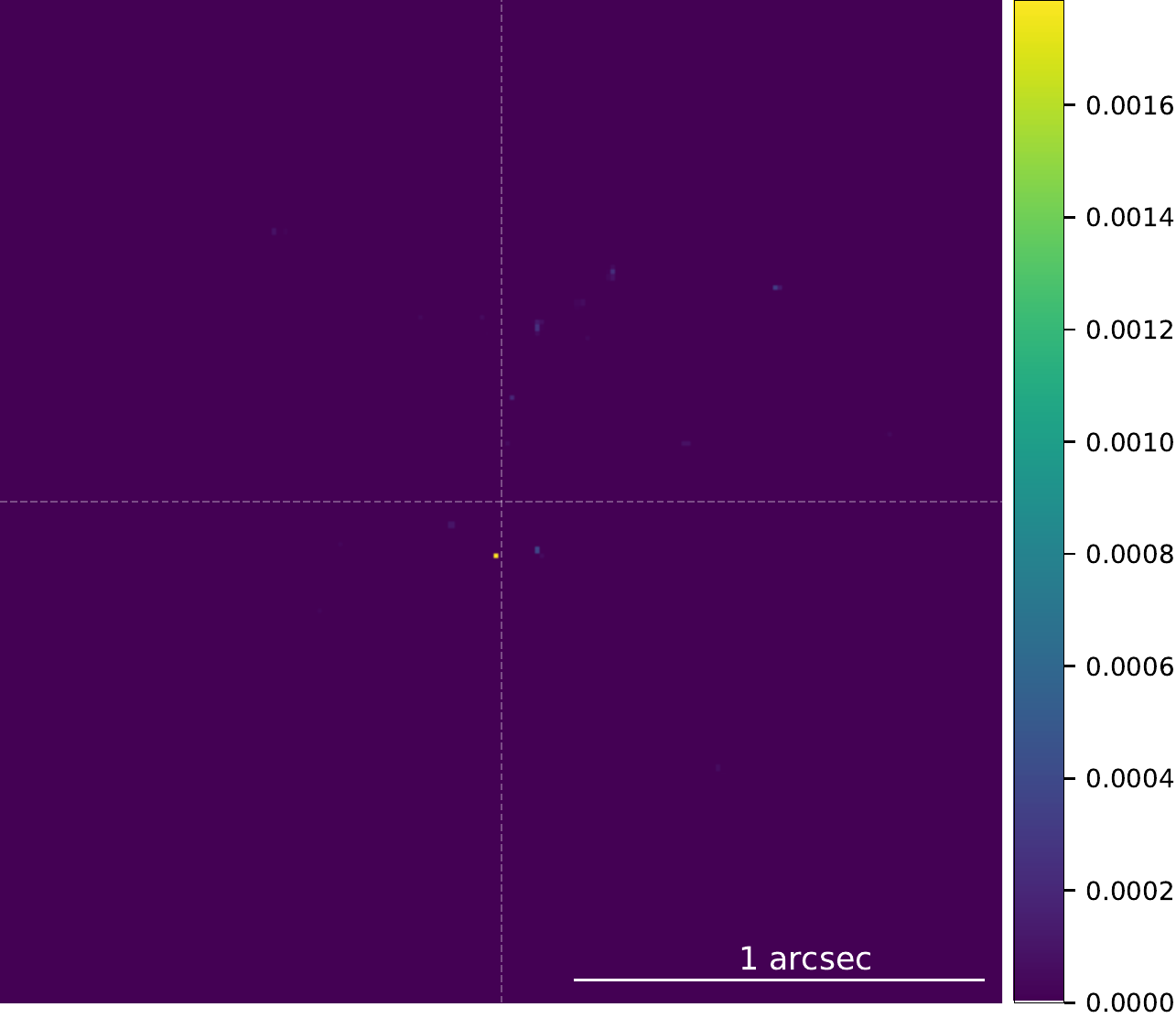}}
        \subfloat[HD71722]{\includegraphics[width=115pt]{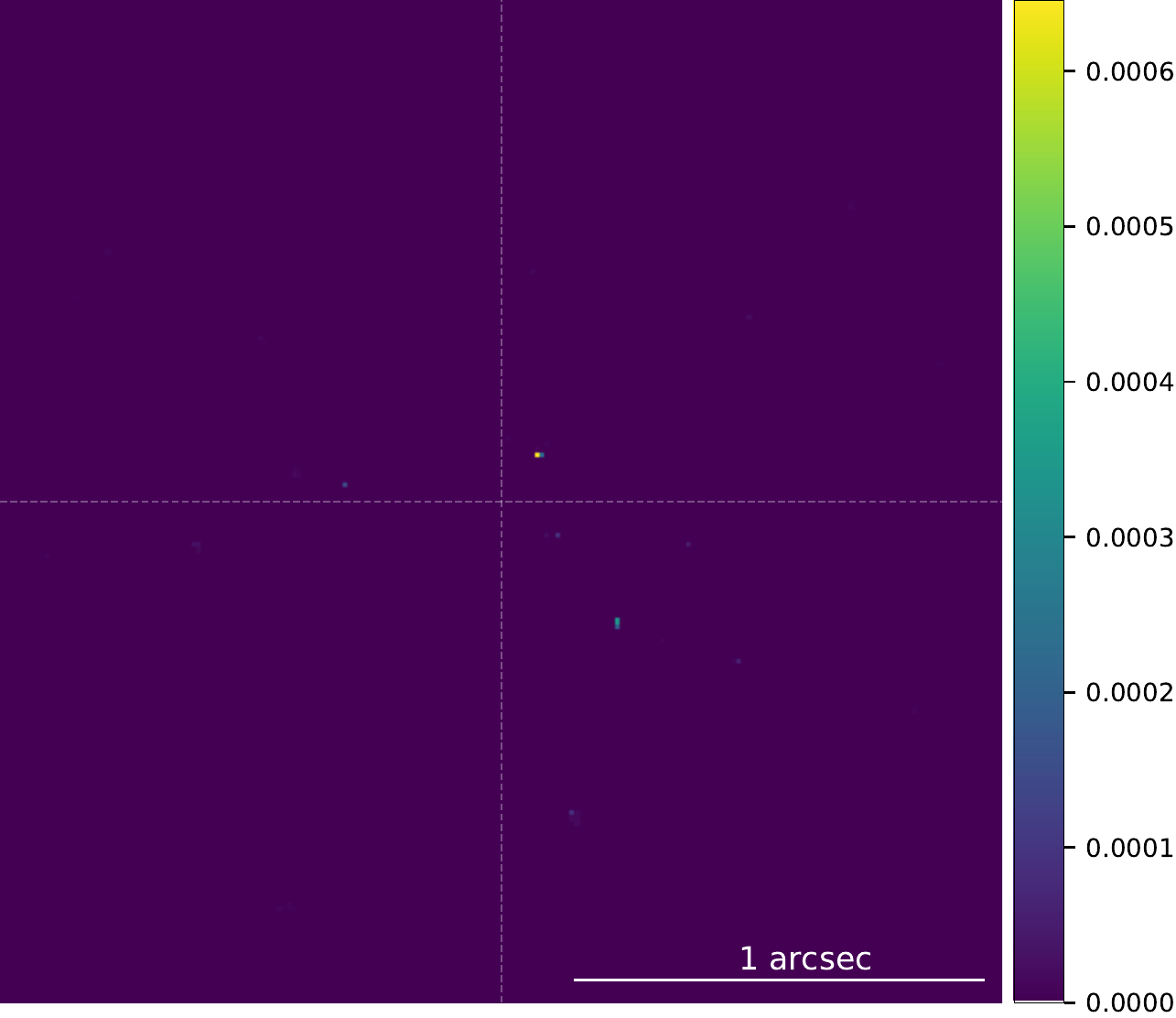}}
        \subfloat[HD73350]{\includegraphics[width=115pt]{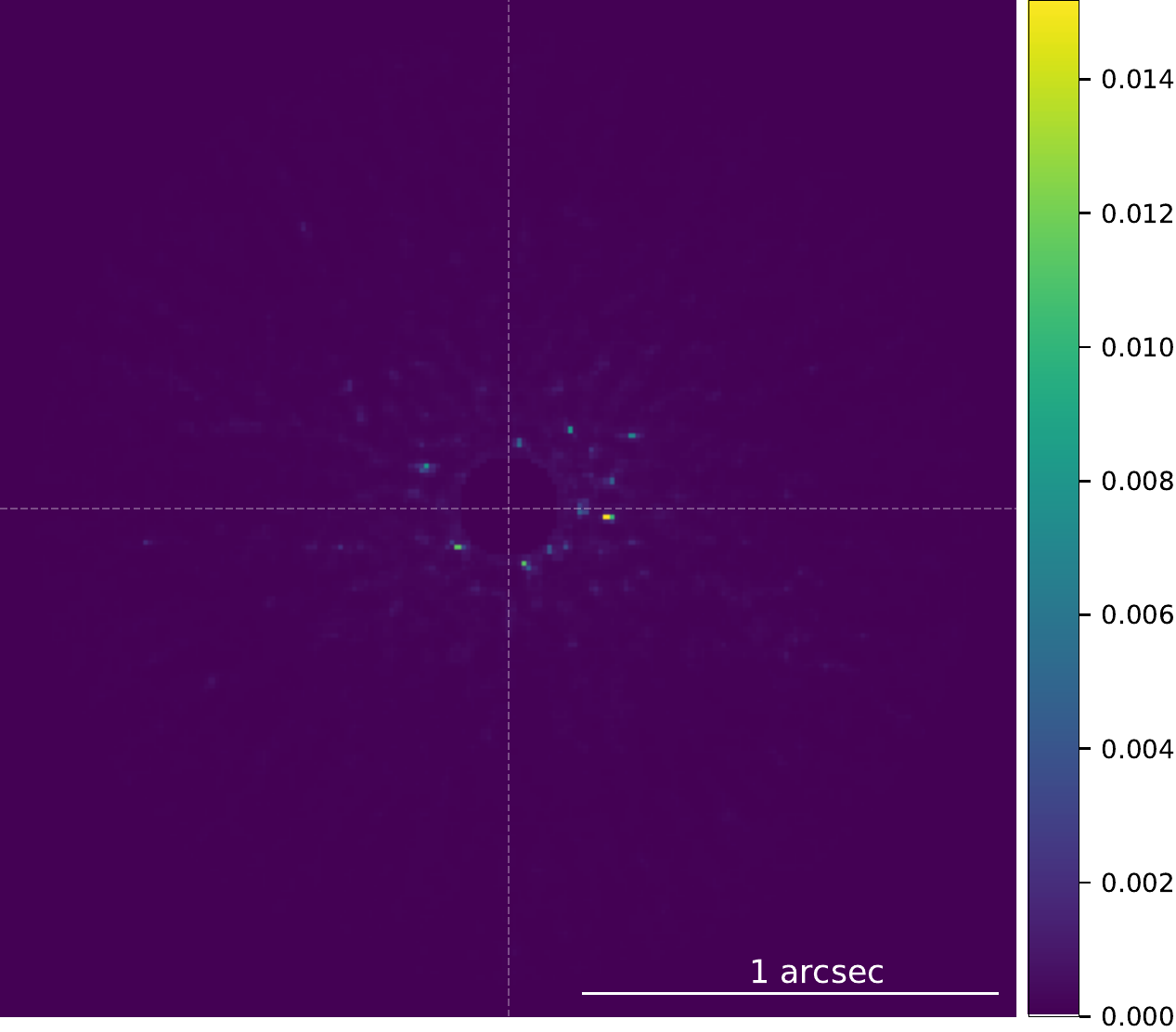}}\\
          \subfloat[HD82943 (2$^{nd}$ epoch)]{\includegraphics[width=115pt]{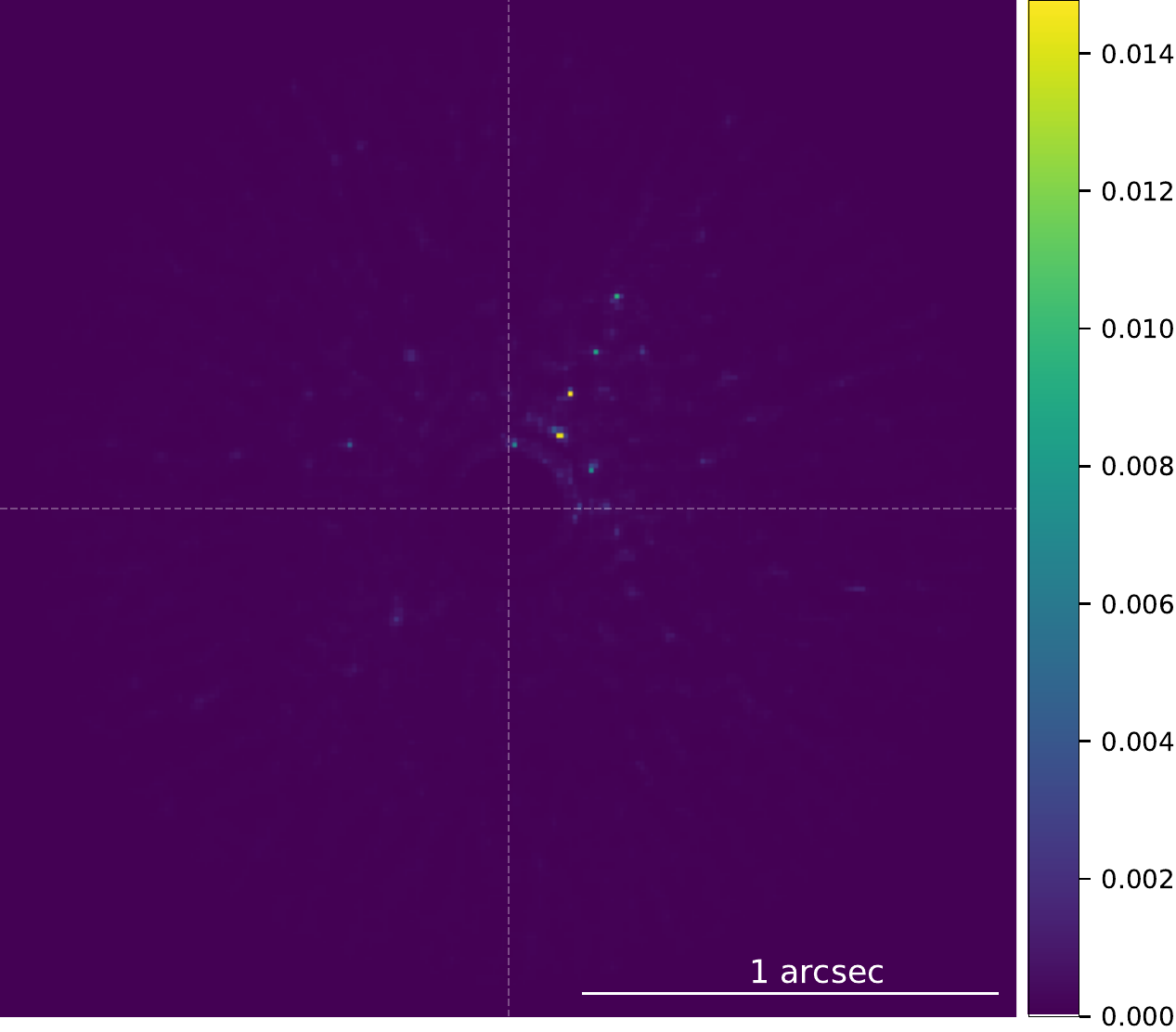}}
    \subfloat[HD82943 (4$^{th}$ epoch)*]{\includegraphics[width=115pt]{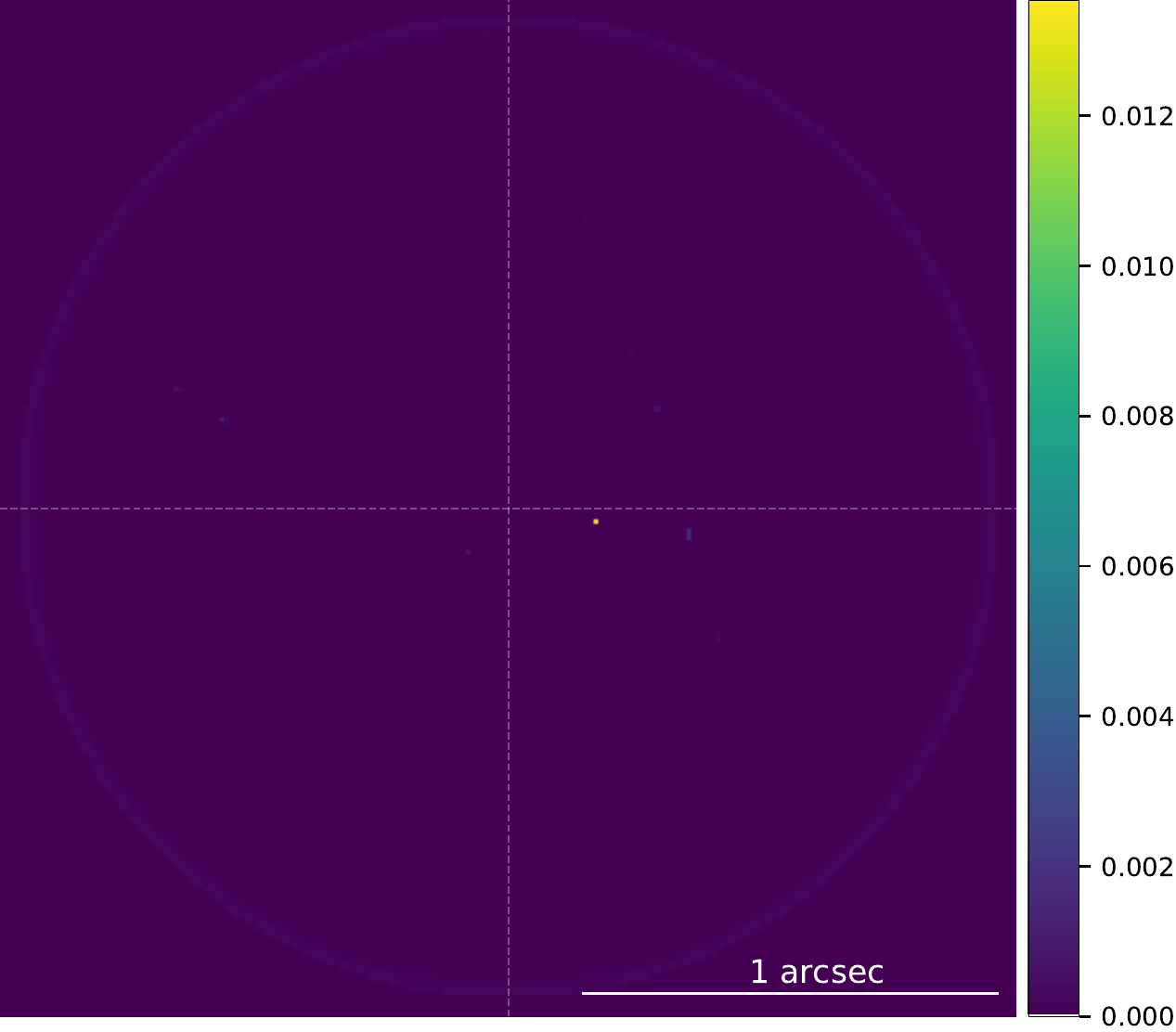}}
        \subfloat[HD84075*]{\includegraphics[width=115pt]{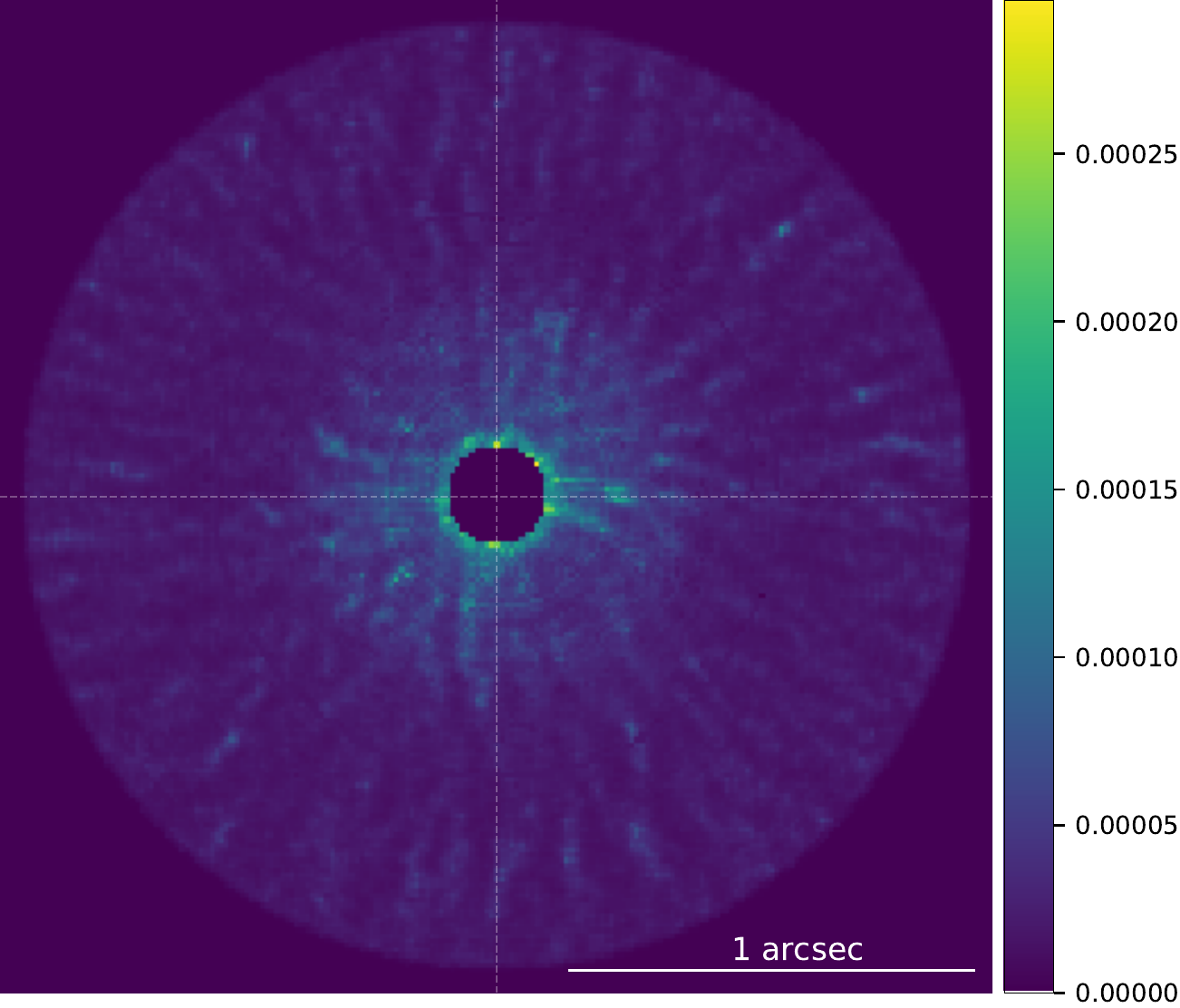}}
        \subfloat[HD120534]{\includegraphics[width=115pt]{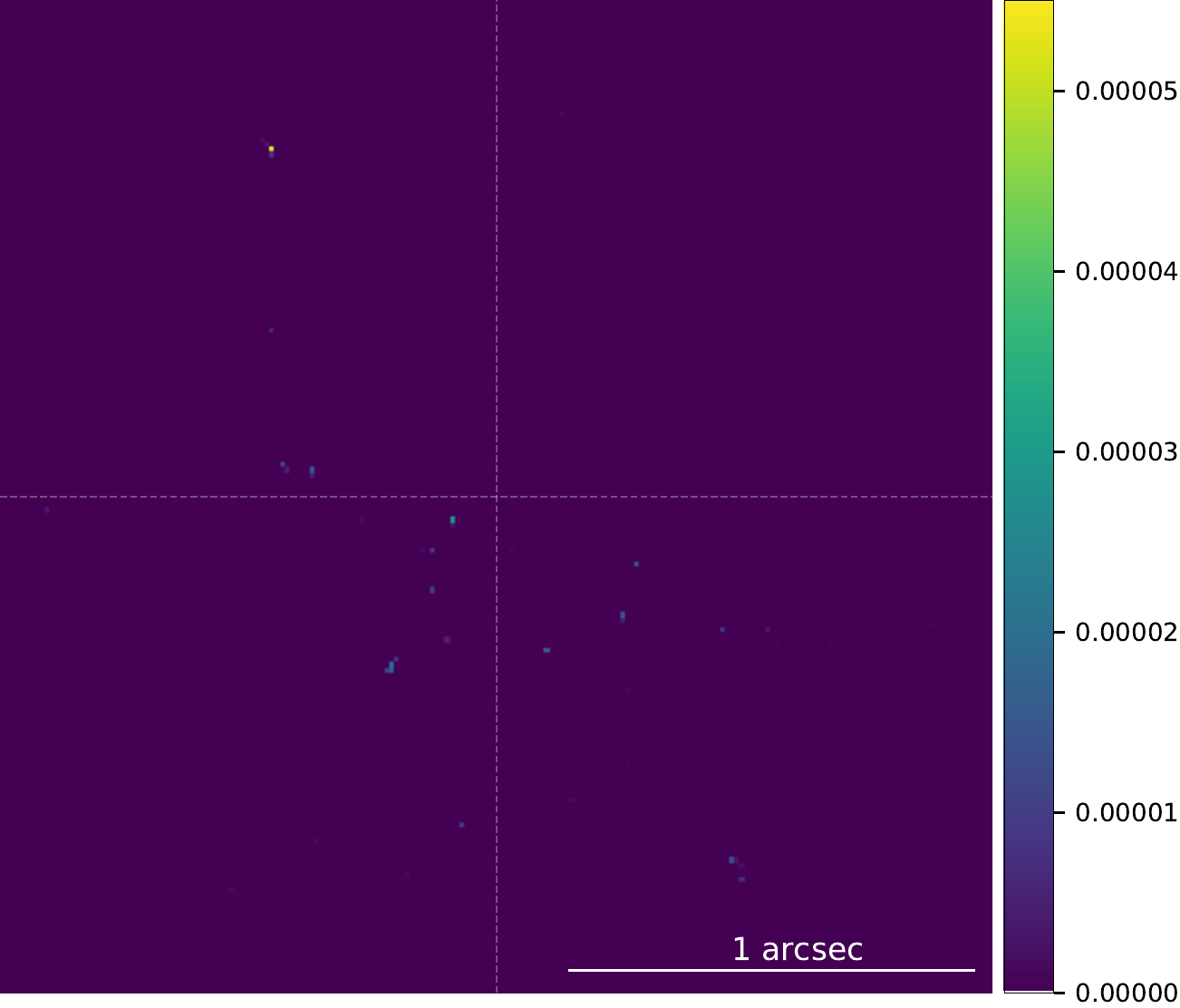}}\\
                \subfloat[HD201219*]{\includegraphics[width=115pt]{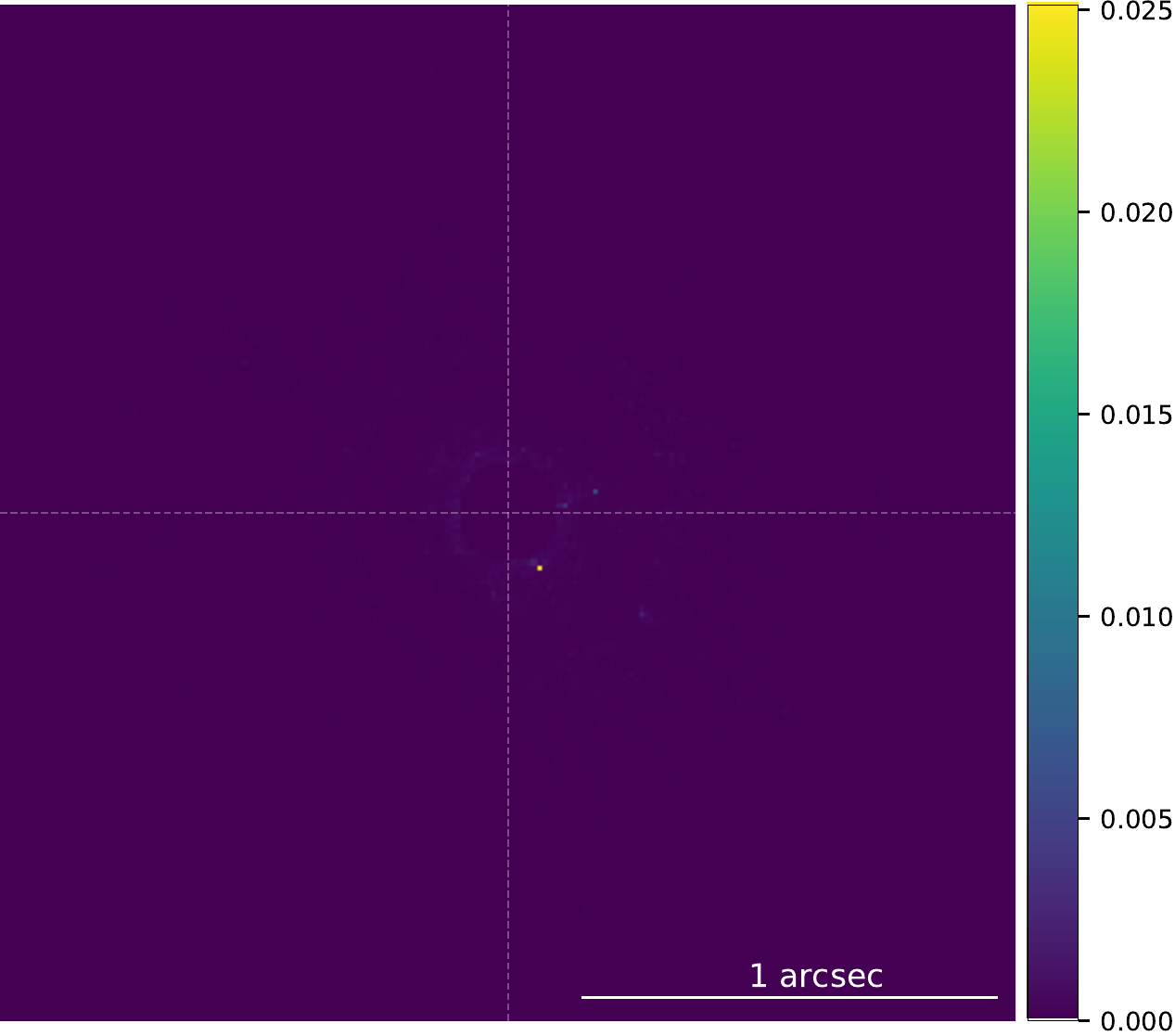}}
    \subfloat[HD192758*]{\includegraphics[width=115pt]{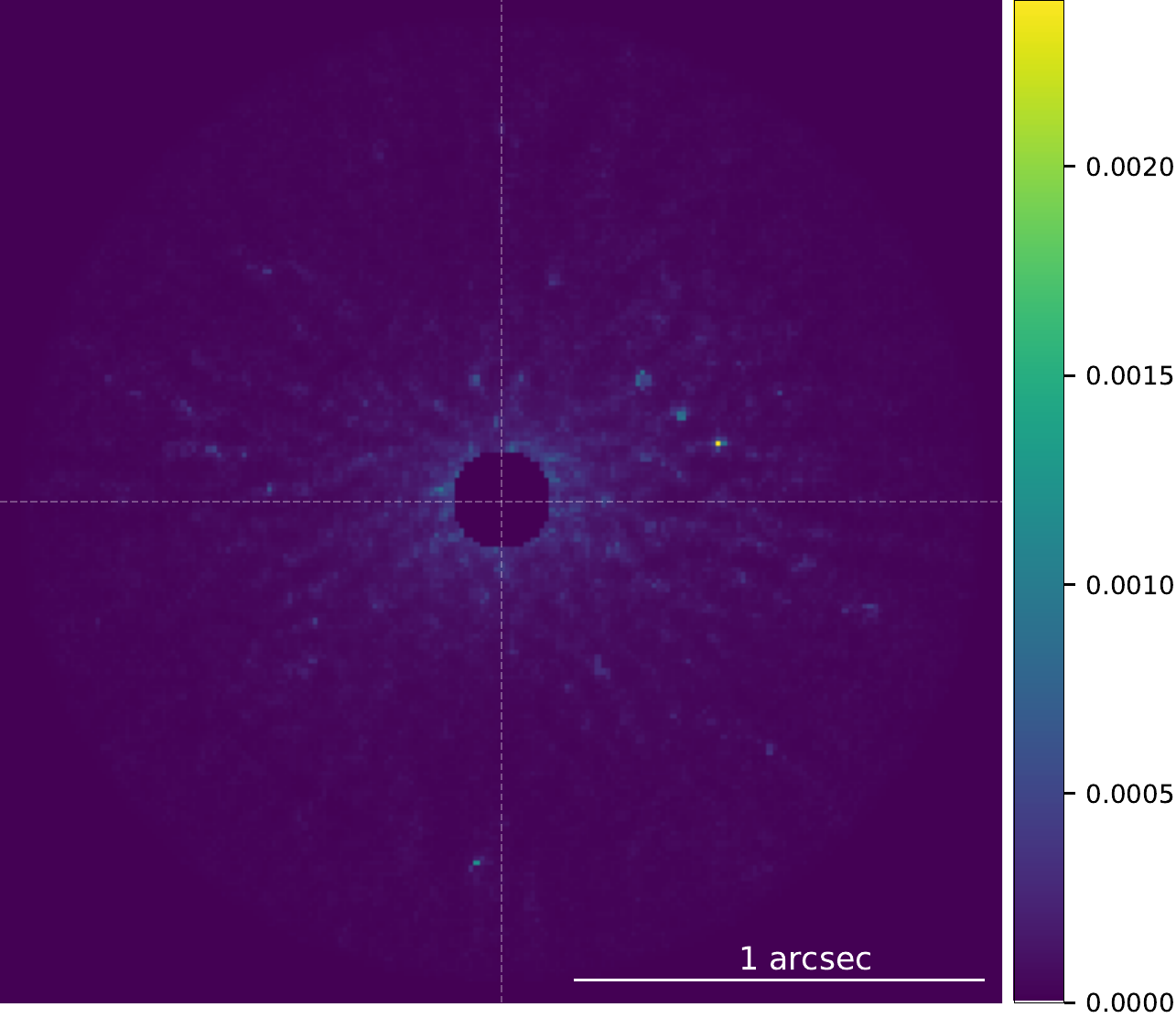}}
        \subfloat[HD192758 (2$^{nd}$ epoch)*]{\includegraphics[width=115pt]{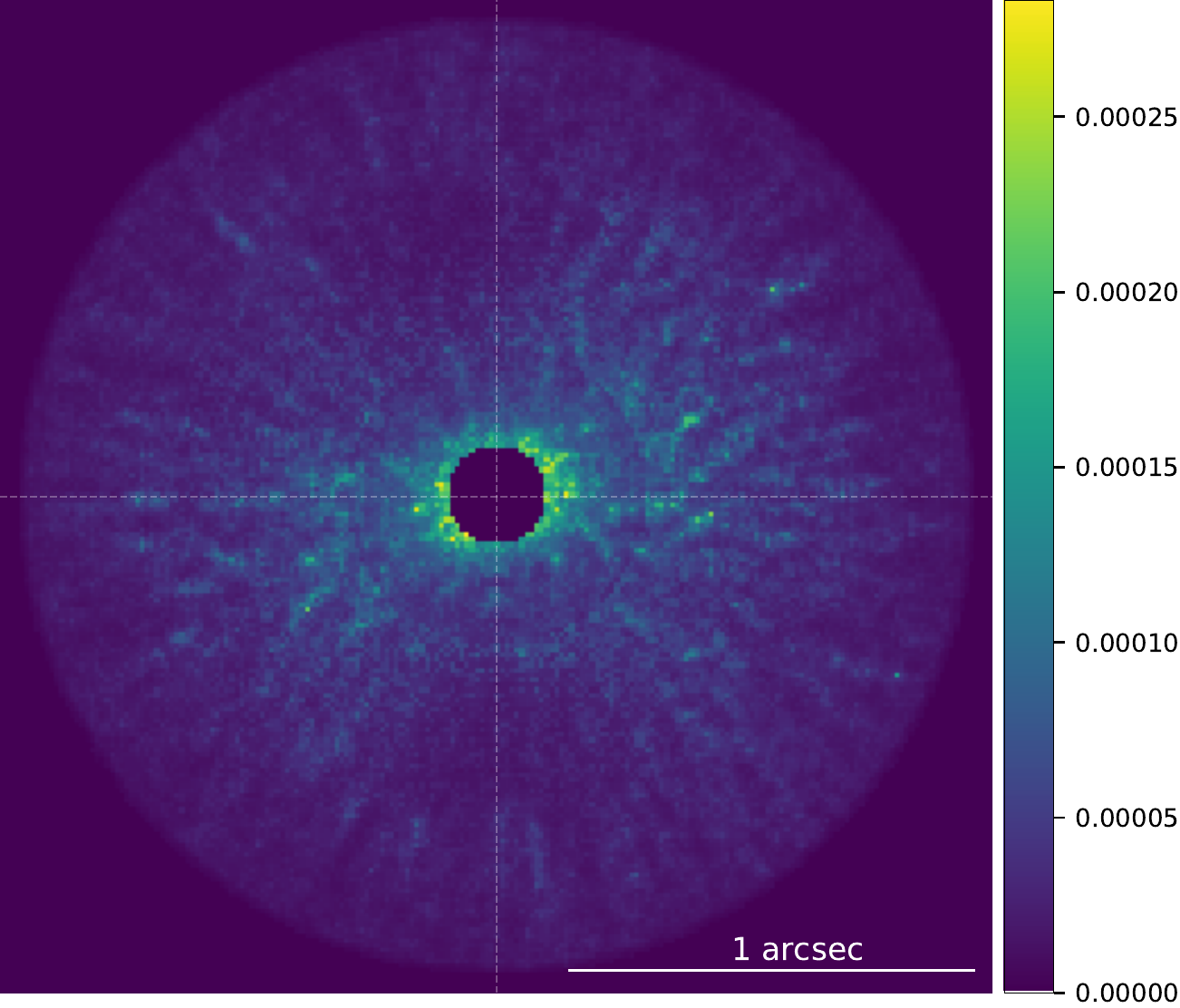}}
    \subfloat[HD218340*]{\includegraphics[width=115pt]{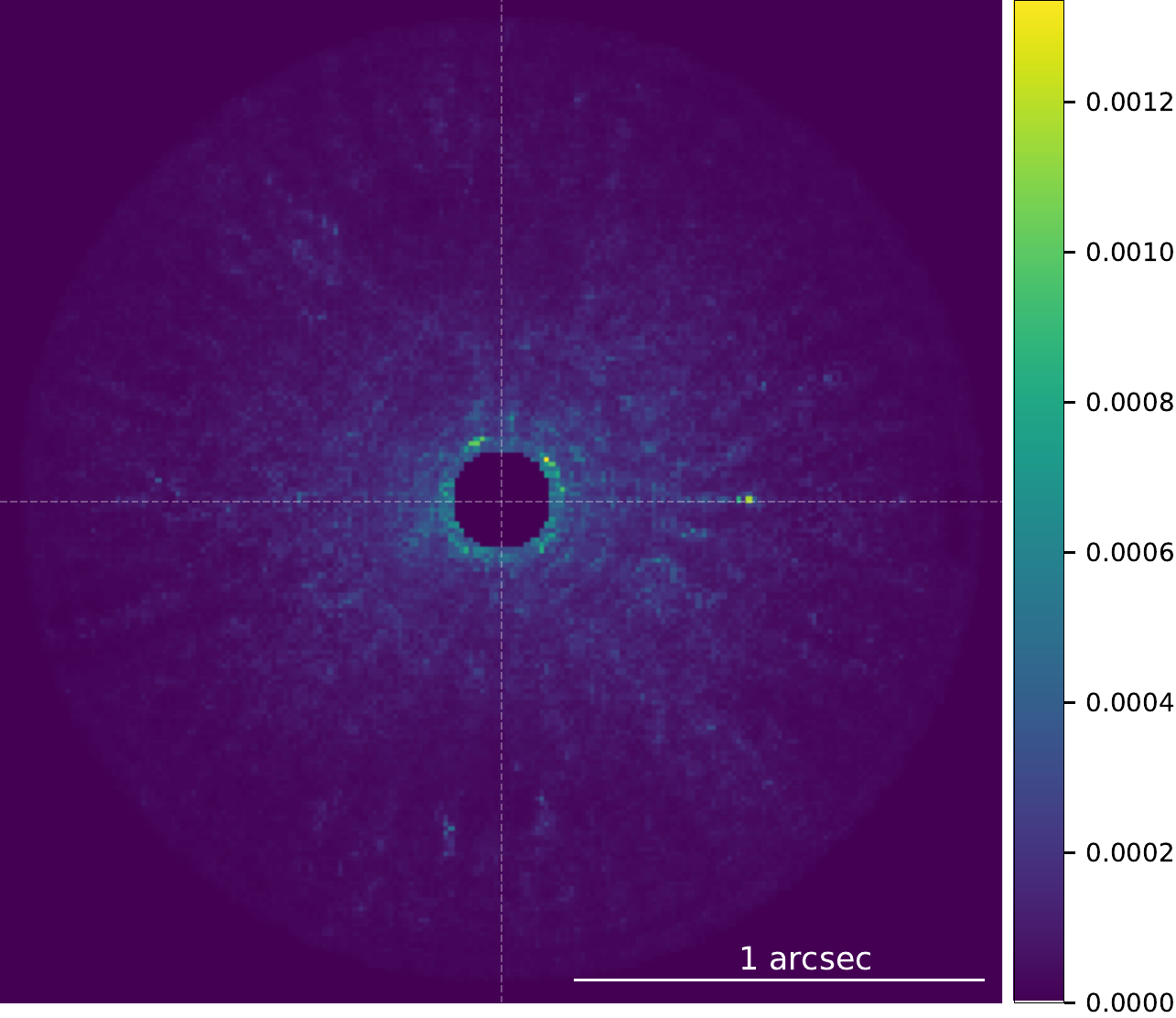}}\\

  \caption{\label{Empty_map2} RSM detection maps generated using Auto-RSM or the optimal parameters obtained with Auto-RSM for the dataset at the center of the clusters (see Table \ref{Clusters})). These detection maps did not lead to the detection of a planetary candidate. The asterisks indicate the targets on which the full Auto-RSM framework was applied.}
\end{figure*}

        \begin{figure*}[!htbp]
\footnotesize
  \centering

          \subfloat[HD120534]{\includegraphics[width=115pt]{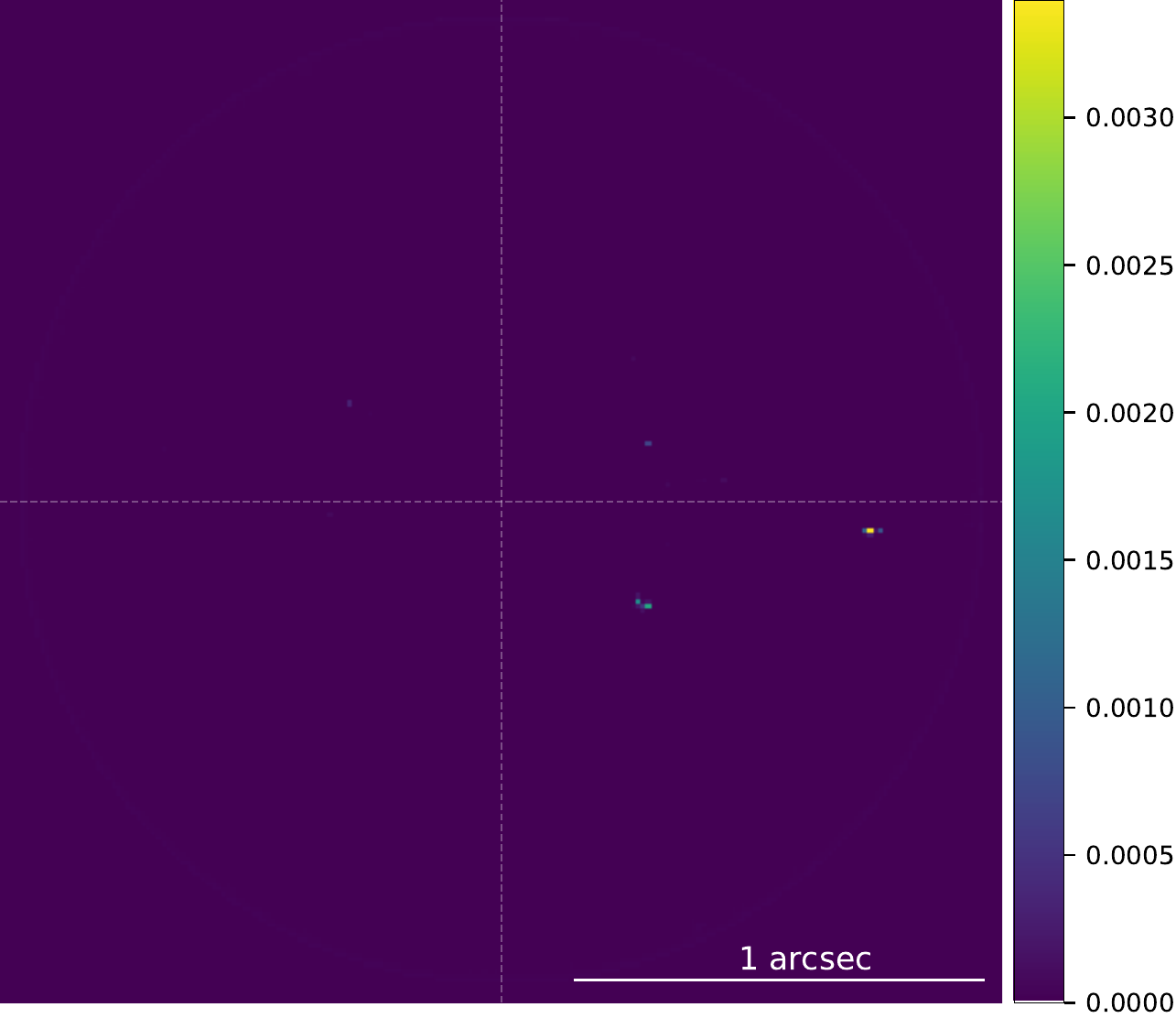}}
    \subfloat[HD122652]{\includegraphics[width=115pt]{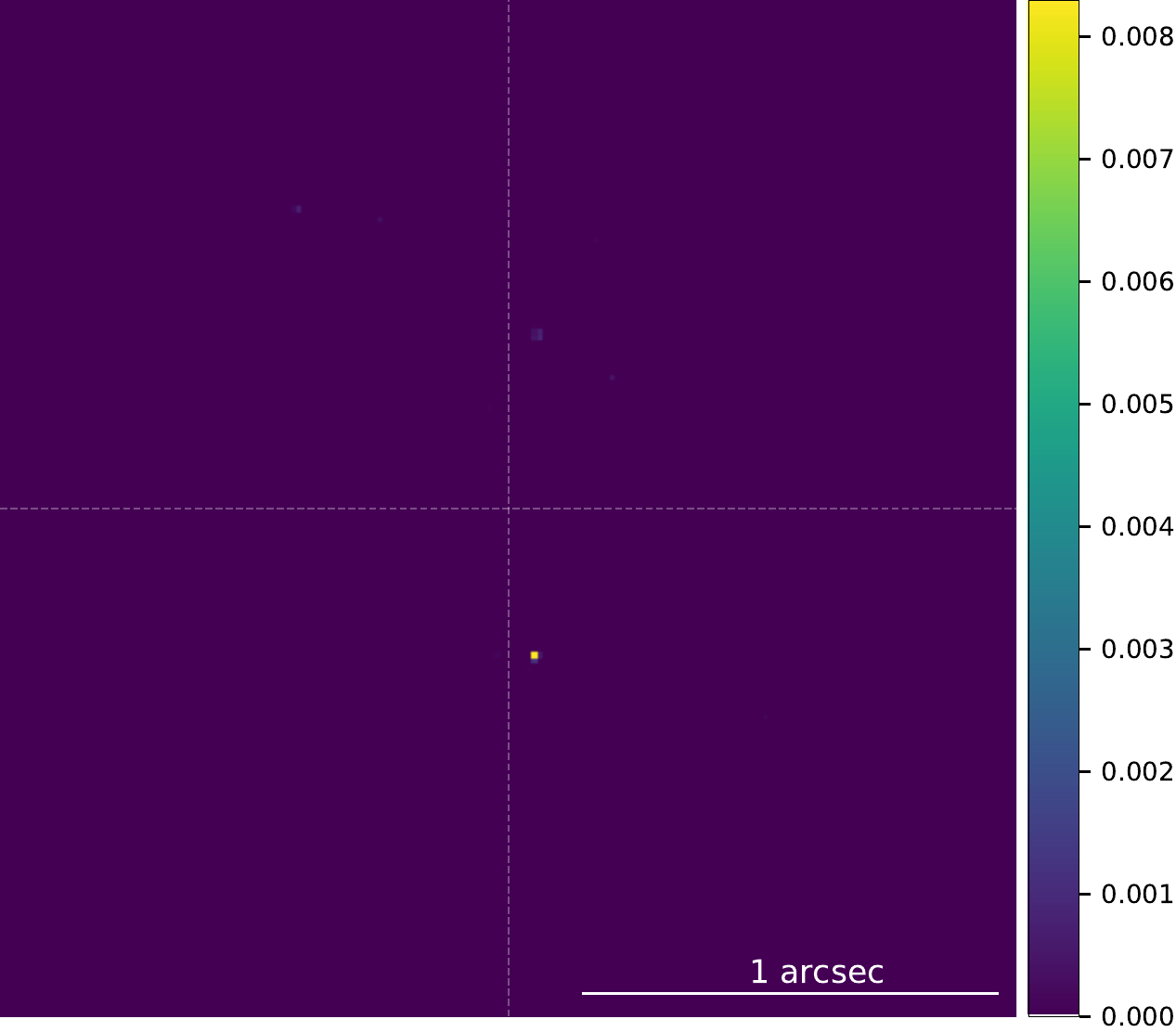}}
        \subfloat[HD122652 (2$^{nd}$ epoch)]{\includegraphics[width=115pt]{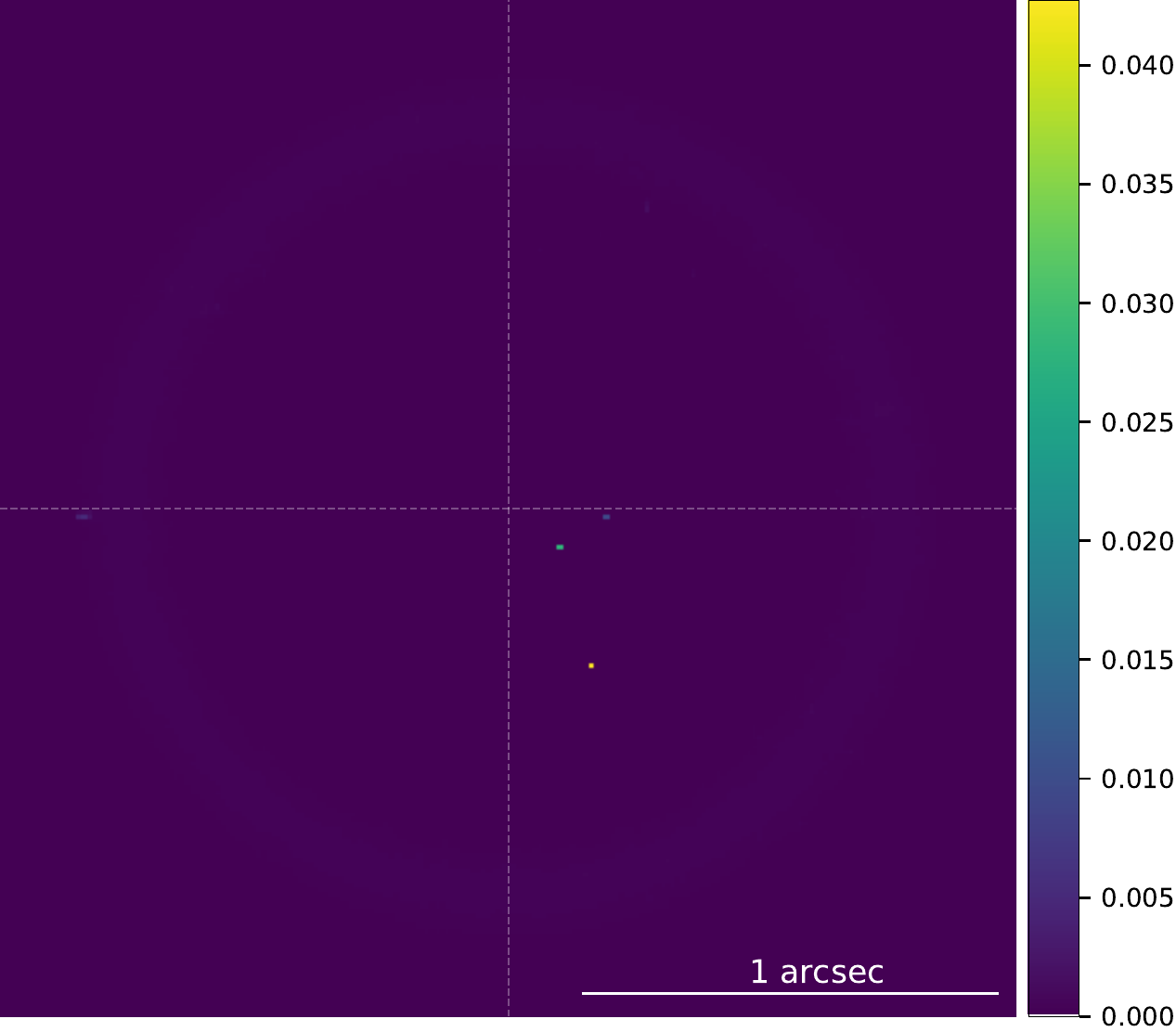}}
        \subfloat[HD133803]{\includegraphics[width=115pt]{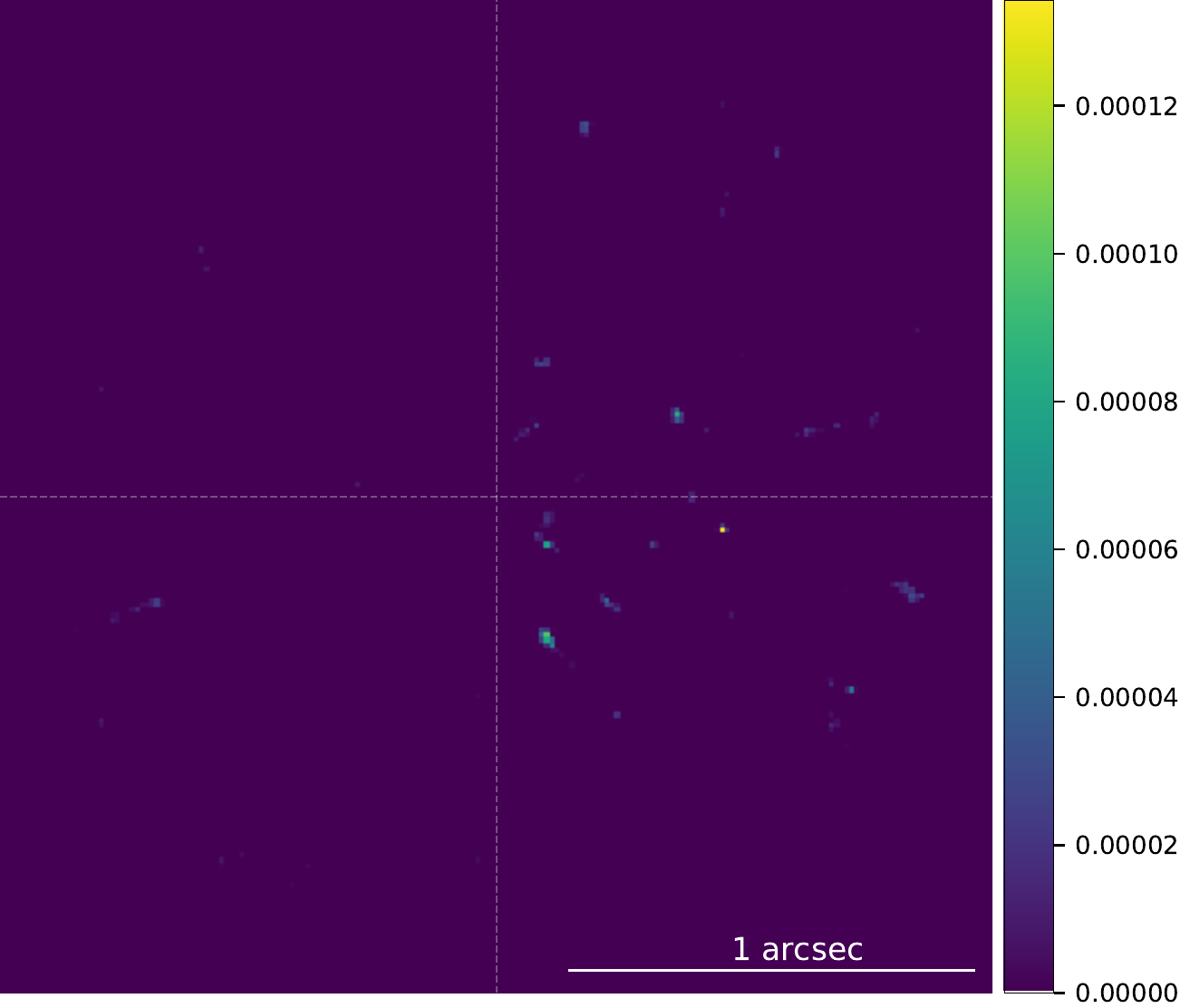}}\\
          \subfloat[HD135599]{\includegraphics[width=115pt]{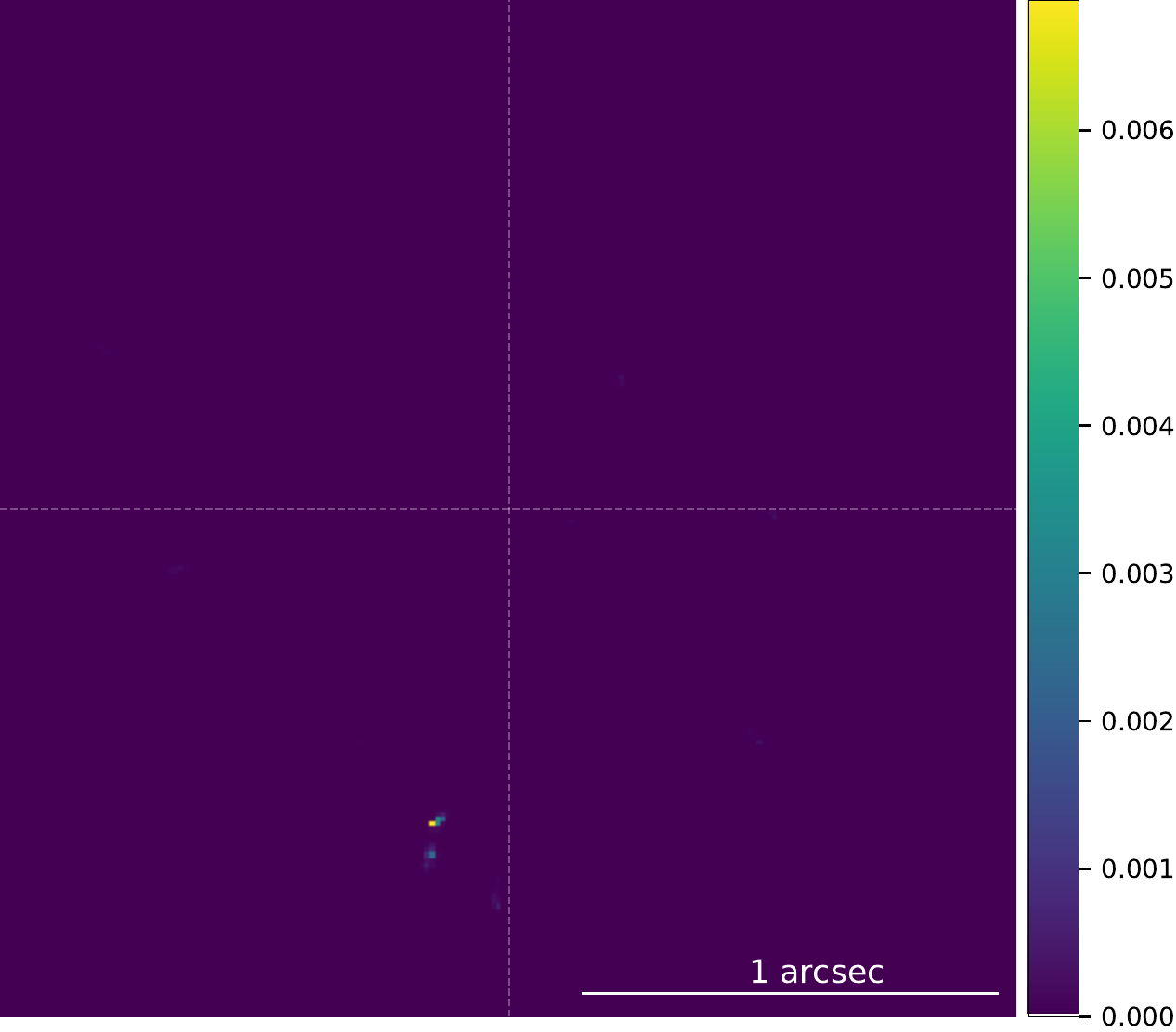}}
    \subfloat[HD138965*]{\includegraphics[width=115pt]{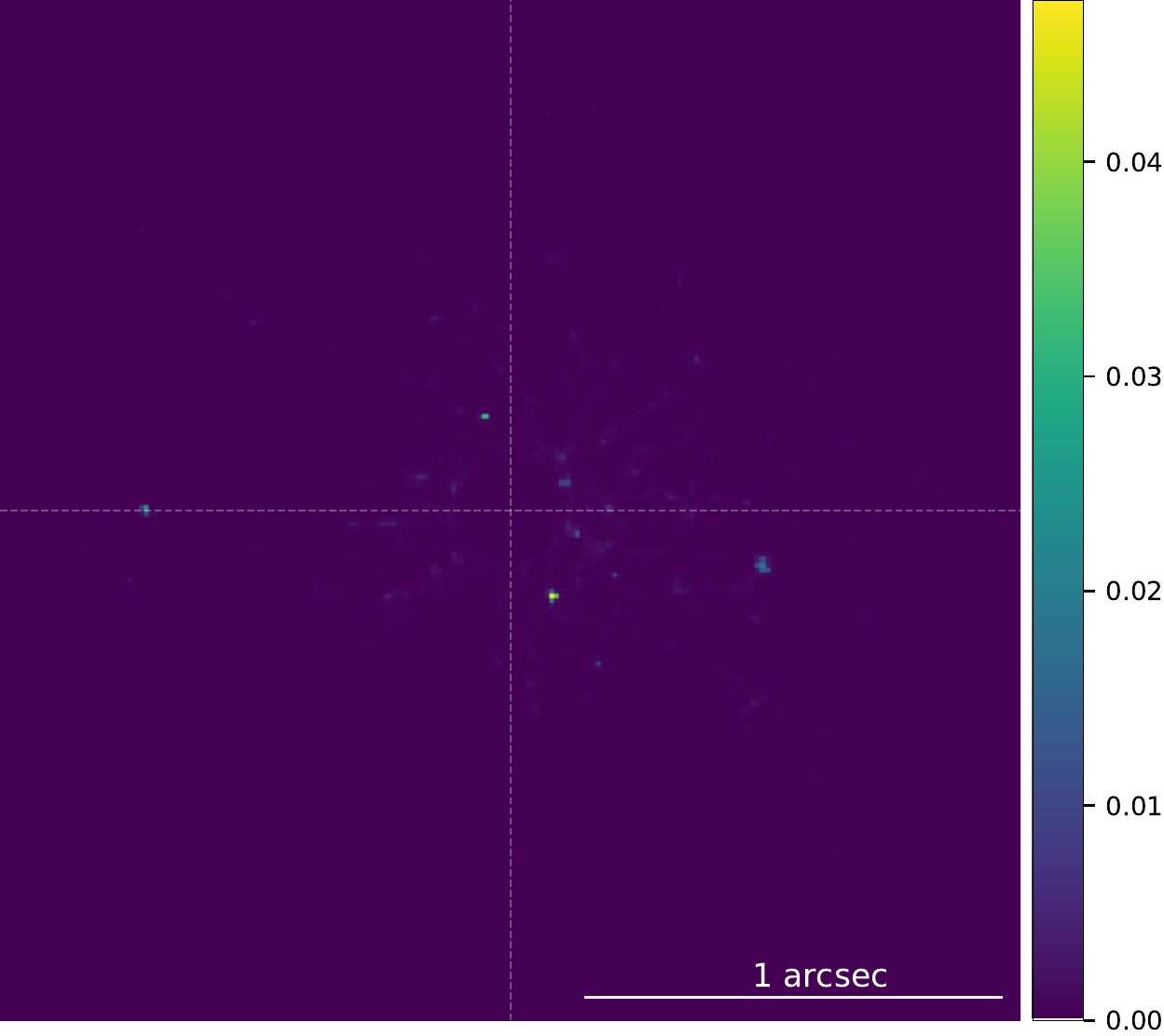}}
        \subfloat[HD145229]{\includegraphics[width=115pt]{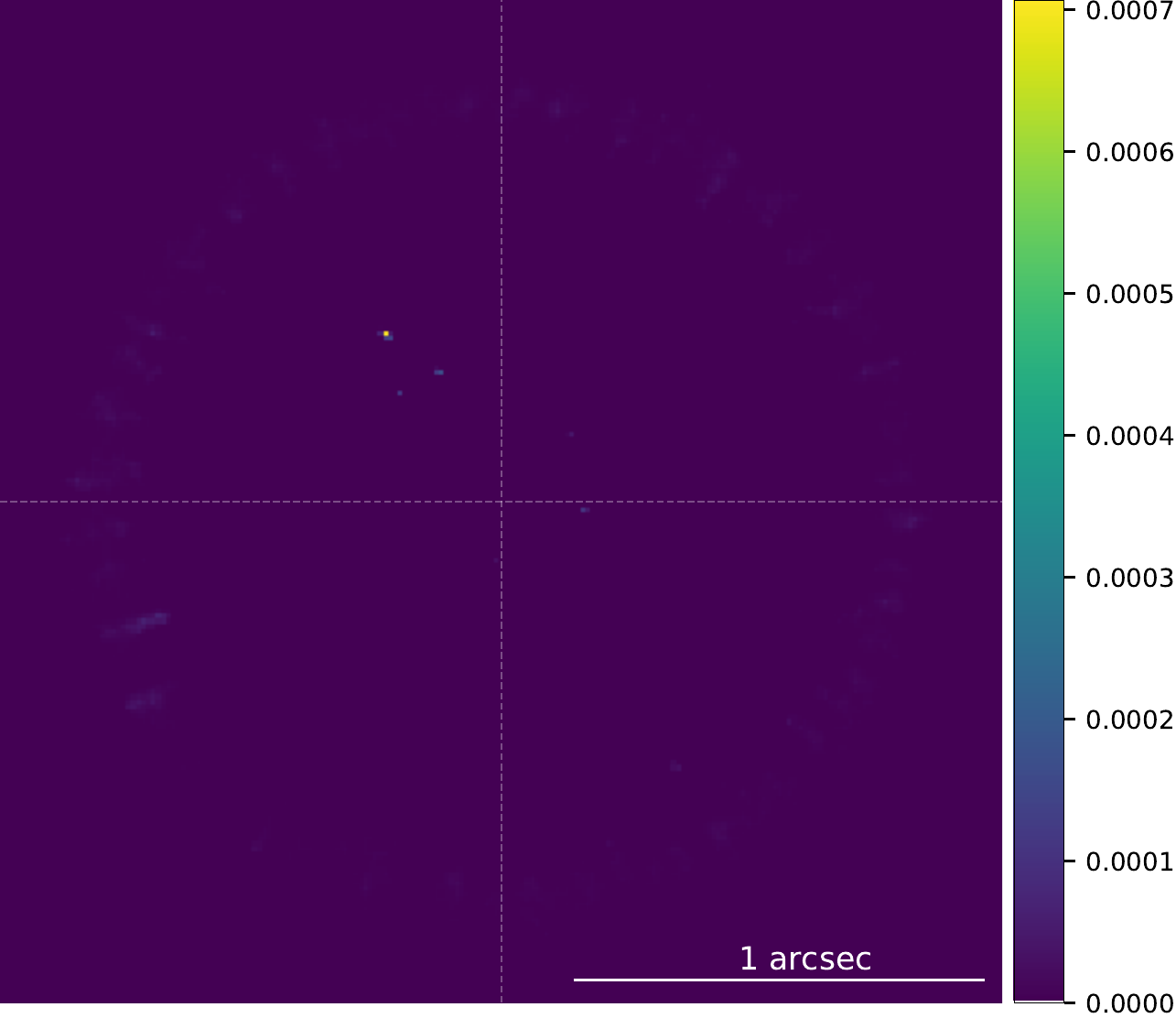}}
        \subfloat[HD145229 (2$^{nd}$ epoch)]{\includegraphics[width=115pt]{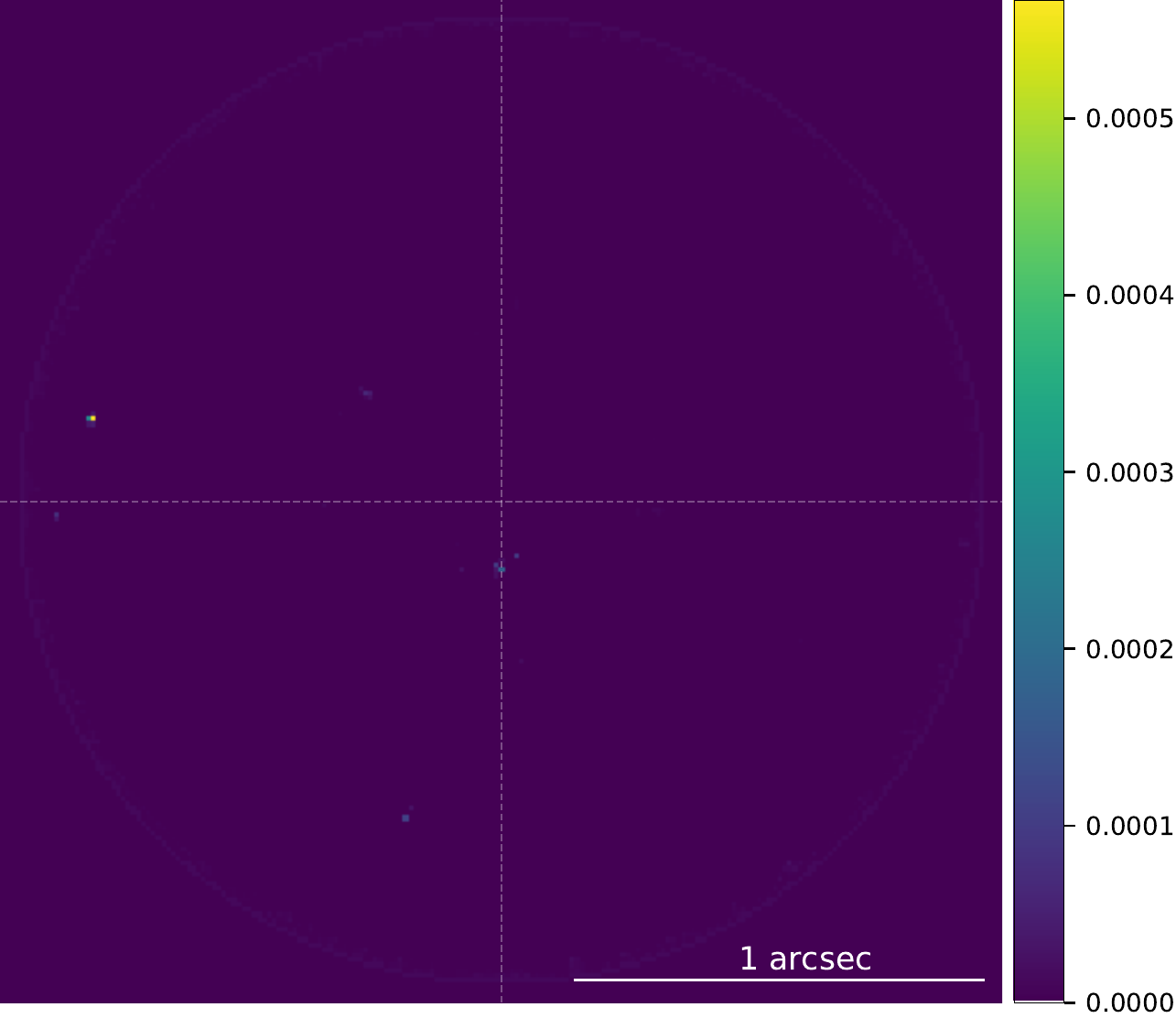}}\\
          \subfloat[HD157728*]{\includegraphics[width=115pt]{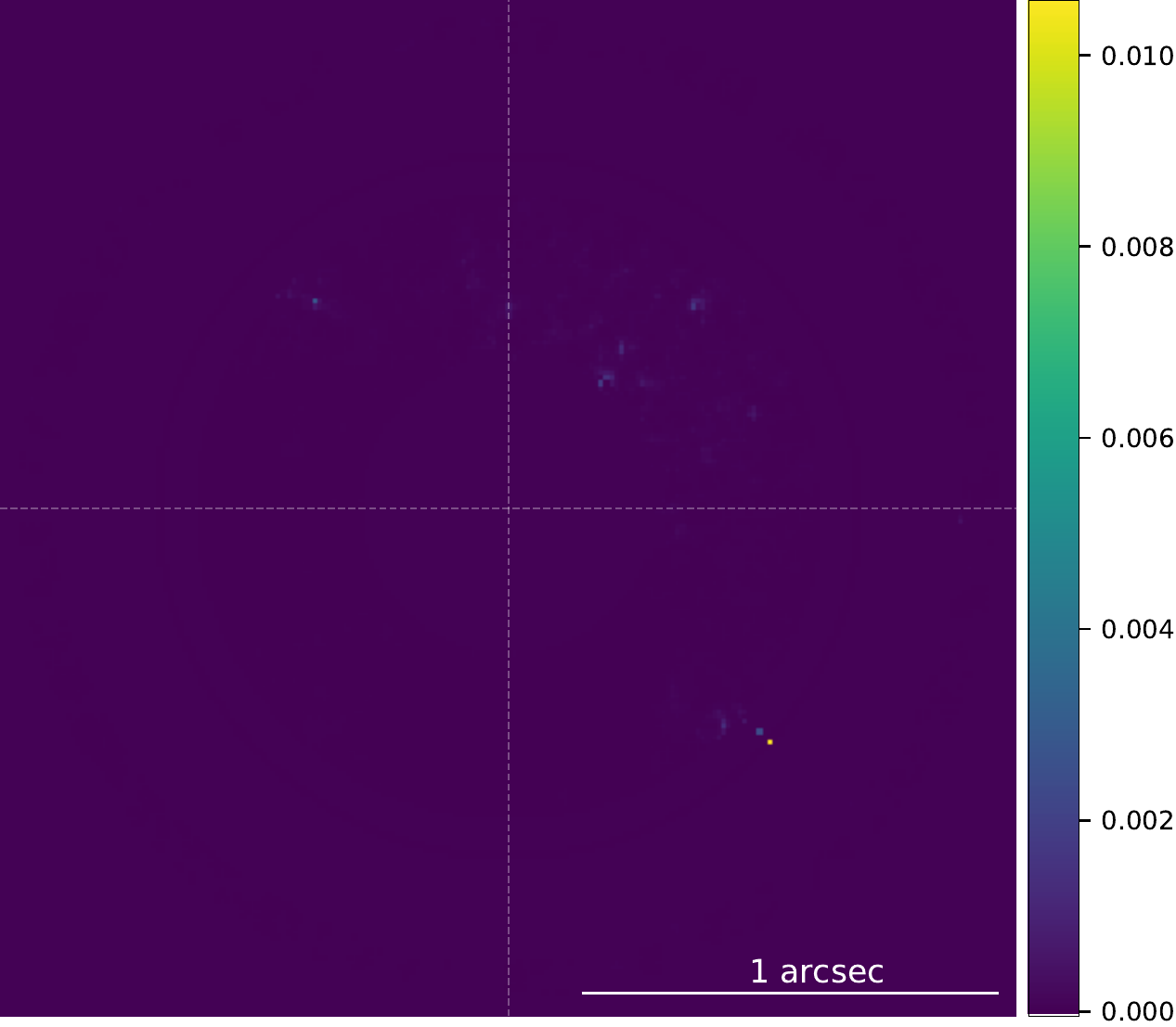}}
    \subfloat[HD164249A]{\includegraphics[width=115pt]{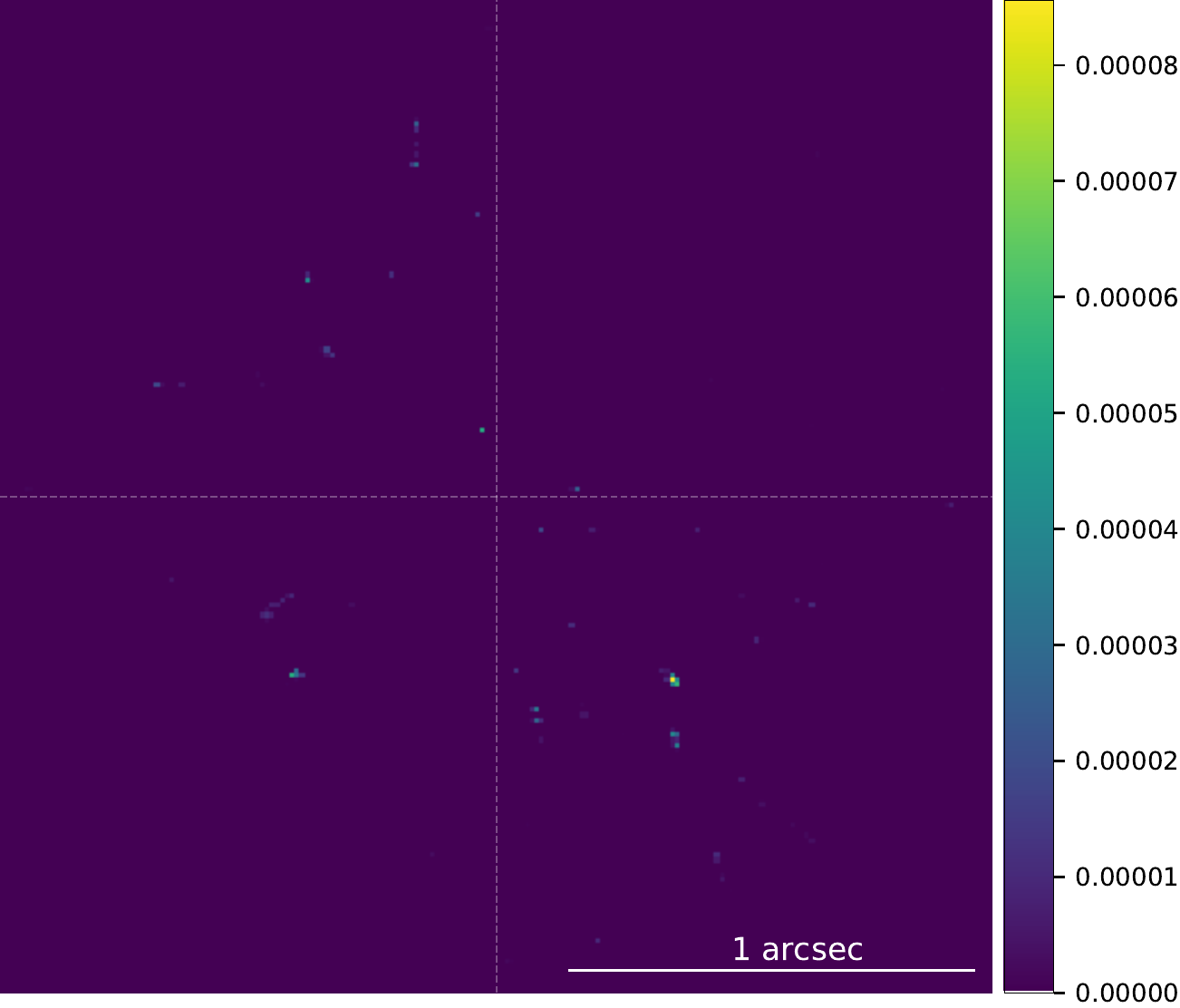}}
        \subfloat[HD172555]{\includegraphics[width=115pt]{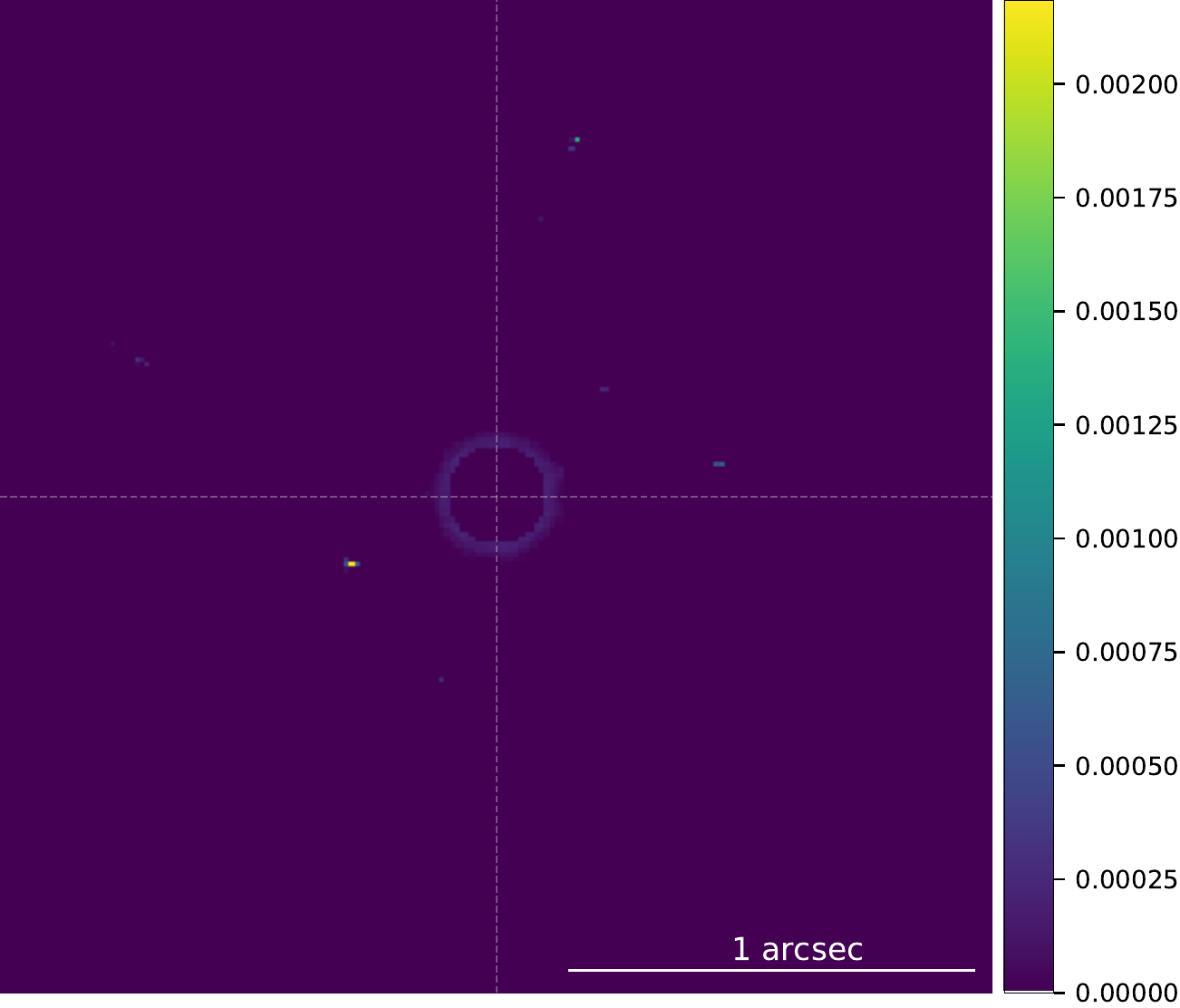}}
        \subfloat[HD181296*]{\includegraphics[width=115pt]{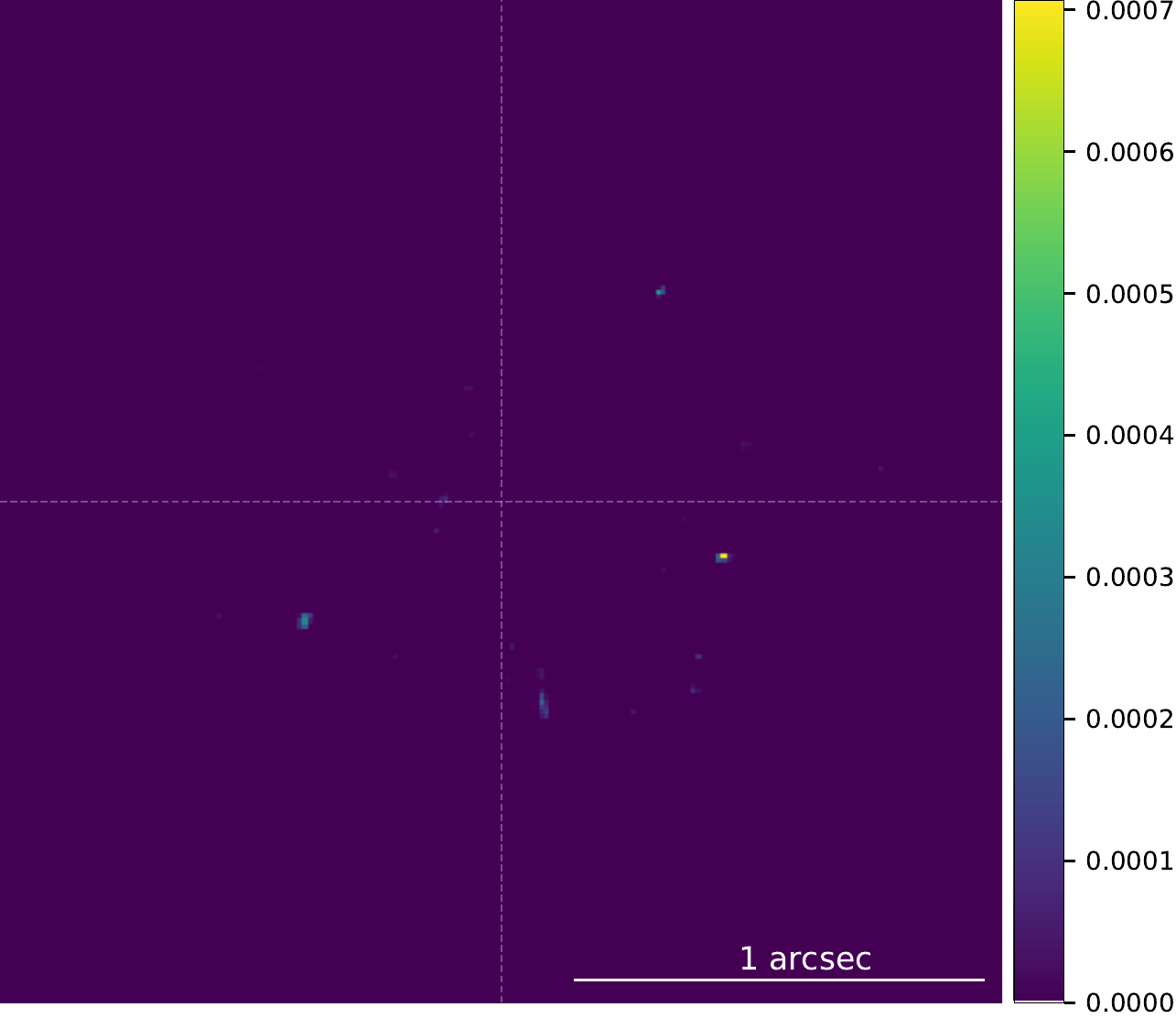}}\\
          \subfloat[HD182681]{\includegraphics[width=115pt]{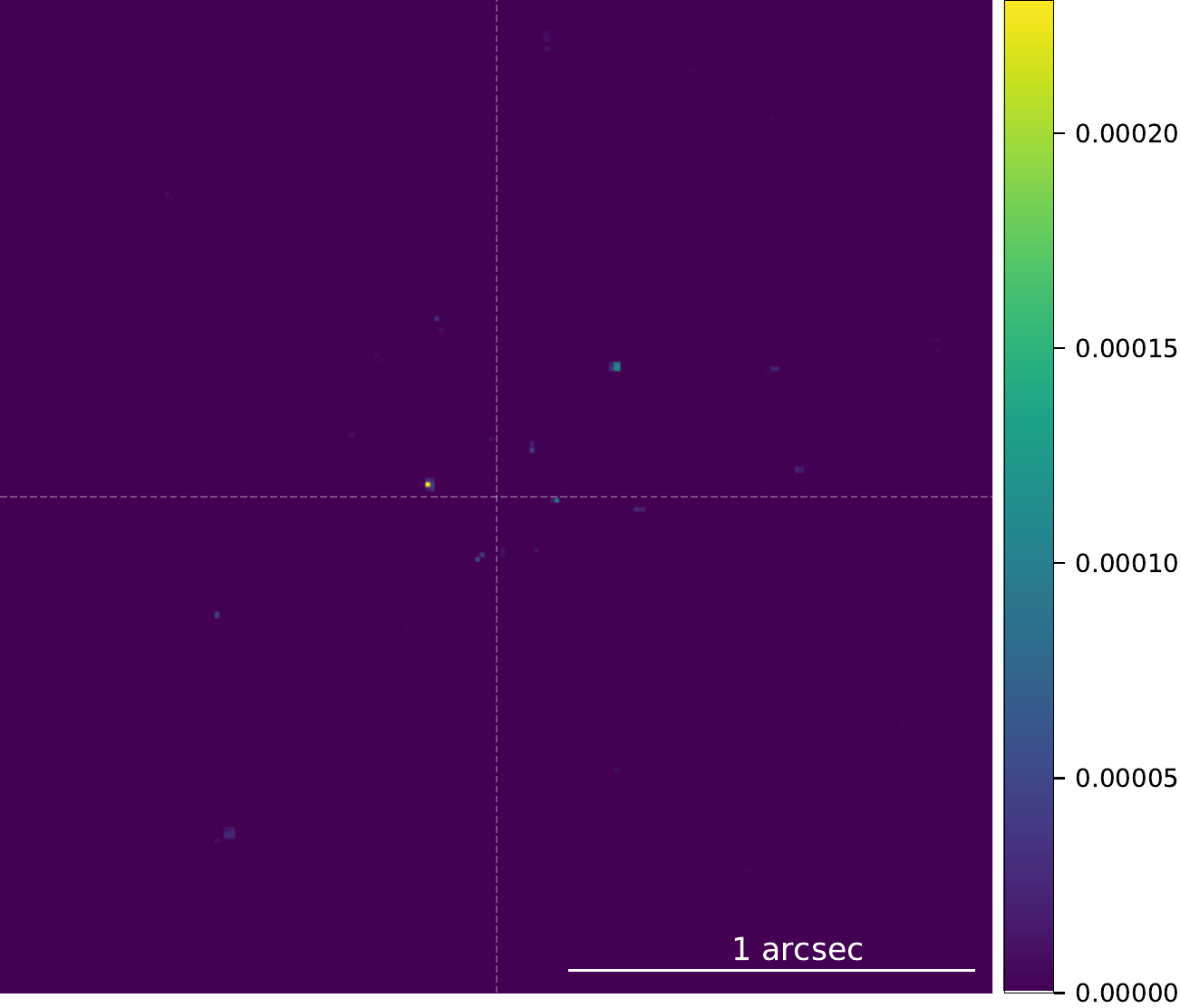}}
    \subfloat[HD205674]{\includegraphics[width=115pt]{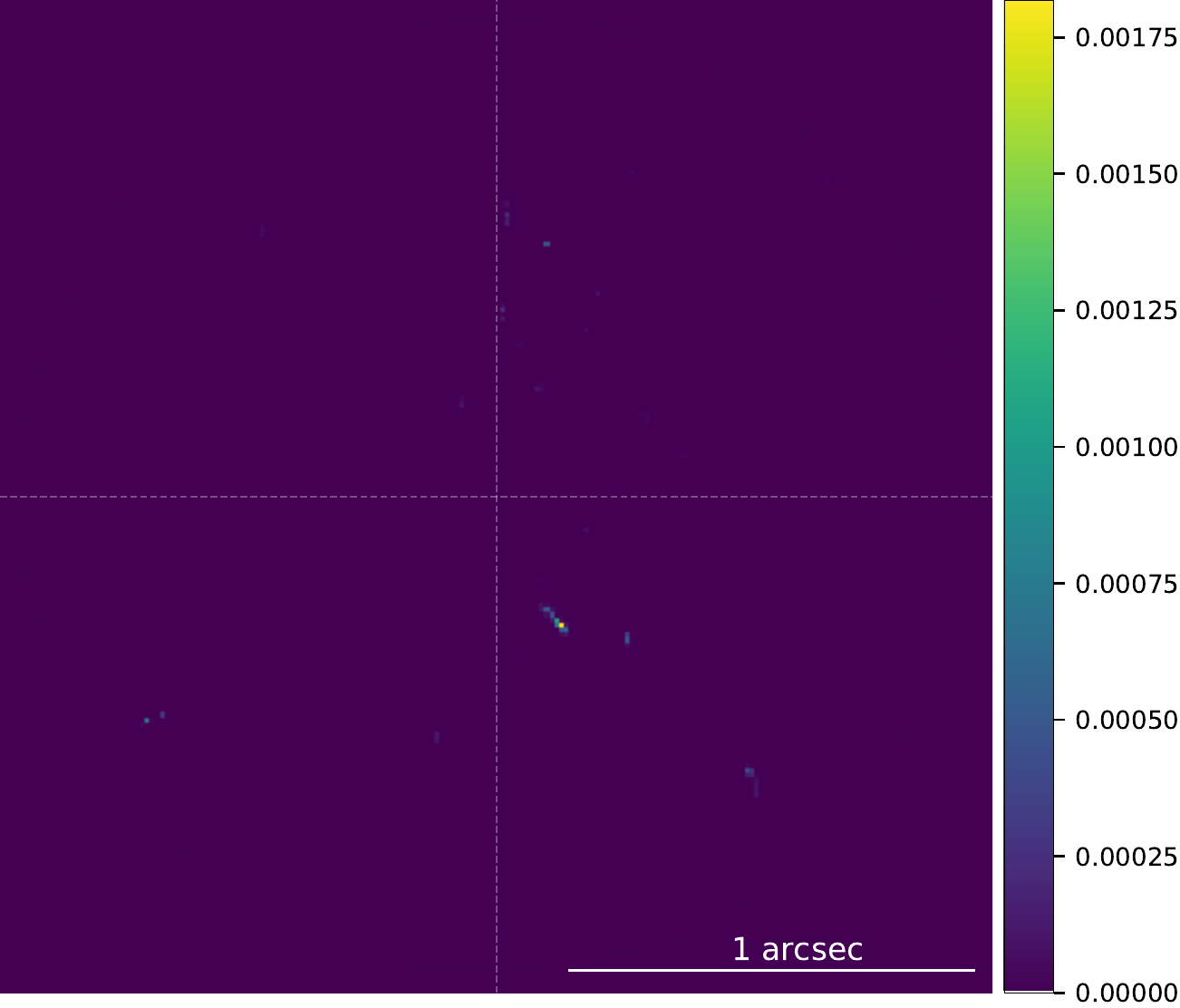}}
        \subfloat[HD221853*]{\includegraphics[width=115pt]{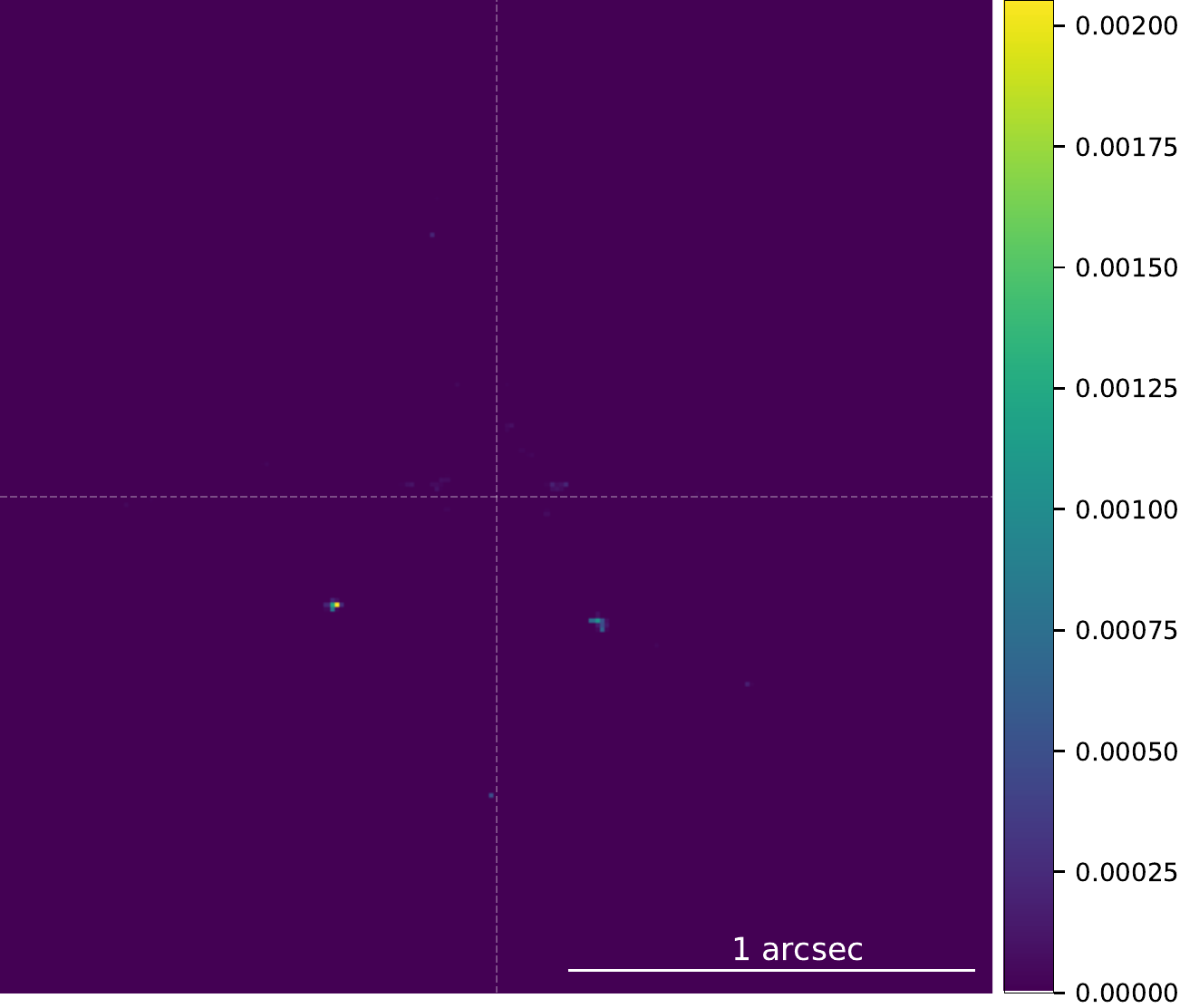}}
        \subfloat[HD274255]{\includegraphics[width=115pt]{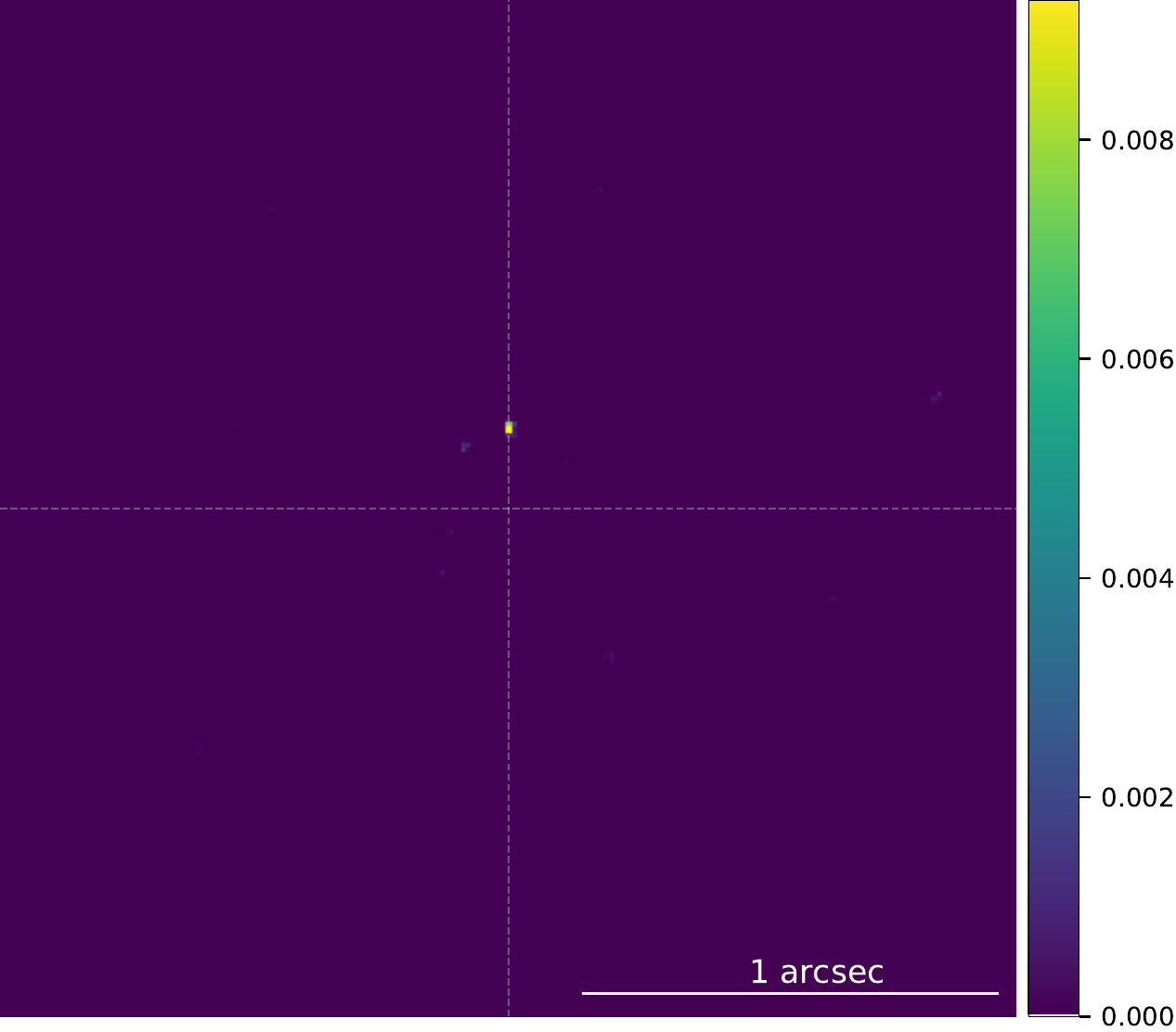}}\\
          \subfloat[HD21997*]{\includegraphics[width=115pt]{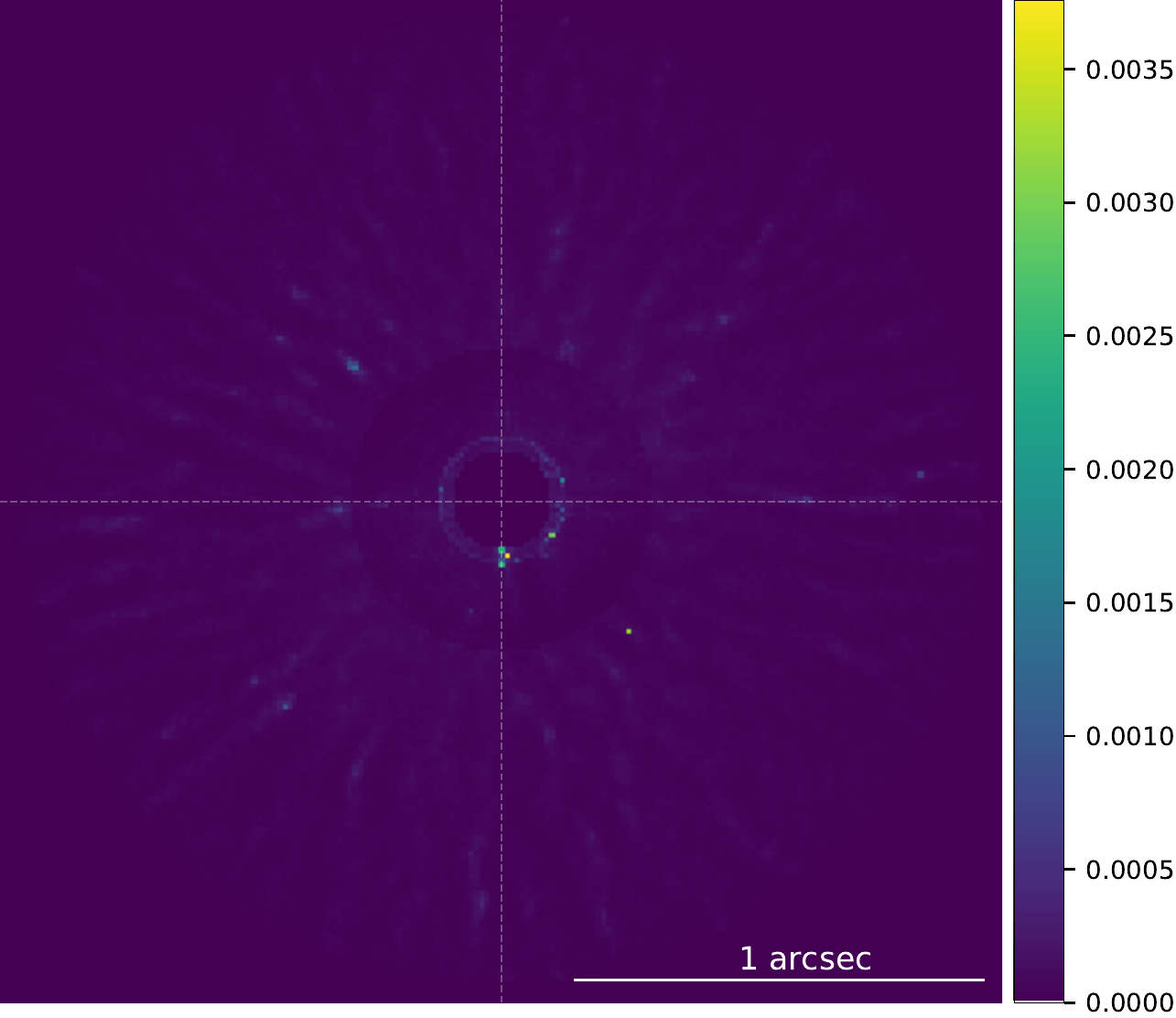}}

  \caption{\label{Empty_map3} RSM detection maps generated using Auto-RSM or the optimal parameters obtained with Auto-RSM for the dataset at the center of the clusters (see Table \ref{Clusters}). These detection maps did not lead to the detection of a planetary candidate. The asterisks indicate the targets on which the full Auto-RSM framework was applied.}
\end{figure*}
\FloatBarrier

\twocolumn
\section{Threshold computation and interpretation}
\label{thresh}
The radially evolving residual noise measure subtracted from the detection map is estimated by taking, for each annulus, the largest value observed in the detection map generated with reversed parallactic angles. A polynomial fit is then applied on the obtained values to limit the influence of potential outliers (see below) and smooth the curve. This radial threshold is finally subtracted from the original detection maps and any negative value is set to zero. This subtraction reduces the background residual noise and therefore eases the detection of potential planetary candidates.

\begin{figure}[t]
\footnotesize
  \centering

    \subfloat[HD122652 (2$^{nd}$ epoch)]{\includegraphics[width=120pt]{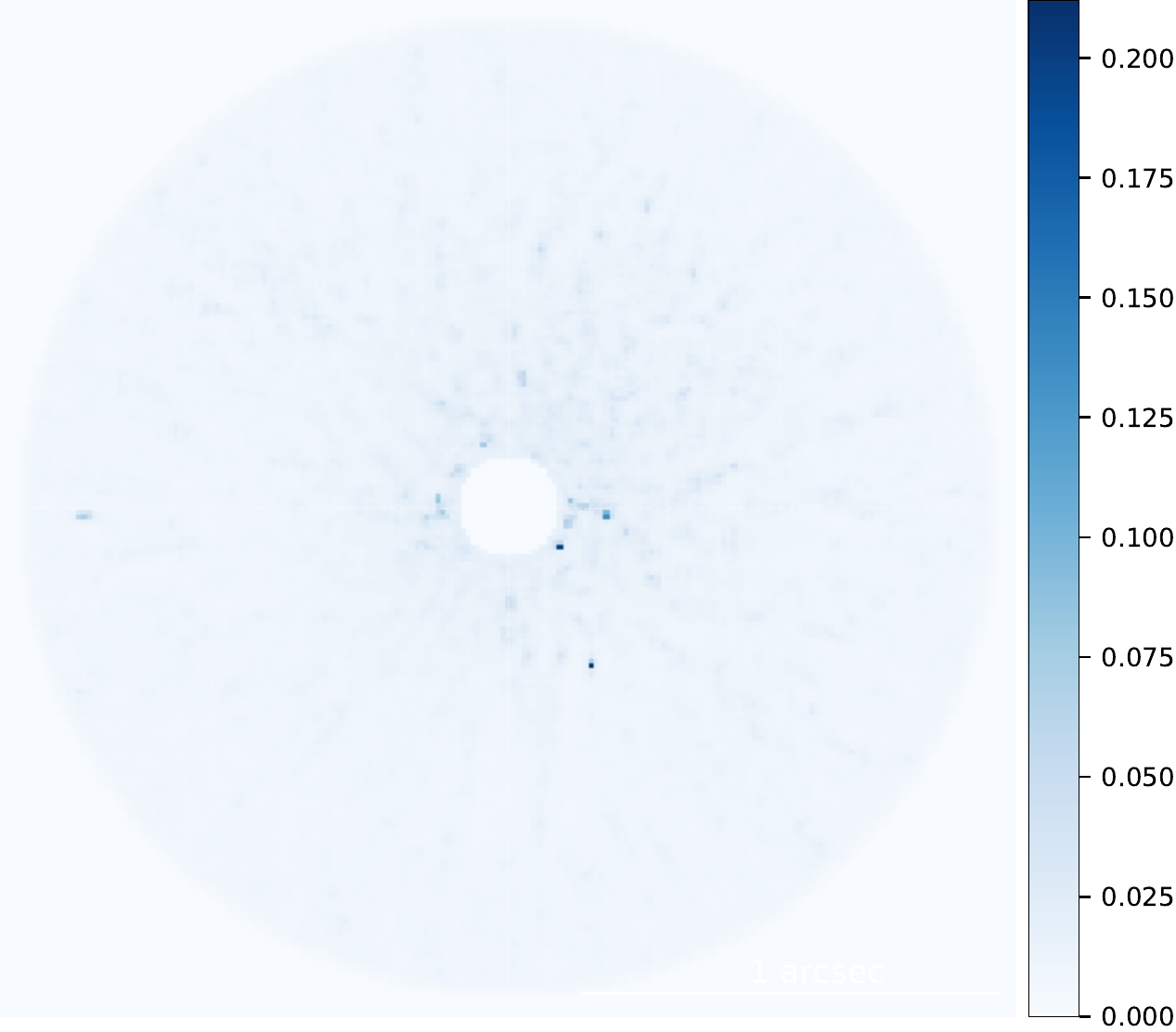}}
        \subfloat[HD122652 (2$^{nd}$ epoch) RPA]{\includegraphics[width=120pt]{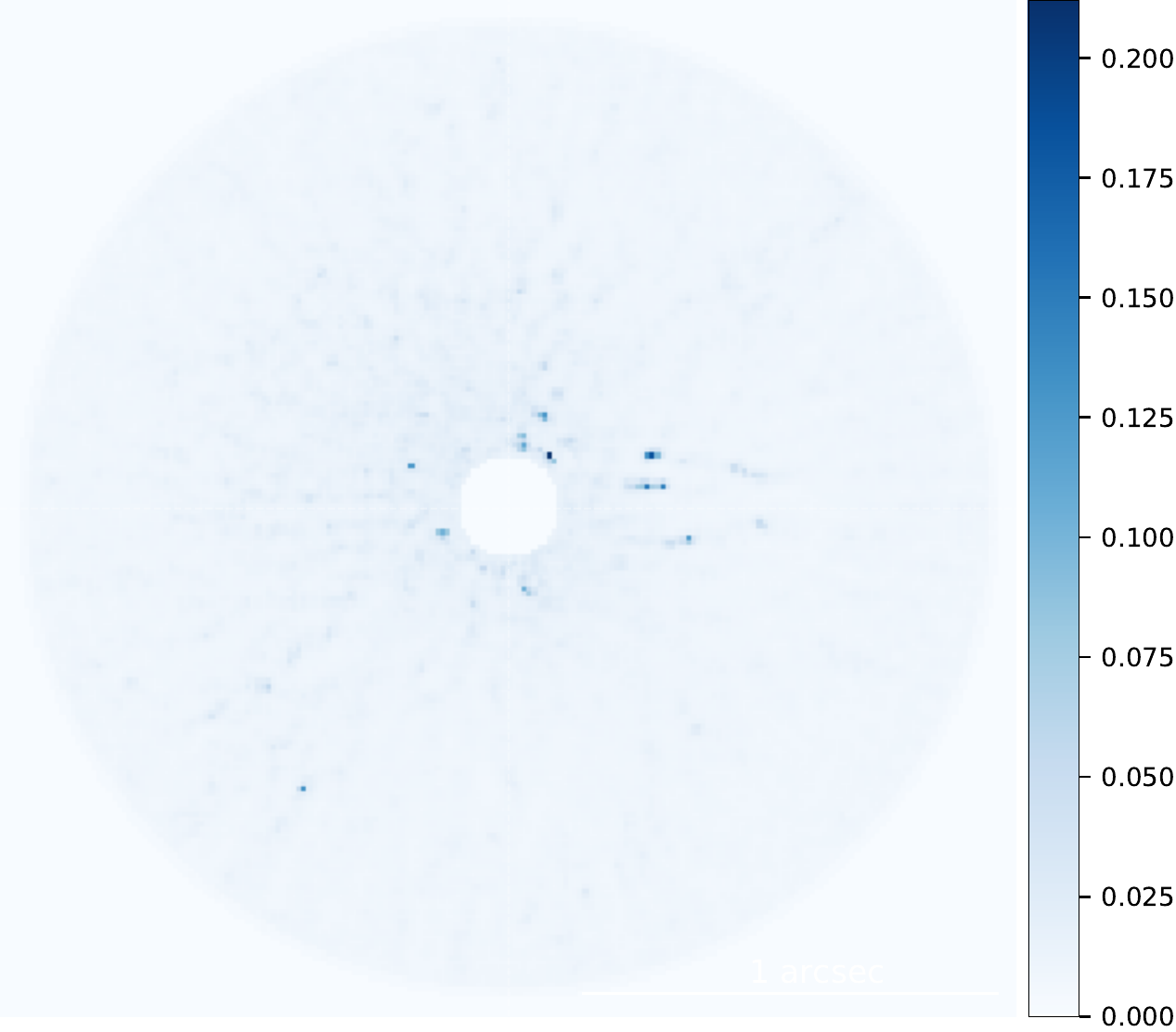}}\\
    \subfloat[HD157728]{\includegraphics[width=120pt]{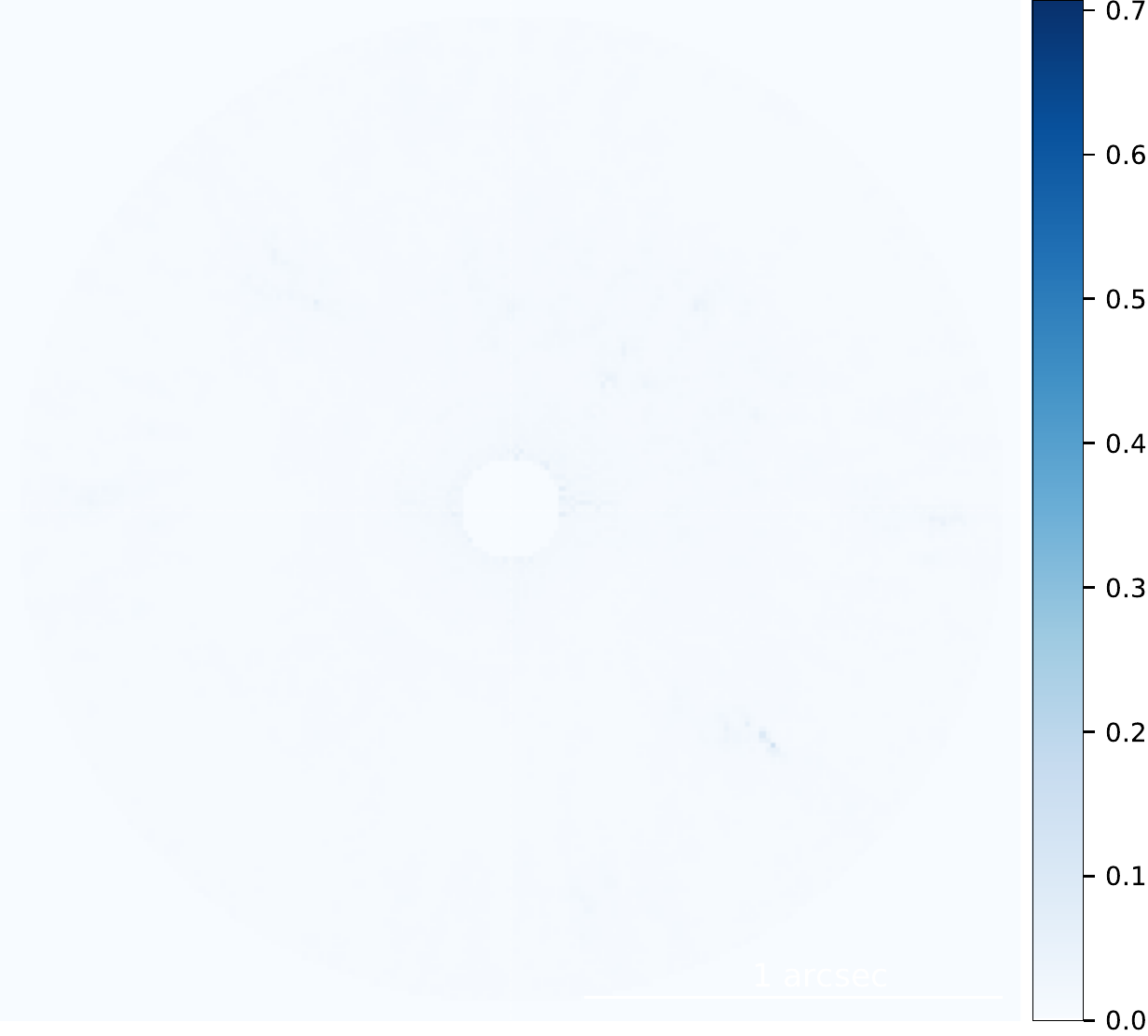}}
        \subfloat[HD157728 RPA]{\includegraphics[width=120pt]{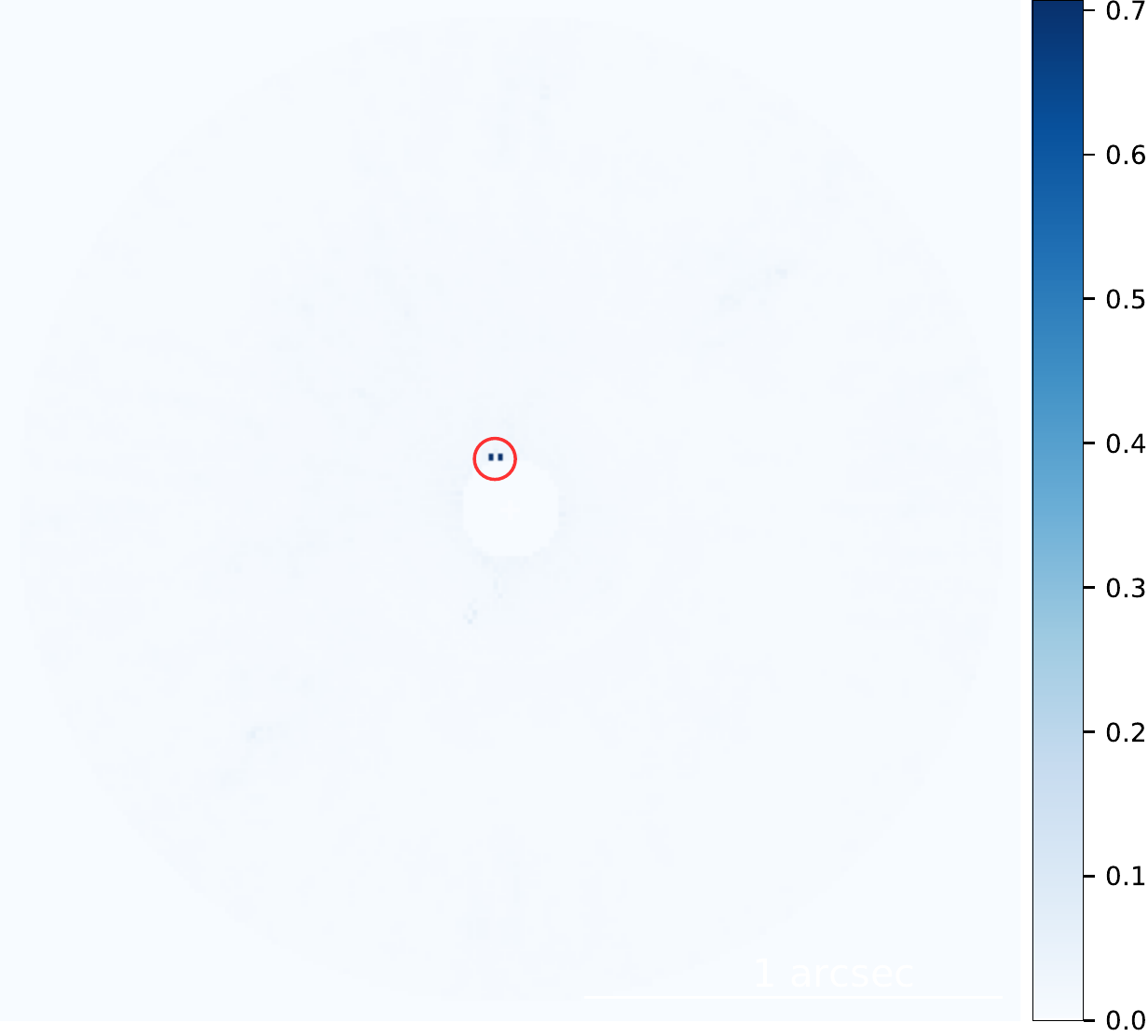}}\\
                
  \caption{\label{Inverted} RSM detection maps with and without the parallactic angles reversed for two datasets (RPA stands for reversed parallactic angles). A square root scale has been selected to highlight residual speckle noise. The bright speckle seen in HD157728 RPA is highlighted by a red circle.}
\end{figure}

This threshold should however not be considered as a sufficient condition to classify any signal above it as a planetary candidate. As can be seen from Figure \ref{Inverted}, bright structures may appear in the detection map generated with the reversed parallactic angles (right), which explains the use of a polynomial fit when estimating the threshold. Most of the time, the residual noise distributions are similar in the two detection maps, as illustrated with HD\,122652 (2$^{nd}$ epoch). But in some cases, very bright artefacts appear in the detection map with reversed parallactic angles although only a weak level of noise is visible in the original detection map (see HD\,157728). Considering all ADI sequences of the SHARDDS survey, around 20\% of the detection maps computed with the reversed parallactic angles show point-like sources or bright structures above a 0.05 threshold, while this percentage falls to 9\% for the original detection maps. It is therefore preferable to avoid using reversed parallactic angles to define a detection threshold. Detection maps generated with reversed parallactic angles may however be used to reduce the level of residual noise in the original detection maps, as described in this appendix.

\FloatBarrier

\newpage
\section{Detection maps for planetary candidates}

This appendix regroups the RSM detection maps obtained with Auto-RSM using either the bottom-up or top-down approaches to select the optimal set of likelihoods cubes (each likelihoods cube corresponding to a PSF-subtraction technique), as well as S/N maps generated via the Auto-S/N approach \citep{Dahlqvist21b} or obtained by averaging the S/N maps generated with APCA, NMF, LLSG, and LOCI, for the two samples containing a potential planetary signal.
                          
        \begin{figure*}[!htbp]
\footnotesize
  \centering
\subfloat[HD206893 RSM Bottom-up]{\includegraphics[width=115pt]{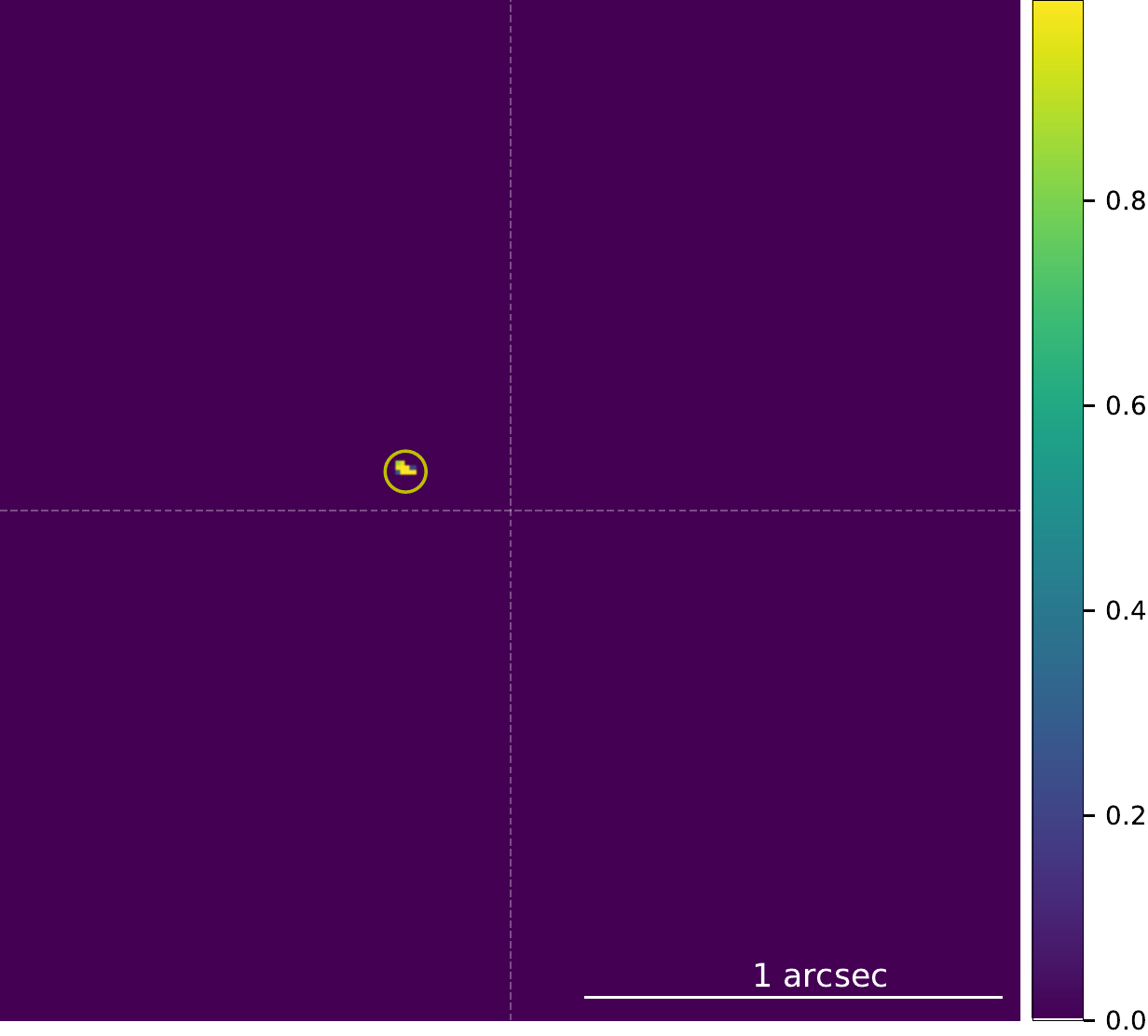}}
          \subfloat[HD206893 RSM Top-down]{\includegraphics[width=115pt]{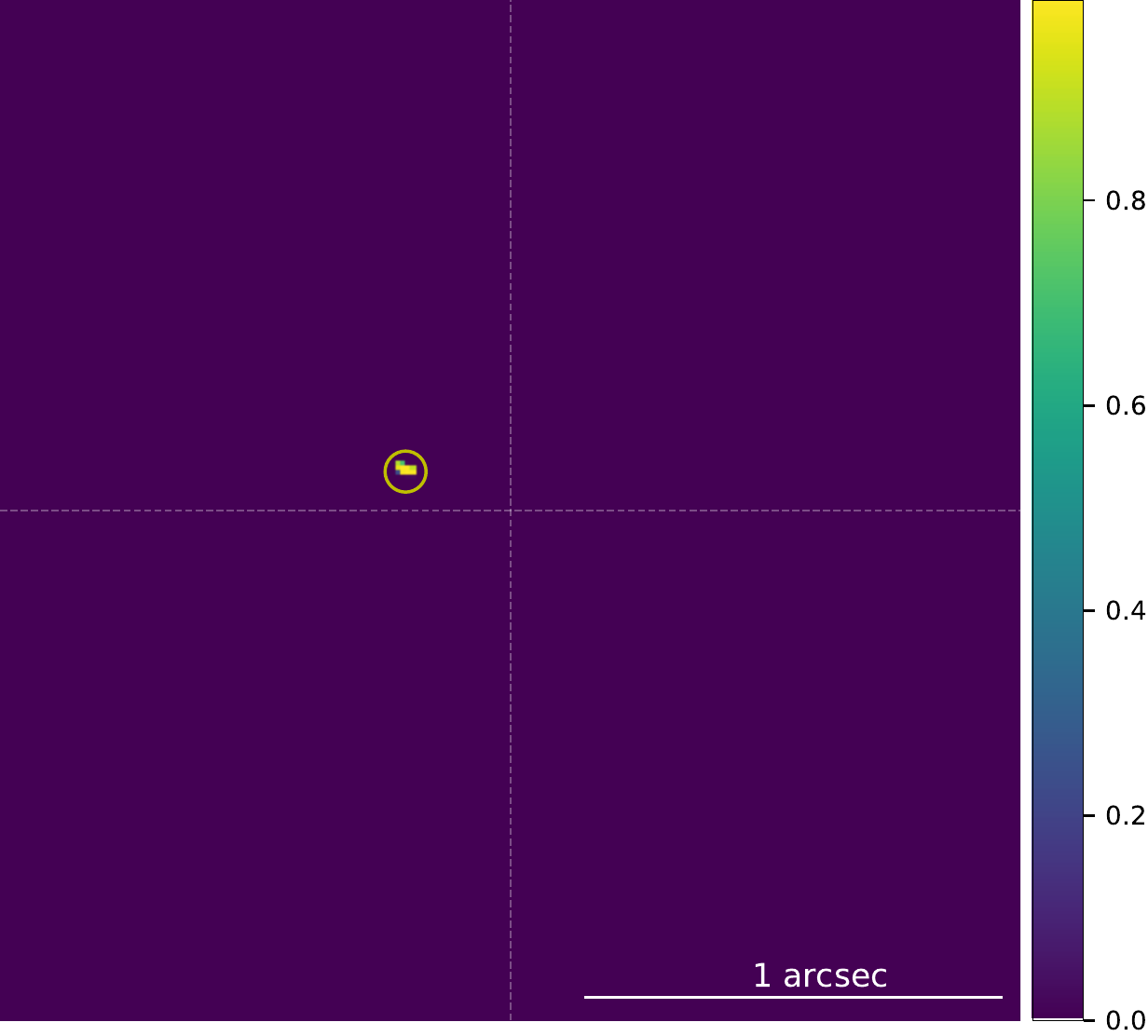}}
    \subfloat[HD206893 optimised S/N]{\includegraphics[width=115pt]{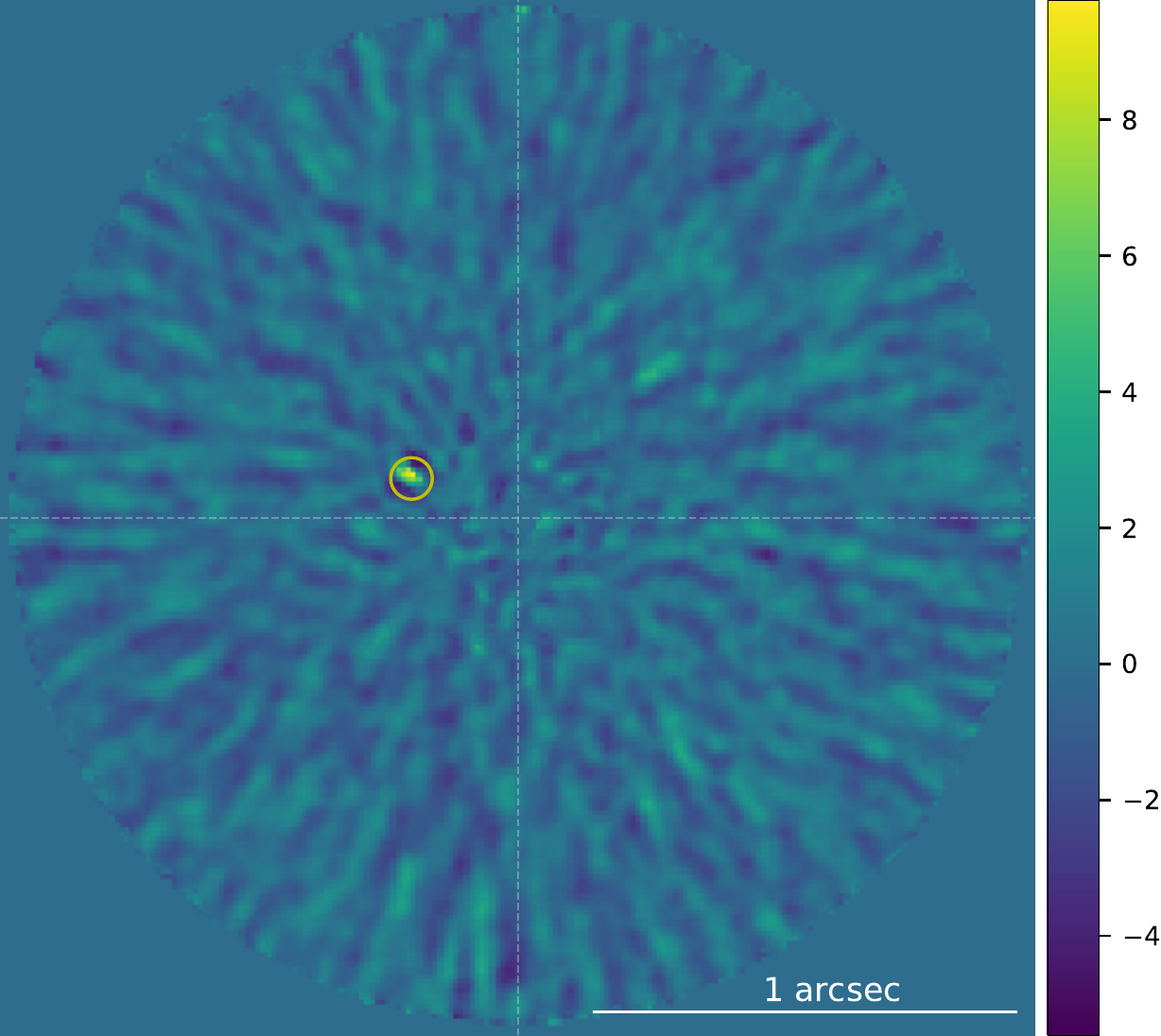}}
        \subfloat[HD206893 average S/N]{\includegraphics[width=115pt]{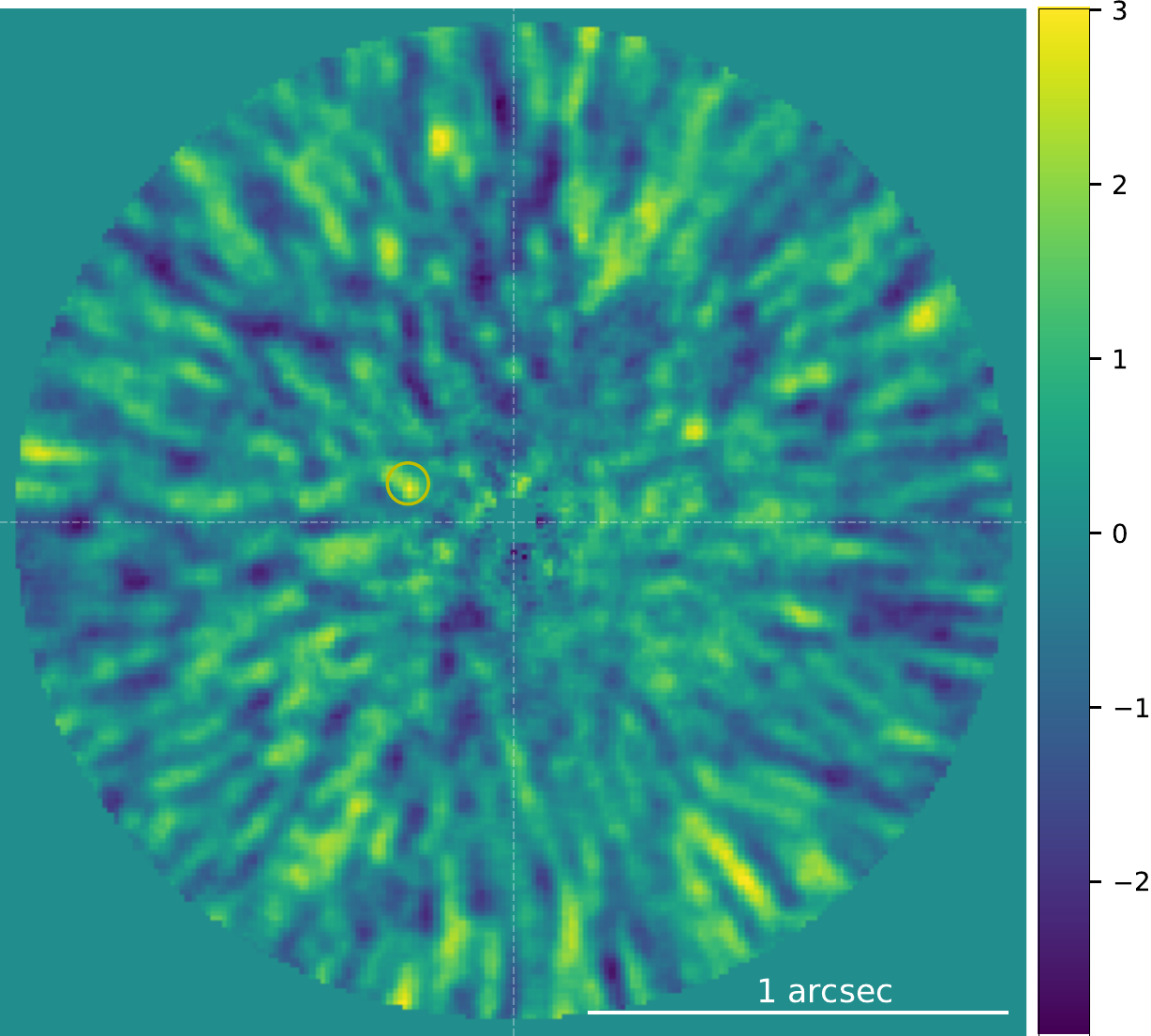}}\\
\subfloat[HD114082 RSM Bottom-up]{\includegraphics[width=115pt]{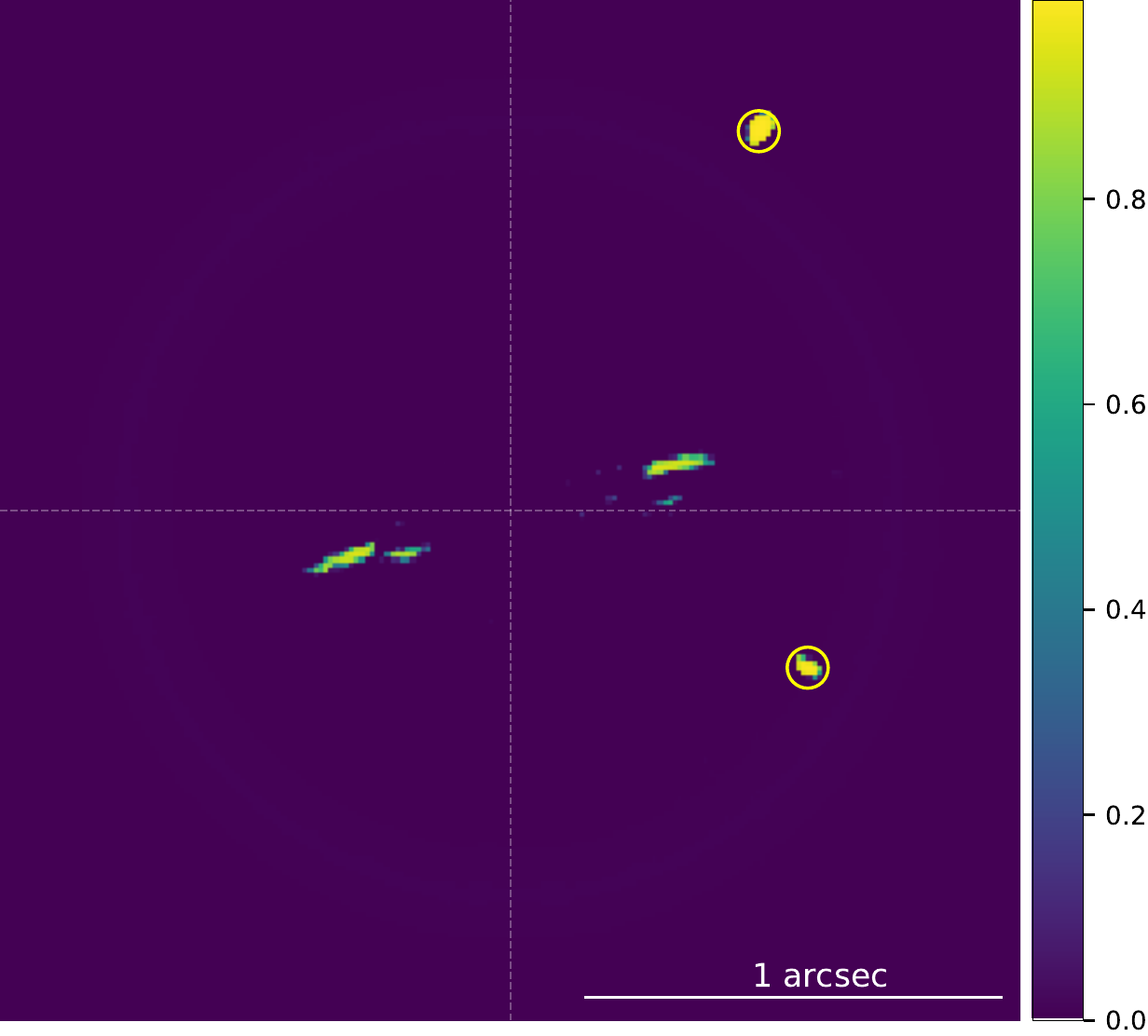}}
          \subfloat[HD114082 RSM Top-down]{\includegraphics[width=115pt]{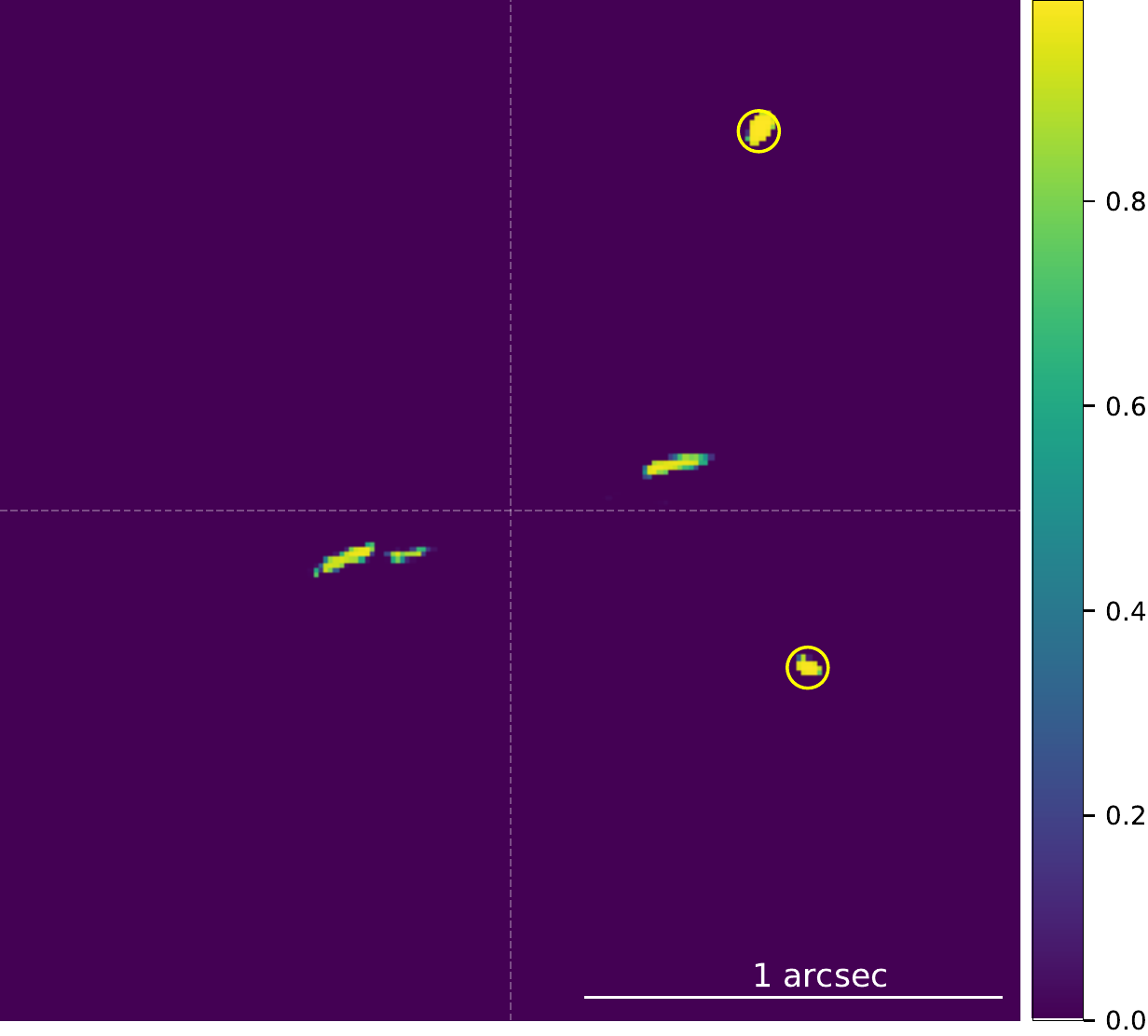}}
    \subfloat[HD114082 optimised S/N]{\includegraphics[width=115pt]{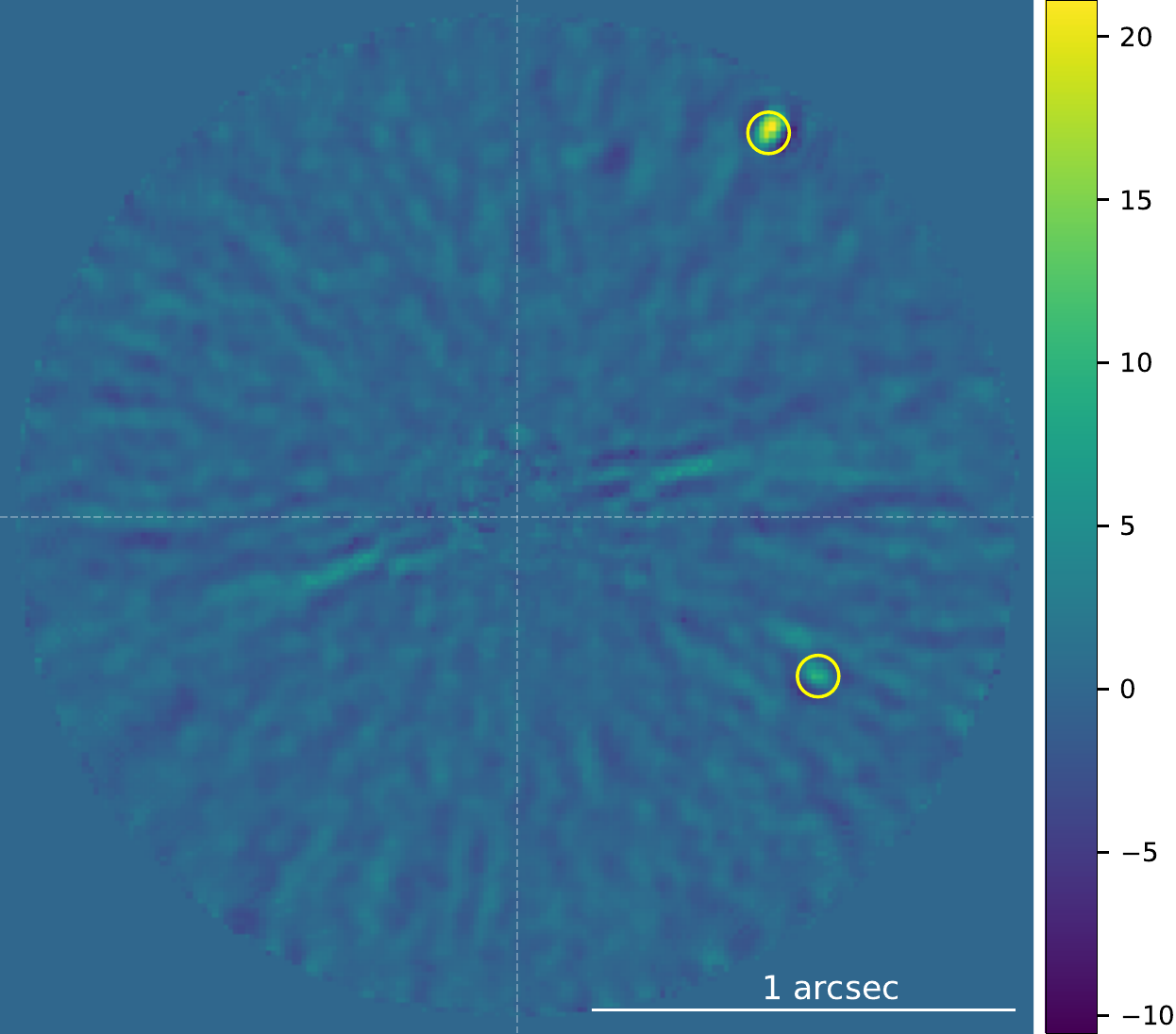}}
        \subfloat[HD114082 average S/N]{\includegraphics[width=115pt]{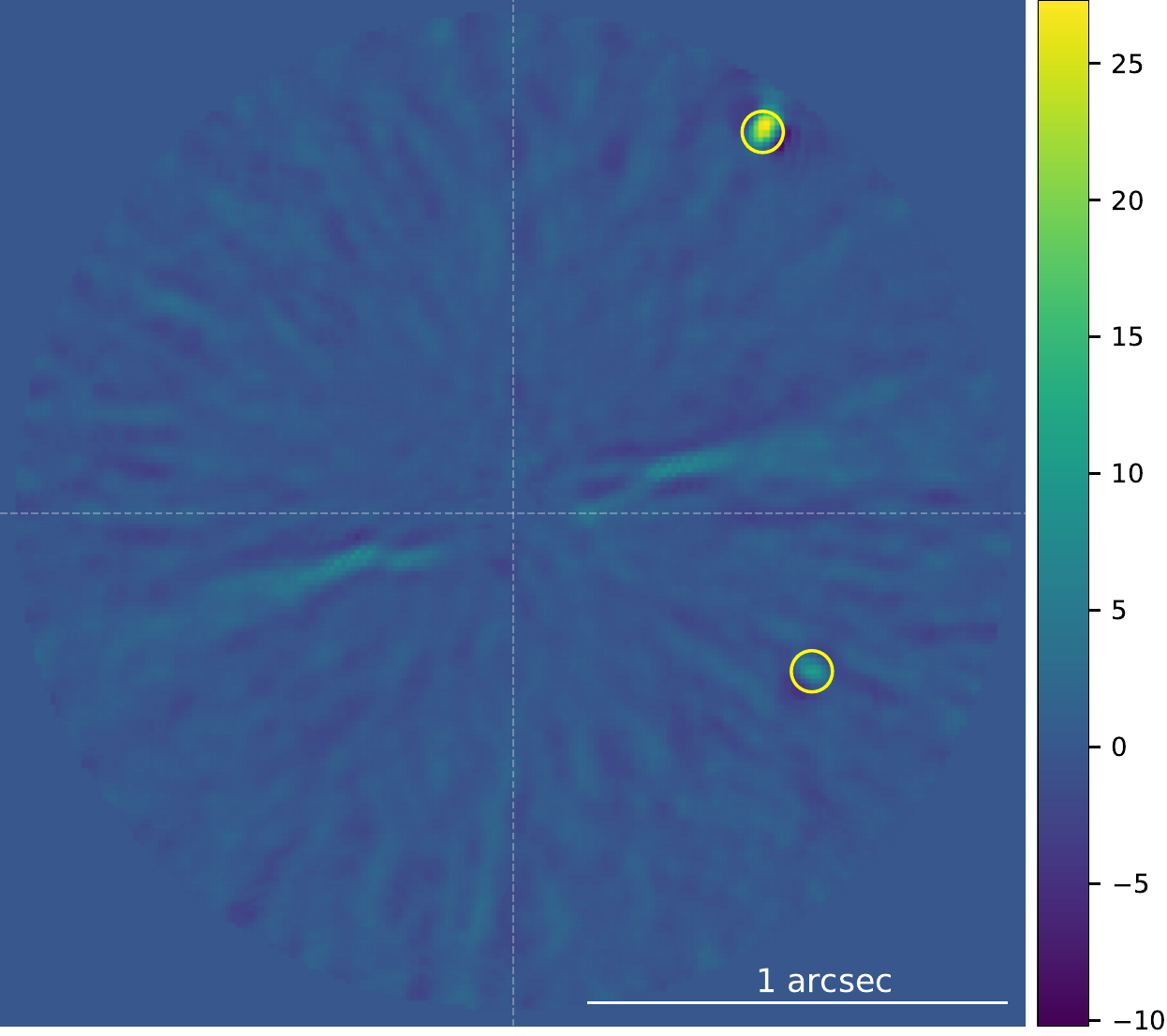}}\\

  \caption{\label{Target_map1} RSM detection maps generated using the Auto-RSM with the bottom-up (left) and top-down (middle left) approaches to select the optimal set of likelihoods cubes (each likelihood cube corresponds to a PSF-subtraction technique), S/N maps generated using the Auto-SNR to select the optimal set of S/N maps (middle right) and S/N maps obtained by averaging the S/N map generated with APCA, NMF, LLSG, and LOCI (right). The expected position of the planetary signal as estimated via the RSM based planetary signal characterisation algorithm is indicated by a yellow circle.}
\end{figure*}
\FloatBarrier

\section{Determination of projected angular separation}
\label{Orbit}

The projected angular separation is computed based on randomly generated orbital elements and on the predefined semi-major axis, relying on Keplerian motion. We first estimate the true anomaly, which is defined as the angle between the direction of the periapsis and the current position vector of the body in the perifocal plane. Its estimation starts by the definition of the mean anomaly, which provides the fraction of the elliptical orbit that was covered since the periapsis expressed in radian $[0,2\pi]$. The mean anomaly is linked to the eccentric anomaly by the following relationship:
\begin{eqnarray}
M=\frac{2\pi}{T} t=E-e\sin(E)\;,
\end{eqnarray}
with $T$ the orbital period and $e$ the eccentricity.
This transcendental equation relating time and eccentric anomaly cannot be directly solved. However, there exists a unique solution for every value of the mean anomaly $M$. We rely on the expansion of $E$ in terms of Bessel functions to relate eccentric anomaly and mean anomaly \citep{Curtis}.
\begin{eqnarray}
E= M + \sum^{ \infty }_{n=1} \frac{2}{n}J_n (ne)\sin(nM)\;,
\end{eqnarray}
with $J_n(x)$ the Bessel function of the first kind. The sum over $n$ is truncated to $N=100$. The true anomaly $\theta$ is then computed via the following relationship:
\begin{eqnarray}
\theta= 2\tan^{-1} \left( \sqrt{\frac{1+e}{1-e}}\tan(\frac{E}{2})\right) \;,
\end{eqnarray}
Once the true anomaly has been estimated, the position vector in the perifocal frame is computed using the elliptic orbit equation:
\begin{eqnarray}
\bm{r}_p= \frac{h^2}{\mu}\frac{1}{1+e\cos(\theta)}(\cos(\theta)\bm{\hat{p}}+\sin(\theta)\bm{\hat{q}})\;,
\end{eqnarray}
where the coordinates are normalised, such as $\hat{p}=[1,0,0]$ and $\hat{p}=[0,1,0]$. Using $h=\sqrt{\mu a ( 1-e^2)}$, we get:
\begin{eqnarray}
\bm{r}_p= \frac{a ( 1-e^2)}{1+e\cos(\theta)}(\cos(\theta)\bm{\hat{p}}+\sin(\theta)\bm{\hat{q}})\;.
\end{eqnarray}
We project this position vector in the equatorial frame via three Euler rotations:
\begin{eqnarray}
\bm{r}_e= \left[ \bm{Q} \right] \bm{r}_p\;,
\end{eqnarray}
with the Euler rotations given by:
\begin{eqnarray}
\left[ \bm{Q} \right] = \left[ \bm{R}_3(w) \right] \left[ \bm{R}_1(i) \right] \left[ \bm{R}_3(\Omega) \right]\;,
\end{eqnarray}
where $i$ is the inclination, $w$ the argument of the periapsis, and $\Omega$ the longitude of the ascending node. The normalised distance to the star is then obtained by computing the norm of the position vector in the equatorial frame:
\begin{eqnarray}
r = \frac{}{} \Vert \bm{r}_e \Vert\;.
\end{eqnarray}
The angular separation expressed in mas is finally defined as the normalised distance to the star multiplied by the semi-major expressed in mas:
\begin{eqnarray}
a_{sep} = r a\frac{1000\times 3600 \times 180}{(206265 \pi d) }\;, 
\end{eqnarray}
with $a$ the semi-major axis expressed in au and $d$ the distance from the star expressed in pc.\\

\newpage
 
\onecolumn
\section{Disks analysed in \citet{Pearce22}}
\label{app_disks}

\noindent There are 21 targets in common between the SHARDDS sample analysed in this paper and the sample of \citet{Pearce22}. We do not consider here Fomalhaut C, part of SHARDDS and in \citet{Pearce22}  because of the very poor quality of the data.  We present these targets in Table \ref{tab_pearce}, with the location of the disk inner radius used in the analysis by \citet{Pearce22} to estimate the planet minimum masses.

\begin{table*}[h]
                        \caption{The 21 common targets between SHARDDS and the sample analysed in \citet{Pearce22}. This table is an extract from Table A.1 in \cite[][see references therein]{Pearce22}}
                        \label{tab_pearce}
\centering

                        \begin{tabular}{lcc}
                        
                        \hline
Target & Disk data  & Disk location and extent$^1$ \\                           
  \hline
HD203 & SED & $29 \pm 6$ \\
HD377 & SED & $60 \pm 10$ \\
HD3003 & SED  & $21 \pm 6$ \\
HD3670 & SED  & $100 \pm 20$ \\
HD9672 & ALMA & $62 \pm 4 \rightarrow 210 \pm 4$ \\
HD10472 & SED & $110 \pm 20$ \\
HD13246 & SED  & $80 \pm 30$ \\
HD16743 & Herschel 100 $\mu$m  & $50 \pm 50 \rightarrow 260 \pm 70$  \\
HD21997 & ALMA  & $68 \pm 4 \rightarrow 120 \pm 4$  \\
HD25457 & SED & $45 \pm 8$ \\
HD37484 & SED & $70 \pm 20$ \\
HD38206 & ALMA & $0 \pm 20, 140^{+30} \rightarrow 190 \pm 30,  320^{+50}$ \\
HD69830 & SED & $0.8 \pm 2$ \\
HD107649 & SED & $15 \pm 3$ \\
HD114082 & SED  & $29 \pm 6$ \\
HD135599 & SED  & $49 \pm 9$ \\
HD172555 & SED & $15 \pm 3$ \\
HD181296 & SED & $81 \pm 10$ \\
HD192758 & Herschel 100 $\mu$m  & $40 \pm 40 \rightarrow 180 \pm 50$ \\
HD218340 & SED & $140 \pm 40$ \\
HD221853 & SED & $47 \pm 9$ \\
\end{tabular}
\tablefoot{
\tablefoottext{1}{The 'Disc location and extent' column describes the location and shape of the debris disc inner and outer edges: if the disc is resolved and fitted with an asymmetric model (case of HD38206), then the column shows the inner edge pericentre, $q_i$, inner edge apocentre, $Q_i$, outer edge pericentre, $q_o$, and outer edge apocentre, $Q_o$, as '$q_i$, $Q_i$ $\rightarrow$ $q_o$, $Q_o$'. Alternatively, if the disc is resolved and fitted with an axisymmetric model, then the column shows the disc inner edge, $a_i$, and outer edge, $a_o$, as '$a_i$ $\rightarrow$ $a_o$'. Finally, if the disc location is estimated from SED data, then only the corrected blackbody radius is shown.}}
                                \end{table*}

\end{appendix}

\end{document}